\documentclass[aip,apl,twocolumn,showpacs,amsmath,amssymb,superscriptaddress,reprint,floatfix]{revtex4-1}
\usepackage{siunitx, graphicx, bm, color, subfigure, dcolumn, threeparttable, booktabs, array, appendix}
\definecolor{lcolor}{RGB}{150,64,163}
\usepackage[colorlinks=true, linkcolor={lcolor}, citecolor={lcolor}, urlcolor={lcolor}]{hyperref}

\allowdisplaybreaks[4]
\usepackage[all]{hypcap}
\newcommand*{\figref}[2][]{%
  \hyperref[{#2}]{%
    Fig.~\ref*{#2}%
    \ifx\\#1\\%
    \else
      \,(#1)%
    \fi
  }%
}

\begin{document}
\title{Coherent and Dissipative Cavity Magnonics}

\author{M. Harder}
\email{mharder14@bcit.ca}
\affiliation{Department of Physics, British Columbia Institute of Technology, Burnaby V5G 3H2, Canada}
\author{B. M. Yao}
\affiliation{State Key Laboratory of Infrared Physics, Shanghai Institute of Technical Physics, Chinese Academy of Sciences, Shanghai 200083, People's Republic of China}
\author{Y. S. Gui}
\affiliation{Department of Physics and Astronomy, University of Manitoba, Winnipeg R3T 2N2, Canada}
\author{C.-M. Hu}
\affiliation{Department of Physics and Astronomy, University of Manitoba, Winnipeg R3T 2N2, Canada}

\begin{abstract}

Strong interactions between magnetic materials and electrodynamic cavities mix together spin and photon properties, producing unique hybridized behaviour.  The study of such coupled spin-photon systems, known as cavity magnonics, is motivated by the flexibility and controllability of these hybridized states for spintronic and quantum information technologies.  In this tutorial we examine and compare both coherent and dissipative interactions in cavity magnonics.  We begin with a familiar case study, the coupled harmonic oscillator, which provides insight into the unique characteristics of coherent and dissipative coupling.  We then examine several canonical cavity magnonic systems, highlighting the requirements for different coupling mechanisms, and conclude with recent applications of spin-photon hybridization, for example, the development of quantum transducers, memory architectures, isolators and enhanced sensing.  
\end{abstract}

\pacs{}

\maketitle
\section{Introduction}
Hybrid systems coupling magnetic excitations to other degrees of freedom are widely used in modern magnetism research.  For example, hybridization enables mutually exclusive states of matter in ferromagnetic-superconducting devices, \cite{Lyuksyutov2005} and enhances the functionality of magnetic semiconductors. \cite{Zakharchenya2005, Maccherozzi2008}  In the field of cavity magnonics coupling between magnons (quantized spin waves) and electrodynamic fields results in hybridized states with a dual spin-photon nature.  As in other hybrid platforms, emergent properties enable new functionality that is unavailable in the uncoupled sub-systems.  Applications of cavity magnonics include memory architectures, \cite{Zhang2015g} non-local spin control, \cite{Bai2017, Xu2019} magnon and photon sensing using exceptional points, \cite{Tserkovnyak2020, Yang2020a, Yu2020b} quantum sensing,\cite{LachanceQuirion2020, Wolski2020} quantum transduction, \cite{Tabuchi2014, Hisatomi2016, Lauk2019} optical to microwave frequency conversion \cite{Hisatomi2016, Lambert2019} and broadband, high isolation non-reciprocity. \cite{Wang2019b, Zhang2020a} Furthermore, cavity magnonics research has provided insight into coupled systems, e.g., the role of dissipative coupling \cite{Harder2018a, Boventer2019, Yao2019b} and exceptional points. \cite{Harder2017, Zhang2017a, Zhang2018b, Zhang2019, Tserkovnyak2020}

Cavity magnonics began with Soykal and Flatt\'{e}'s 2010 prediction of a large, quantum-coherent, magnon-photon interaction in a ferromagnetic nanomagnet \cite{Soykal2010} and the subsequent experimental observation of hybridization at low temperatures by Huebl et al. in 2013. \cite{Huebl2013a}$^{,}$\footnote{Though in hindsight coherent spin-photon coupling can be traced back at least as early as the work of Artman and Tannenwald in 1953 \cite{Artman1953}} Following these discoveries the theoretical and experimental foundations of coherent magnon-photon coupling were established between $\approx$ 2013 - 2017.  Some key early developments include the investigation of strong coupling in the quantum limit, \cite{Tabuchi2014a} the coupling of magnons and a superconducting qubit, \cite{Tabuchi2014} observation of room temperature hybridization, \cite{Zhang2014} the electrical detection of hybridization, \cite{Bai2015} investigations of optomagnonics, \cite{Zhang2015b} and the demonstration of optical to microwave conversion. \cite{Hisatomi2016}  More detail on early cavity magnonic development is available in the reviews of Refs. \citenum{Harder2018, Maksymov2018, LachanceQuirion2019, Li2020, Hu2020, Zhang2021, Barman2021}.

Following the early advances of coherent cavity magnonics, the field grew rapidly between 2017 - 2020.  Amongst the important results of this period were the observation of dissipative coupling, \cite{Harder2018a} the realization of strong coupling in nanoscale ferromagnetic metals, \cite{Hou2019, Li2019a} the exploration of exceptional points and the role of $\mathcal{PT}$ symmetry, \cite{Harder2017, Zhang2017a, Zhang2019, Cao2019, Yuan2020c} experimental studies of strong coupling in antiferromagnets,\cite{Yuan2017, Bialek2019, Everts2019} and the advancement of quantum magnonics, enabling, for example, the quantum sensing of magnons.\cite{LachanceQuirion2020, Wolski2020} These discoveries opened many new doors for cavity magnonics, which should provide years of fruitful discovery.  An outlook will be provided in \autoref{sec:outlook}, and for additional perspectives on the future of cavity magnonics, see Refs. \citenum{LachanceQuirion2019, Hu2020} and \citenum{Barman2021}.

Cavity magnonics is inherently diverse, taking inspiration from spintronics, hybrid quantum systems, optomechanics and optomagnonics.  This multifaceted nature is a strength, but can also be daunting to non-experts.  Fortunately there is an insightful, universal system which clarifies many properties of spin-photon hybridization: the harmonic oscillator.  In the hope of making cavity magnonics accessible to a broad audience, we begin this tutorial with an examination of coherent and dissipative coupling in the harmonic oscillator.  Using the insights gained from these case studies we then examine key features of spin-photon hybridization, taking a detailed look at canonical cavity magnonic devices and applications.  For further related reading we recommend the following review resources by topic: Coherent Cavity Magnonics: Refs. \citenum{Harder2018, Maksymov2018} and \citenum{Zhang2021}; Quantum Magnonics: Ref. \citenum{LachanceQuirion2019}; Dissipative Cavity Magnonics: Ref. \citenum{Wang2020}; Future Directions and Applications: \citenum{Hu2020, Li2020} and \citenum{Barman2021}; Optomagnonics: Refs. \citenum{Kusminskiy2019} and \citenum{KusminskiyBook}; Optomechanics: Refs. \citenum{Aspelmeyer2014} and \citenum{AspelmeyerBook}; Quantum Optics/Cavity QED: Refs. \citenum{HarocheBook} and \citenum{WallsBook}; Cavity Magnonics in the Context of Hybrid Quantum Systems: Ref. \citenum{Clerk2020}; Cavity Magnonics in the Context of Magnonics: Ref. \citenum{Barman2021}.

\begin{widetext}

\begin{figure} [!h]
\begin{center}
\includegraphics[width=15cm]{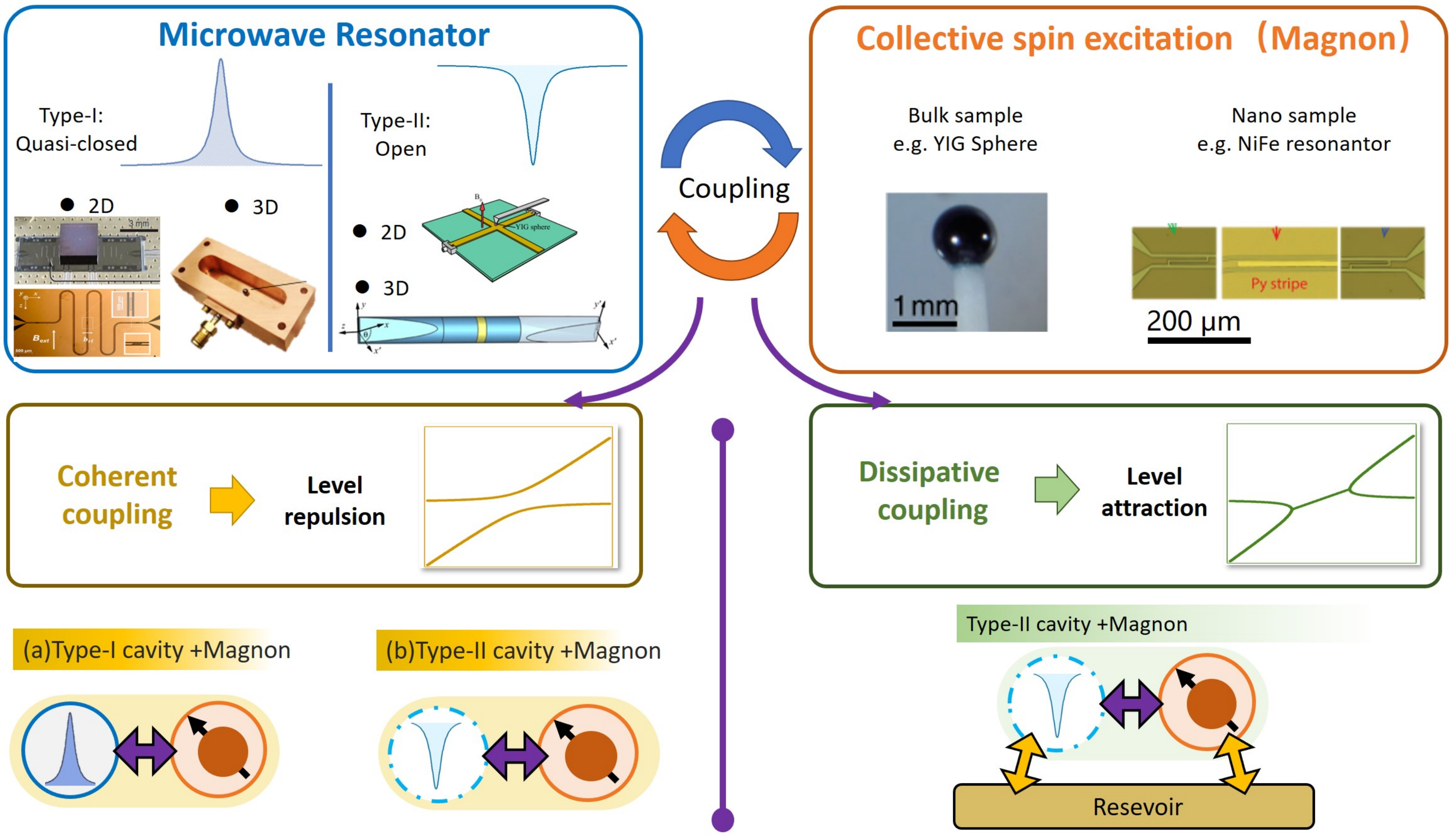}
\caption{Cavity magnonics studies hybridized spin-photon states formed by the coupling of magnetic excitations to light.  Hybridization can be realized using either quasi-closed or open microwave resonators coupled to either bulk or nanostructured magnetic devices.  In a quasi-closed resonator, such as a 3D cavity or a gap-coupled resonator, the intrinsic dissipation (e.g., conductive losses and magnon damping), is much larger than extrinsic dissipation (e.g., radiative losses).  In an open resonator, such as a side-coupled resonator, extrinsic dissipation is comparable to, and often larger than, the intrinsic dissipation.   Two unique forms of hybridization, level repulsion, due to coherent coupling, and level attraction, due to dissipative coupling, lead to the rich physics and applications of cavity magnonics.  This figure summarizes the most common platforms used to achieve coherent and dissipative hybridization.  Although both forms of coupling generally coexist in the same system, coherent coupling typically dominates in quasi-closed resonators, and the tunability of open resonators is often used to suppress coherent coupling, thus highlighting dissipative behavior.  In this figure, the peaks and dips indicate the transmission spectra generally observed in each platform.  Reproduced with permission from Hu, in \emph{Solid State Physics}, edited by R. L. Stamps and B. Camley, Vol. \textbf{71}, Chap. 4, pp. 117-121.  Copyright 2020 Elsevier.}
\label{fig:overview}
\end{center}
\end{figure}

\end{widetext}
\subsection{What is Cavity Magnonics?}
Cavity magnonics studies the emergent physics of magnetically ordered materials coupled to light, with two broad objectives: (1) Gain physical insight into the physics of hybridized systems; and (2) Exploit unique emergent behaviour for technological development, focusing on spintronics, magnonics and hybrid quantum systems.  Really, only two components are required for a cavity magnonic experiment: (1) A magnetically ordered material; and (2) An electrodynamic cavity.  Therefore a basic cavity magnonic system is actually quite simple, typically consisting of the ferrimagnetic insulator yttrium-iron-garnet (YIG), mounted inside a microwave cavity, and probed via microwave transmission/reflection measurements.  The widespread use of YIG is due to its characteristically small damping rate and the commercial availability of large ($V_s > 1 \text{ mm}^2$, number of spins, $N_s > 10^{17}$), high quality (Gilbert damping, $\alpha_G \approx 1 \times 10^{-4}$), low cost ($\approx 100$ USD) samples.  However the cavity magnonic platform has diversified rapidly.  For example, strong coupling has now been achieved using the ferromagnetic metal permalloy, important for nanoscale spintronic development, \cite{Hou2019, Li2019a} and several antiferromagnets. \cite{Yuan2017, Everts2019, Bialek2019, Shi2020}  From the cavity perspective, all that is required is a well defined resonance;\footnote{If plane electromagnetic waves are used, bulk magnetic polaritons will be produced. See, for example, Refs. \citenum{Mills1974} and \citenum{AlbuquerqueBook}} a variety of unique 3D cavity designs \cite{Zhang2014, Tabuchi2014a, Goryachev2014, Haigh2015b, MaierFlaig2016, Harder2016b, Castel2018} and 2D resonators \cite{Huebl2013a, Bhoi2014, Yao2017, Harder2018, Rao2020} have been used.  Often cavity modes with quality factor $Q \gtrsim 1000$ are desired to help achieve high cooperativity.  However this is an application specific requirement, not a necessity of cavity magnonics.  In fact, the large extrinsic damping of open systems is interesting in its own right as we will see in \hyperref[sec:basePendulums]{Secs. IIC} and \ref{sec:physicsTraveling}.  

\hyperref[fig:overview]{Figure 1} summarizes the resonator and sample configurations often used in cavity magnonics.  We note that throughout this tutorial we will use the words cavity and resonator interchangeably, and refer to quasi-closed and open resonators.  In a quasi-closed resonator such as a 3D cavity or a gap-coupled resonator, the intrinsic dissipation (e.g., conductive losses and magnon damping), is much larger than extrinsic dissipation (e.g., radiative losses).  In an open resonator, such as a side-coupled resonator, extrinsic dissipation is comparable to, and often larger than, the intrinsic dissipation.  There are two unique forms of spin-photon coupling, coherent coupling, characterized by level repulsion, and dissipative coupling, characterized by level attraction.  In general hybridized states will be formed by a mixture of coherent and dissipative spin-photon coupling.  However, in quasi-closed resonators, coherent coupling generally dominates, while the tunability of open resonators allows one to suppress coherent coupling in order to study and use purely dissipative behavior.  \autoref{fig:overview} schematically summarizes the most common platforms used to achieve coherent and dissipative coupling.  Here, the peaks and dips indicate the transmission spectra typically observed in quasi-closed and open resonators, respectively.   

Finally, although microwave transmission/reflection measurements are commonly used to probe the cavity magnonic system, other experimental techniques have also proven useful, such as the electrical detection of hybridized spin current \cite{Bai2015, MaierFlaig2016} and inelastic scattering measurements via Brillouin light-scattering. \cite{Klingler2016}  Experimental methods are discussed in more detail in \hyperref[sec:physics]{Secs. III} and \ref{sec:applications}.
\subsection{Coherent vs. Dissipative Coupling}
Coherent coupling in cavity magnonics is characterized by level repulsion in the dispersion and an attraction in the linewidth.  The term coherent references the fixed spin-photon phase relationship in the hybridized modes that results from the dipole interaction between the spin ensemble and cavity magnetic field.  This was the first form of coupling discovered in cavity magnonics and plays an important role in applications involving transduction, allowing efficient energy exchange between spin and photon.  In the time-domain, this spin-photon energy exchange is evidenced by Rabi oscillations which are often encountered in the context of hybrid quantum systems. \cite{WallsBook, Zhang2014}  Importantly, however, since this general behaviour applies to both classical and quantum systems, we can use the classical example of coupled pendulums to understand certain key features.

Dissipative coupling on the other hand is characterized by level attraction in the dispersion and a repulsion in the linewidth.  This indirect coupling is mediated through a reservoir leading to a complex effective spin-photon coupling which provides an additional source of dissipation, hence the terminology.  Dissipative coupling has important applications to non-reciprocal transport and enhanced sensing techniques.  More generally, dissipative systems are an important, and widespread, example of non-Hermitian physics, \cite{Ashida2020} commonly discussed in the context of open quantum systems.  However again, many common characteristics of non-Hermiticity appear in both classical and quantum systems alike, allowing us to gain insight into dissipative behaviour from the familiar example of coupled pendulums.

Generally, in cavity magnonics, both coherent and dissipative coupling will occur simultaneously.  Yet, as summarized above, dissipative and coherent coupling are distinctly different and produce different physical phenomena.  For this reason we will introduce the key properties of each form of coupling separately, through a variety of model systems, in \hyperref[sec:harmonic]{Secs. II} and \ref{sec:physics}.  In short, we will see that coherently coupling is an intrinsic interaction that is independent of dissipation; i.e., coherent coupling survives even in ideal systems without dissipation.  In contrast, base-mediated dissipative coupling stems from the correlated dissipation of the individual oscillators.  Therefore, the dissipation is precisely the origin of the base-mediated coupling.  With both interactions present, a strength of the cavity magnonic platform is its tunability.  As we will see in \autoref{sec:physics}, cavity magnonic systems can be engineered so that either coherent or dissipative coupling dominates.  Each form of coupling has unique properties, and, therefore, leads to unique applications.  For example, coherent coupling is typically used for transduction applications,\cite{Tabuchi2015, Hisatomi2016} while dissipative coupling may be used for enhanced sensing techniques.\cite{Cao2019, Zhang2019}  Furthermore, the interplay of coherent and dissipative coupling is also important, for example in enabling non-reciprocal microwave transmission. \cite{Wang2019b}
\subsection{Strong Coupling}
Coupling strength and damping both play an important role in cavity magnonics.  In the strongly coupled regime the coupling rate $J$ greatly exceeds the magnetic and cavity loss rates, $\alpha$ and $\beta$, respectively.  This is best characterized by the cooperativity, $C = J^2/(\alpha \beta) > 1$. \footnote{$J, \alpha$ and $\beta$ may be measured in units of frequency, or defined as dimensionless quantities by normalizing to the cavity frequency, $\omega_c$.  In the latter case, $\alpha$ would be the Gilbert damping parameter and $\beta = 1/2Q$ is related to the cavity quality $Q$.}  $C$ can be used to characterize both coherent and dissipative interactions although the physical meaning is different.  In the case of coherent coupling $C > 1$ heuristically means that information can be transferred between the magnon and photon faster than it is lost from either system.  However, for dissipative interactions, it indicates that the reservoir coupling exceeds the intrinsic loss rates.  The high controllability of $J$ and $\beta$ allows strong coupling to be achieved in many ways, for example via high $Q$ cavities, \cite{Huebl2013a, Zhang2014, Tabuchi2014, Yao2017} large spin numbers, \cite{Zhang2014, Bourhill2015} or a large mode filling factor. \cite{Hou2019, Li2019a}

The most striking signature of strong coupling is a modified spin-photon dispersion relation.  At low cooperativity the eigenfrequencies of the spin-photon system are the magnetic resonance frequency $\omega_m$, and the cavity frequency $\omega_c$.  Experimentally, $\omega_m$ can be controlled by an external bias magnetic field $H$ (think ferromagnetic resonance) while $\omega_c$ is typically fixed by the cavity geometry.  However, $\omega_c$ is independent of $H$ and $\omega_m$ is independent of $\omega_c$.  Therefore, there is a degeneracy at, and only at, $\omega_c = \omega_m$, where the eigenfrequencies match but the eigenmodes are still independent.  At large cooperativities, $\approx C > 1$, this is no longer true.  Instead, in the case of a coherent interaction, level repulsion is observed, characterized by an avoided crossing in the dispersion and a crossing in the damping.  On the other hand, dissipative coupling results in level attraction, with a merging of the two dispersion branches and a gap in the damping.  A more detailed discussion of the coupling signatures can be found in \autoref{sec:physics}.

\section{Insights From the Harmonic Oscillator}\label{sec:harmonic}
In this section, we examine the canonical system of coupled harmonic oscillators, highlighting the characteristics of coherent and dissipative coupling with an eye on behaviour observed in cavity magnonics.  Coupled harmonic oscillators serve as an instructive toy model in which we can easily examine the influence of nonlinear forces, damping and coupling, and which can be fully solved using simple mathematical techniques.  To explore the range of coupling behaviour observed in cavity magnonics we consider three case studies: (1) Spring coupled pendulums, demonstrating coherent coupling; (2) Dashpot coupled pendulums, demonstrating direct dissipative coupling; and (3) Base-mediated coupled pendulums, highlighting the role of a reservoir in the dissipative interaction.
\subsection{Spring Coupled Pendulums: A Case-Study in Coherent Coupling} \label{sec:spring}

\begin{figure} [!t]
\begin{center}
\includegraphics[width=8cm]{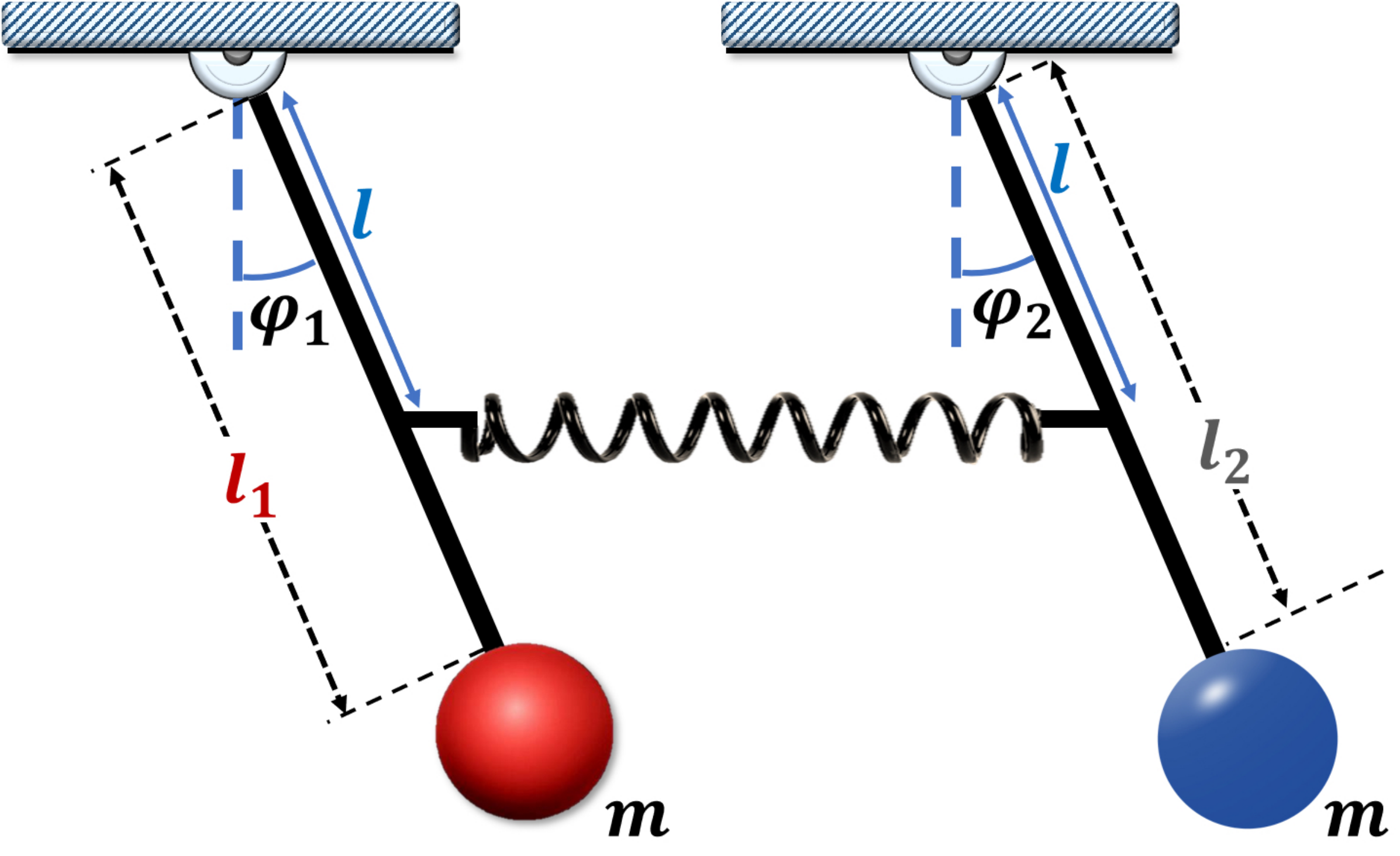}
\caption{Spring coupled simple pendulums.  This system is characterized by coherent coupling of the oscillations, which results in level repulsion of the hybridized modes.}
\label{fig:coherentPendulum}
\end{center}
\end{figure}

Interacting harmonic oscillators have long been used to model strong coupling and the physics of polaritons. \cite{Huang1951, Mills1974, AlbuquerqueBook}  For the cavity magnonic system this approach is closely connected to RLC circuit models,\cite{Alzar2001, Bhoi2014, Tay2018} often used in microwave engineering, \cite{PozarBook} and to the Tavis-Cummings model, \cite{Tavis1968, Garraway2011} as we will explore in \autoref{sec:coherent}.  To establish an oscillator model of strong coupling, consider two equal mass ($m$), spring coupled, simple pendulums as illustrated in \autoref{fig:coherentPendulum}.  The pendulums are connected to fixed pivot points by massless rods of length $l_{1}$ and $l_2$ respectively, and therefore each pendulum has a unique uncoupled oscillation frequency, $\omega_{1,2}=\sqrt{g/l_{1,2}}$ ($g$, the acceleration of gravity).  A  massless spring (spring constant $k$) connects the pendulums a distance $l$ from the pivot points, resulting in coherent coupling between the two pendulums, as discussed more in \hyperref[Sec:appendixB]{Appendix B}.

The equations of motion for the spring coupled pendulums are most easily solved in a reference frame that rotates at the average frequency $\omega_{r\!e\!f}=(\omega_1+\omega_2)/2$.  As detailed in \hyperref[Sec:appendixA]{Appendices A} and \ref{Sec:appendixB}, the complex eigenfrequencies of the hybridized modes are found to be
\begin{align}\label{Eq:eigenfrequency}
\widetilde{\omega}_{\pm} - \omega_{r{\!}e{\!}f}&=\dfrac{1}{2}[(J_1-i\lambda_1)+(J_2-i\lambda_2)]\nonumber\\
&\pm\frac{1}{2}\sqrt{[\Delta-(J_1-i\lambda_1)+(J_2-i\lambda_2)]^2+4J_1J_2},
\end{align}
where $\lambda_{1,2}$ are the intrinsic damping rates of the pendulums and $J_{1,2}=kl^2/(2m\omega_{1,2}l_{1,2}^2)$ is the coherent coupling strength. The influence of coupling is most noticeable near zero frequency detuning, $\Delta = \omega_1 - \omega_2 \approx 0$, in which case $J = J_1 \approx J_2$, and therefore it is convenient to write,
\begin{equation}
\widetilde{\omega}_{\pm} \approx\dfrac{1}{2}\left[\widetilde{\omega}_1 + \widetilde{\omega}_2+2J
\pm\sqrt{(\widetilde{\omega}_1-\widetilde{\omega}_2)^2+4J^2}\right]. \label{eq:eigenvalue_spring}
\end{equation}
Here, $\widetilde{\omega}_{1,2} = \omega_{1,2} - i\lambda_{1,2}$ denotes the complex oscillation frequencies of the uncoupled oscillators, including both the frequency $\omega_{1,2}$, and linewidth $\lambda_{1,2}$.  Often, it is convenient to write Eq. \eqref{eq:eigenvalue_spring} in the form $\widetilde{\omega}_{\pm} = \omega_{\pm} - i \Delta \omega_{\pm}$, where $\text{Re}\left(\widetilde{\omega}_{\pm}\right) = \omega_\pm$ and $-\text{Im}\left(\widetilde{\omega}_{\pm}\right) = \Delta\omega_\pm$ are real valued functions which describe the hybridized oscillation frequency and linewidth respectively. \footnote{Analytic expressions for $\omega_{\pm}$ and $\Delta \omega_{\pm}$ can be found in Appendix A of Ref. \citenum{Harder2018}.}

\hyperref[fig:springResult]{Figures 3 (a)} and \hyperref[fig:springResult]{3 (b)}, respectively, show the hybridized frequency and linewidth as a function of the detuning $\Delta$, plotted according to Eq. \eqref{eq:eigenvalue_spring} in the strong coupling regime where $C = J^2/\left(\lambda_1 \lambda_2\right) > 1$.  The horizontal and diagonal dashed lines in panel (a) indicate the uncoupled oscillation frequencies $\omega_1$ and $\omega_2$, respectively, while the horizontal dashed lines in panel (b) are the uncoupled dissipation rates, $\lambda_{1,2}$ with $\lambda_2 > \lambda_1$.  A blue-shifting of the oscillation frequencies is evident in panel (a) by the asymptotic behaviour at large $\Delta$. However, more strikingly, the pendulums are strongly hybridized in both the frequency and linewidth, with the new eigenfrequencies deviating significantly from their uncoupled values.

In the frequency dispersion, hybridization produces level repulsion, characterized by an avoided crossing between the upper ($\omega_+$) and lower ($\omega_-$) branches.  The deviation from the unhybridized behaviour is greatest at $\Delta = 0$ where the Rabi-like gap \cite{WallsBook, Zhang2014} between the upper and lower branches is directly proportional to the coupling strength, $\omega_+ - \omega_- = 2J$.  The coherent nature of these modes is also most easily identified at $\Delta = 0$, where the lower branch, $\omega_- \left(\Delta = 0\right) =\omega_1$, corresponds to an in-phase oscillation of the two pendulums,\footnote{When the two pendulums move in-phase at zero detuning ($\Delta = 0; \omega_1 = \omega_2$) there is no coupling force between them, and therefore, the lower normal mode must oscillate at the natural frequency of the pendulums.  Therefore, the lower normal mode lies above the horizontal dashed line in \figref[a]{fig:springResult}.  The coupling term in cavity magnonics is subtly different, leading to one normal mode with a lower frequency at $\Delta = 0$. This is analogous to bonding and antibonding states \cite{ThorntonClassicalBook, GoldsteinBook}} while the upper branch, $\omega_+ \left(\Delta = 0\right) = \omega_1+2J$, corresponds to a 180$^\circ$ out-of-phase motion.\footnote{For a derivation of these eigenmodes see Appendix C of Ref. \citenum{Harder2018}.}  Physically, the coupling between the two pendulums has broken the degeneracy at $\Delta = 0$, raising the energy of the out-of-phase, upper branch.  The scale of the degeneracy breaking is set by the coupling strength, which enables coherent energy exchange between the two pendulums.  This is most apparent in the time domain, as discussed later in this section.

\begin{figure} [t]
\begin{center}
\includegraphics[width=8.5cm]{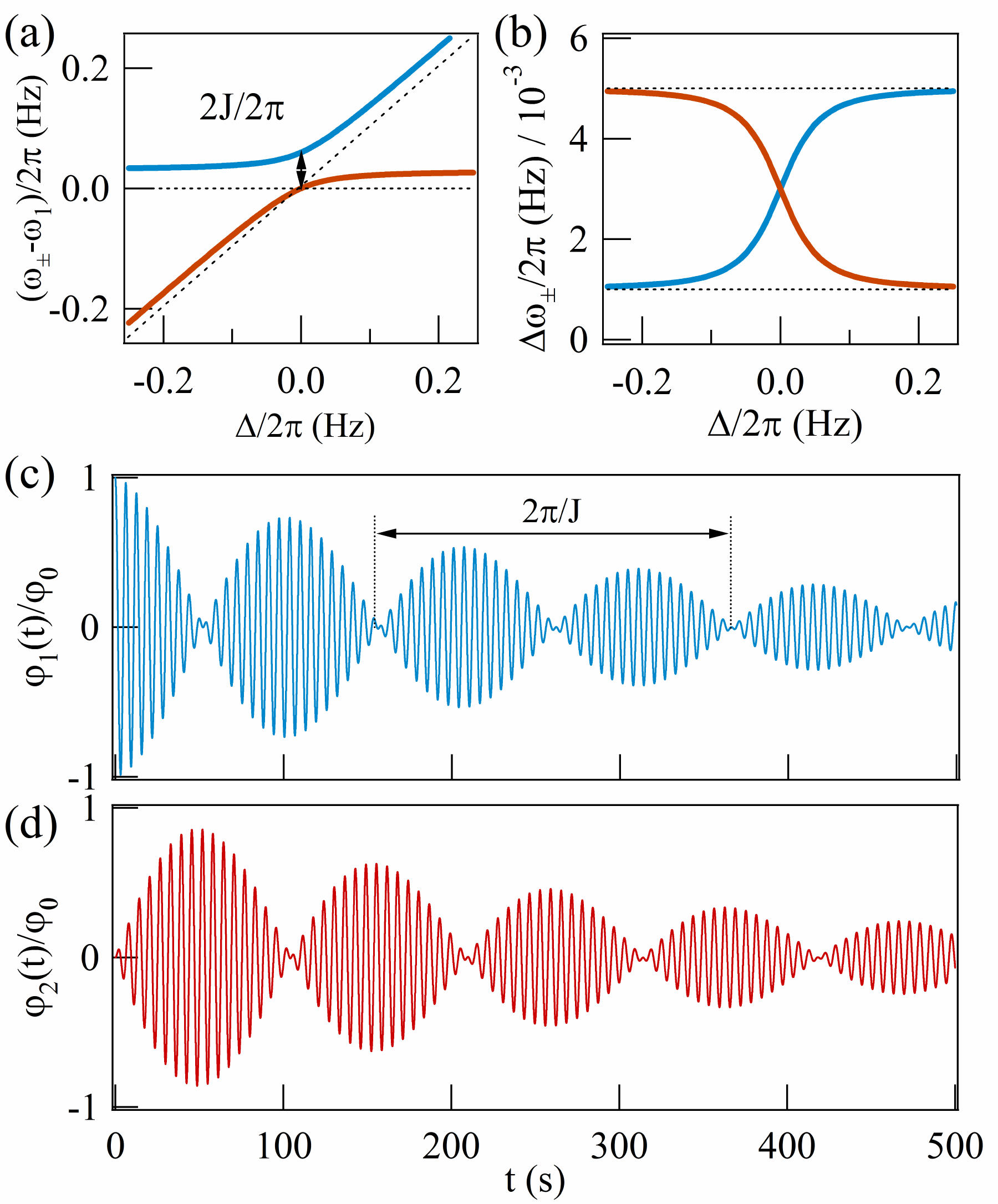}
\caption{(a) Hybridized frequency $\omega_\pm$ and (b) linewidth $\Delta\omega_\pm$ of coherently coupled pendulums, calculated according to Eq. \eqref{eq:eigenvalue_spring} in the strong coupling regime, $J^2/\left(\lambda_1 \lambda_2\right) > 1$.  Panel (a) illustrates level repulsion and panel (b) illustrates linewidth attraction, which are both characteristic of coherently coupled systems.  Blue (red) curves correspond to the $\widetilde{\omega}_+$ ($\widetilde{\omega}_-$) branch.  The horizontal and diagonal dashed lines in (a) show the frequency of the uncoupled pendulums, while the horizontal dashed lines in (b) indicate the damping limits set by the intrinsic damping of the pendulums.  (c) $\varphi_1$ and (d) $\varphi_2$ calculated according to Eqs. \eqref{eq:pendTEv1} and \eqref{eq:pendTEv2}, respectively, showing the time evolution signature of coherent coupling.  Throughout this figure $J/\omega_1 = 0.03$ and $\lambda_2/\omega_1 = 5 \lambda_1/\omega_1 = 0.005$.  For the time domain calculation, $\omega_1 = \omega_2$.}
\label{fig:springResult}
\end{center}
\end{figure}

Hybridization also impacts the linewidth, as can be seen from \figref[b]{fig:springResult}.  For coherent coupling, the linewidth evolution is attractive, with a crossing at $\Delta = 0$ where $\Delta \omega_+ = \Delta \omega_- = \left(\lambda_1 + \lambda_2\right)/2$.  Furthermore, even though the damping of each mode is a function of detuning, the total damping is constrained such that $\Delta\omega_+ + \Delta\omega_- = \lambda_1 + \lambda_2$ for all $\Delta$.  This reflects the fact that the spring-coupled pendulums form a closed system, and while coupling allows energy exchange between the two pendulums, it does not introduce an additional damping channel.  This constraint sets asymptotic limits on $\Delta\omega_{\pm}$ as $|\Delta| \to \infty$, ensuring that $\Delta\omega_{\pm}$ is bounded above and below by $\lambda_{1,2}$.

The dynamics of the coupled eigenmodes is governed by two time scales: a fast, sinusoidal oscillation at $\omega_{r{\!}e{\!}f}$, and a slow drift in the overall amplitude $A_{1,2}$ and phase $\theta_{1,2}$, which can be written as
\begin{equation}
\label{eq:pendPerturb2}
\varphi_{1,2}(t)=A_{1,2}(t)\cos[\omega_{r\!e\!f} t+\theta_{1,2}(t)].
\end{equation}
The slowly varying amplitude and phase can be determined by time averaging over the fast oscillations (see Appendix \ref{Sec:appendixA}).  Near $\Delta \approx 0$ and for $J\ll\omega_{1,2}$,
\begin{subequations}
\label{eq:pendEnvSimp}
\begin{align}
\frac{dA_{1,2}}{dt}&=-\lambda_{1,2} A_{1,2} + A_{2,1}J\sin(\theta_{2,1}-\theta_{1,2}), \\
\frac{d(\theta_1-\theta_2)}{dt}&=-\Delta+\frac{(A_1^2-A_2^2)J\cos(\theta_1-\theta_2)}{2A_1A_2}.
\end{align}
\end{subequations}
According to Eq. \eqref{eq:pendEnvSimp}, the amplitude of the uncoupled pendulums will decay exponentially with a time constant determined by the damping, while the phase difference increases linearly at a constant rate of $\Delta$.  However the evolution of the coupled pendulums is very different.  At $\Delta = 0$ in the steady state, meaning when $d(\theta_1-\theta_2)/dt=0$, but $not$ necessary when $d\theta_1/dt=d\theta_2/dt=0$, either $A_1^2=A_2^2$ or $\cos(\theta_1-\theta_2)=0$. The first condition corresponds to in-phase ($A_1 = A_2$) or $180^\circ$ out-of-phase ($A_1 = -A_2$) motion of the two pendulums, i.e., the canonical eigenstates, while the second condition corresponds to $90^\circ$ out-of-phase oscillations. This fundamental $90^\circ$ phase shift provides a sensitive measure of coherent coupling in cavity magnonics. \cite{Bai2015, Harder2018a}

With initial conditions $\varphi_1(t=0)=\varphi_0$ and $\varphi_2(t=0)=0$, the time evolution of the strongly coupled pendulums at $\Delta = 0$ is determined from Eqs. \eqref{eq:pendPerturb2} and \eqref{eq:pendEnvSimp} to be
\begin{subequations}
\begin{eqnarray}
\varphi_1&\simeq&\varphi_0e^{-(\lambda_1+\lambda_2)t/2}\left\{\cos(\omega_1t)+\cos\left[\left(\omega_1+2J\right)t\right]\right\}/2 \nonumber \\
&=&\varphi_0e^{-(\lambda_1+\lambda_2)t/2}\cos(Jt)\cos\left[\left(\omega_1+J\right)t\right], \label{eq:pendTEv1}\\
\varphi_2&\simeq&\varphi_0e^{-(\lambda_1+\lambda_2)t/2}\left\{\cos(\omega_1t)-\cos\left[\left(\omega_1+2J\right)t\right]\right\}/2 \nonumber \\
&=&\varphi_0e^{-(\lambda_1+\lambda_2)t/2}\sin(Jt)\sin\left[\left(\omega_1+J\right)t\right].
 \label{eq:pendTEv2}
\end{eqnarray}
\end{subequations}
It should be noted the phase evolution of the individual pendulums is determined from Eq. \eqref{Eq: TA_method_phase} to be $d\theta_1/dt=d\theta_2/dt=J$.

Equations \eqref{eq:pendTEv1} and \eqref{eq:pendTEv2} are plotted in \hyperref[fig:springResult]{Figs. 3 (c)} and \hyperref[fig:springResult]{3 (d)}, respectively, revealing a beating pattern with fast oscillations at frequency $\omega_1 + J$ modulated by the Rabi-like frequency $J$.  This modulation frequency determines the rate of energy transfer between the two subsystems, as can be observed by the 90$^\circ$ phase difference of $\varphi_{1}$ and $\varphi_{2}$.  In general, this phase difference will be observed if the initial conditions are not symmetric or antisymmetric, i.e., if the system is not initially an eigenmode.  Physically, this is due to energy conservation during coupling, which will not allow the system to reach its ground state (in-phase motion) if it does not start in this state.  Furthermore, both pendulums decay at the same rate, $\left(\lambda_1 + \lambda_2\right)/2$, which is the average value of the linewidth evolution, as seen in \figref[b]{fig:springResult}.  This confirms that the coupling does not open new decay channels.
\subsection{Dashpot Coupled Pendulums: A Case-Study in Dissipative Coupling} \label{sec:dashpot}
\begin{figure} [!b]
\begin{center}
\includegraphics[width=8.5 cm]{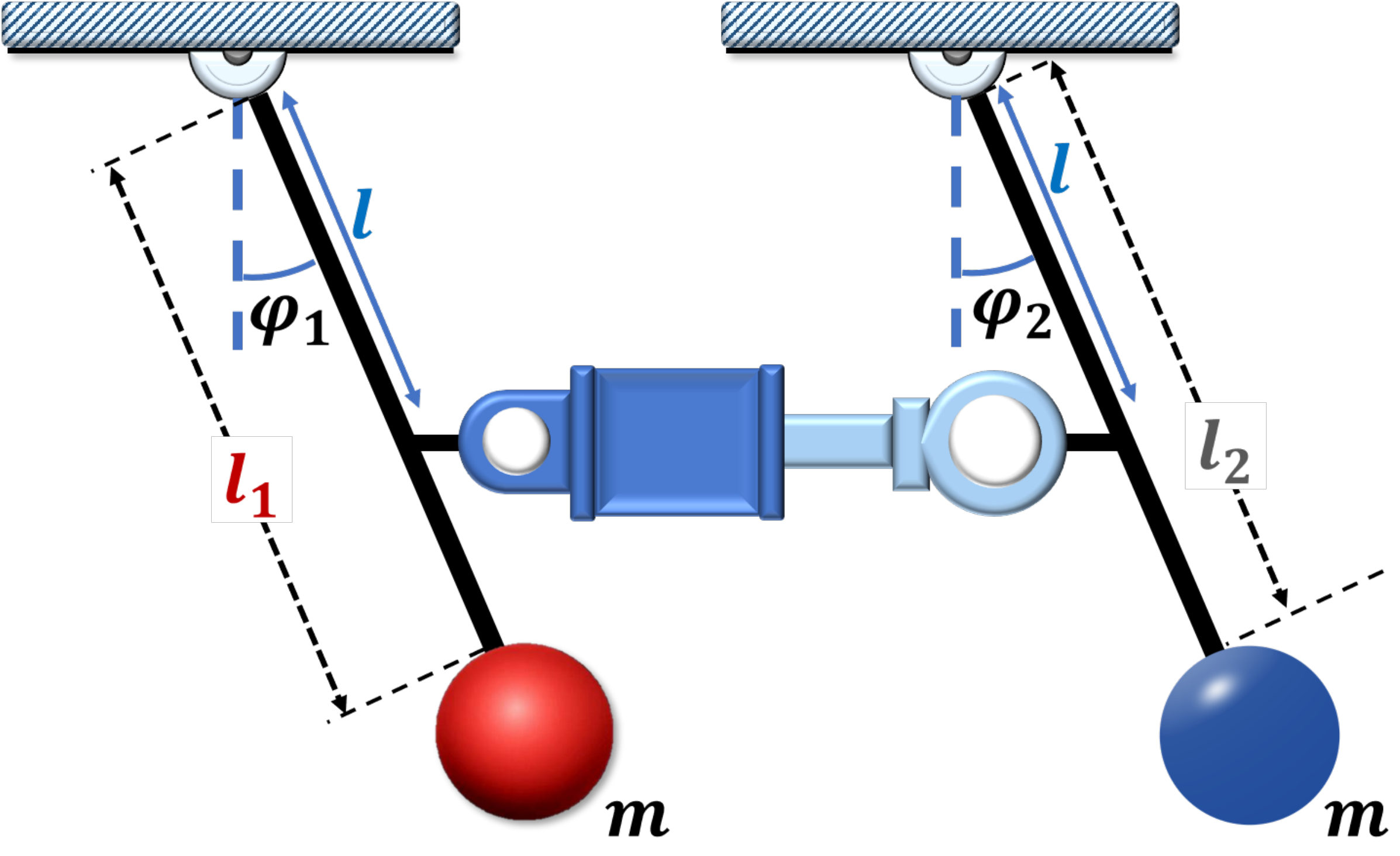}
\caption{Dashpot coupled simple pendulums.  In this system the oscillators are dissipatively coupled, resulting in level attraction.}
\label{fig:dissipativePendulum}
\end{center}
\end{figure}

In \autoref{fig:dissipativePendulum}, two pendulums of equal mass $m$ are coupled by a massless dashpot, i.e., a damper in a viscous fluid which resists motion.  For this system there is no potential energy associated with coupling.  Instead the dashpot introduces a velocity proportional friction force. As discussed in \hyperref[Sec:appendixC]{Appendix C}, this setup leads to dissipative coupling between the two pendulums with coupling strength $\Gamma_{1,2}=\nu/l_{1,2}^2\ll \omega_{1,2}$, where $\nu$ is the kinematic viscosity of fluid in the dashpot.

Near zero detuning $\Gamma = \Gamma_1 \approx \Gamma_2$ and taking the same approach as in \autoref{sec:spring} leads to the complex eigenfrequencies of the dashpot coupled system,
\begin{equation}\label{Eq:eigenvalue_dashpot}
\widetilde{\omega}_{\pm}= \dfrac{1}{2}\left[\widetilde{\omega}_1 + \widetilde{\omega}_2 - 2i\Gamma
\pm\sqrt{(\widetilde{\omega}_1-\widetilde{\omega}_2)^2-4\Gamma^2}\right].
\end{equation}
Again, $\widetilde{\omega}_{1,2} = \omega_{1,2}-i\lambda_{1,2}$ and we can write the complex hybridized eigenfrequencies in the form $\widetilde{\omega}_\pm = \omega_\pm - i \Delta \omega_\pm$.

\begin{figure} [t]
\begin{center}
\includegraphics[width=8.5cm]{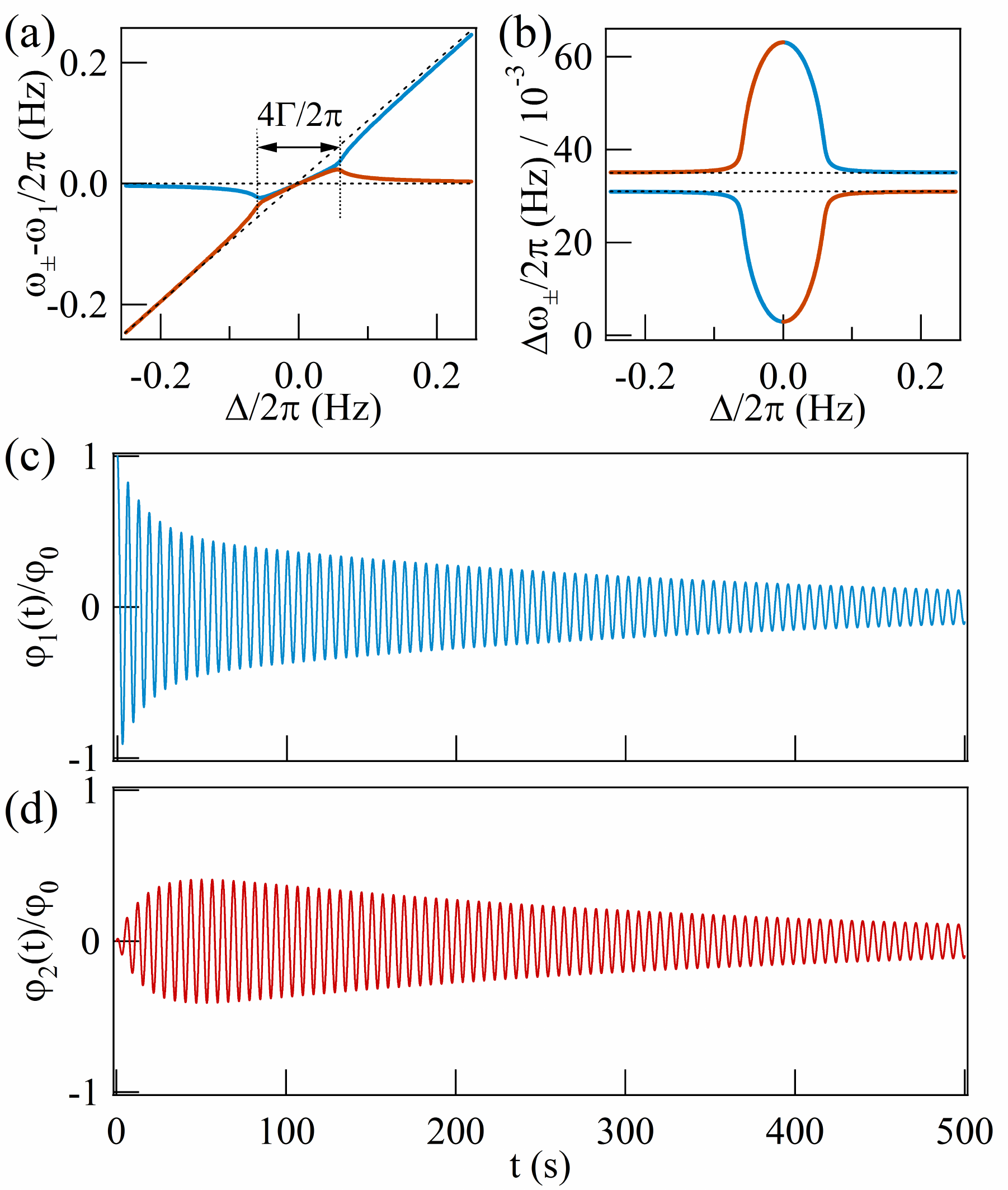}
\caption{(a) Hybridized frequency $\omega_\pm$ and (b) linewidth $\Delta\omega_\pm$ of dissipatively coupled pendulums, calculated according to Eq. \eqref{Eq:eigenvalue_dashpot}, demonstrating level attraction and linewidth repulsion respectively.  Blue (red) curves correspond to the $\widetilde{\omega}_+$ ($\widetilde{\omega}_-$) branch.  The horizontal and diagonal dashed lines in (a) show the frequency of the uncoupled pendulums, while the horizontal dashed lines in (b) indicate the damping enhancement $\Gamma+\lambda_1$  and $\Gamma+\lambda_2$ due to dissipative coupling.  (c) $\varphi_1$ and (d) $\varphi_2$ calculated according to Eqs. \eqref{eq:pDTEv1} and \eqref{eq:pDTEv2}, respectively, showing the time evolution signature of dissipative coupling.  Throughout this figure $\Gamma/\omega_1 = 0.03$ and $\lambda_2/\omega_1 = 5 \lambda_1/\omega_1 = 0.005$.  For the time domain calculation $\omega_1 = \omega_2$.}
\label{fig:dashpotResult}
\end{center}
\end{figure}

The dispersion and linewidth of the dashpot coupled pendulums are plotted in \hyperref[fig:dashpotResult]{Figs. 5 (a)} and \hyperref[fig:dashpotResult]{5 (b)} respectively.  The behaviour is strikingly different than the coherent spring coupled system.  First, the eigenfrequencies in \figref[a]{fig:dashpotResult} now coalesce, merging at $\Delta = 0$ in a phenomena known as level attraction.  Therefore the dissipative coupling can no longer be characterized by the Rabi-like gap.  Instead, the onset of level attraction is characterized by two exceptional-like points, corresponding to a bifurcation in the dispersion where the two eigenmodes become degenerate. \cite{Bernier2017}  The gap between these exceptional points is proportional to the dissipative coupling rate, as shown in \figref[a]{fig:dashpotResult}.  Exceptional points are tied to the topological structure of the complex eigenfrequencies and underly a number of interesting physical effects, such as non-reciprocal transport, \cite{Doppler2016, Xu2016, Wang2019c} which will be discussed in \autoref{sec:applications}.

Dissipative coupling is also characterized by damping repulsion, which creates a forbidden linewidth gap between $\Gamma+\lambda_1$ and $\Gamma+\lambda_2$, as shown in \figref[b]{fig:dashpotResult}.  As damping is a non-conservative, irreversible process, dissipative coupling can only accelerate decay, hence the dissipative coupling enters as an additional damping term in Eq. \eqref{Eq:eigenvalue_dashpot}.  Although the dissipative coupling effectively suppresses the damping rate of the hybridized mode with lower damping ($\lambda_1$ here), the lowest total damping rate of the hybridized modes is fundamentally limited by the intrinsic damping rate, $\min(\lambda_1,\lambda_2)$.  This minimum damping state corresponds to in-phase motion of the two pendulums, when the dashpot range of motion is minimized and therefore the frictional force is zero.\footnote{The dissipatively coupled oscillators have a minimum damping state at resonance, when the two oscillators move in-phase.  At this condition, the amplitude is largest.  Analogous physical behavior has been discussed for level attraction in the two-tone system. \cite{Grigoryan2017, Boventer2019} In that system, the amplitude enhancement at resonance also has an interference like character and is dependent on both the amplitude and phase of the real and imaginary coupling strength components \cite{Boventer2019}}  On the other hand the maximum damping state corresponds to the maximal dashpot motion when the the two pendulums oscillate 180$^\circ$ out-of-phase.  Although the physical phenomena associated with dissipative and coherent coupling are very different, from a mathematical point of view these two interactions are equivalent under frequency and damping exchange in the complex frequency plane; i.e., for coherent coupling, the hybridized frequencies are repelled while the linewidths are attracted and the opposite is true for dissipative coupling.

Similar to coherent coupling,  the time evolution of $\varphi_{1,2}$ is predominately oscillatory.  Therefore, the method of averages \cite{BogliubovBook, MinorskyBook, StrogatzBook} can again be applied to determine that the amplitude and phase are governed by the equations
\begin{subequations}
\label{Eq: First order dashpot}
\begin{align}
\frac{dA_{1,2}}{dt}&=-(\lambda_{1,2}+\Gamma_{1,2})A_{1,2}+A_{2,1}\Gamma_{1,2}\cos(\theta_{2,1}-\theta_{1,2}), \\
\frac{d(\theta_1-\theta_2)}{dt}&=-\Delta-\left(\frac{A_1\Gamma_2}{A_2}+\frac{A_2\Gamma_1}{A_1}\right)\sin(\theta_1-\theta_2).\label{Eq: First order dashpotc}
\end{align}
\end{subequations}
The additional decay terms, $\Gamma_{1,2} A_{1,2}$, will enhance the decay rate as seen in \figref[b]{fig:dashpotResult}.  \footnote{While the enhanced decay rate is seen in \figref[b]{fig:dashpotResult}, since $\Delta = 0$ in \hyperref[fig:dashpotResult]{Figs. 5 (c)} and \hyperref[fig:dashpotResult]{5 (d)}, the effect is compensated by the coupling and does not appear in these panels.}  In contrast, the spring-coupled resonance frequency was blueshifted, an example of the frequency-damping symmetry between dissipative and coherent coupling.

According to Eq. \eqref{Eq: First order dashpotc}, the phase difference between the pendulums in the steady state, at zero detuning, is 0$^\circ$ or 180$^\circ$. Although both states coexist in the coupled system, the 180$^\circ$ out-of-phase motion decays rapidly, leading to in-phase synchronization of the two pendulums at late times.  The sharp contrast to the 90$^\circ$ phase-shift in the coherent system, combined with the phase sensitivity of microwave transmission measurements, provides a way to distinguish the coupling mechanism in cavity magnonic systems.

Solving Eqs. \eqref{Eq: First order dashpot} in the strong coupling limit, $\Gamma^2/\lambda_1\lambda_2 \gg 1$, with $\Gamma = \Gamma_1 = \Gamma_2$ and the initial conditions $\varphi_1(t=0)=\varphi_0$, $\varphi_2(t=0)=0$
\begin{subequations}
\begin{eqnarray}
\varphi_1&\simeq&\varphi_0e^{-(\lambda_1+\lambda_2)t/2}(1+e^{-2\Gamma{t}})\cos(\omega_1t)/2, \label{eq:pDTEv1}\\
\varphi_2&\simeq&\varphi_0e^{-(\lambda_1+\lambda_2)t/2}(1-e^{-2\Gamma{t}})\cos(\omega_1t)/2. \label{eq:pDTEv2}
\end{eqnarray}
\end{subequations}
As in the case of coherent coupling, the average damping $(\lambda_1+\lambda_2)/2$ sets the overall decay rate of the oscillations.  However, unlike coherent coupling, $\Gamma$ also acts as a source of damping.  Since $\Gamma \gg \lambda_{1,2}$, the decay rate set by the coupling greatly exceeds the intrinsic losses, and a rapid decrease in $\varphi_1$ is accompanied by a rapid increase in $\varphi_2$, as shown in \hyperref[fig:dashpotResult]{Figs. 5 (c)} and \hyperref[fig:dashpotResult]{5 (d)}.  The quick decay of the dissipative coupling term leads to rapid synchronization, $\varphi_1=\varphi_2=\varphi_0e^{-(\lambda_1+\lambda_2)t/2}\cos(\omega_1t)/2$, where the two pendulums have identical amplitude and phase.  Physically, this is an example of two-tone decay, where an initial rapid decay to the lowest energy state is followed by a slower, collective decay due to intrinsic damping.  In contrast, coherent coupling is characterized by a two-tone oscillation, which produces beating with an overall decay envelope dictated by the collective damping.  This contrast exemplifies the frequency-damping symmetry between dissipative and coherent systems, ensuring that synchronization will occur independently of initial conditions provided the oscillatory motion is not damped too quickly, i.e., as long as the collective decay rate does not exceed the rate of dissipative coupling.  In general if the initial angles are $\varphi_{1,2}^0$, the amplitude of the synchronized modes will be $\left(\varphi_1^0 + \varphi_2^0\right)/2$, and therefore synchronization will occur unless $\varphi_2^0=-\varphi_1^0$, in which case both pendulums are rapidly damped before synchronization can occur.

Note that the synchronization discussed here does not involve self-sustained oscillators,\cite{PikovskyBook} as the cavity magnonic system is externally driven.  While self-sustained oscillators are often necessary in a synchronized system, there are other factors that can influence synchronization. For example, in the case of dissipative cavity magnonics, isochronous oscillators can be synchronized through dissipative coupling.\cite{PikovskyBook}

\begin{figure} [!b]
\begin{center}
\includegraphics[width=6cm]{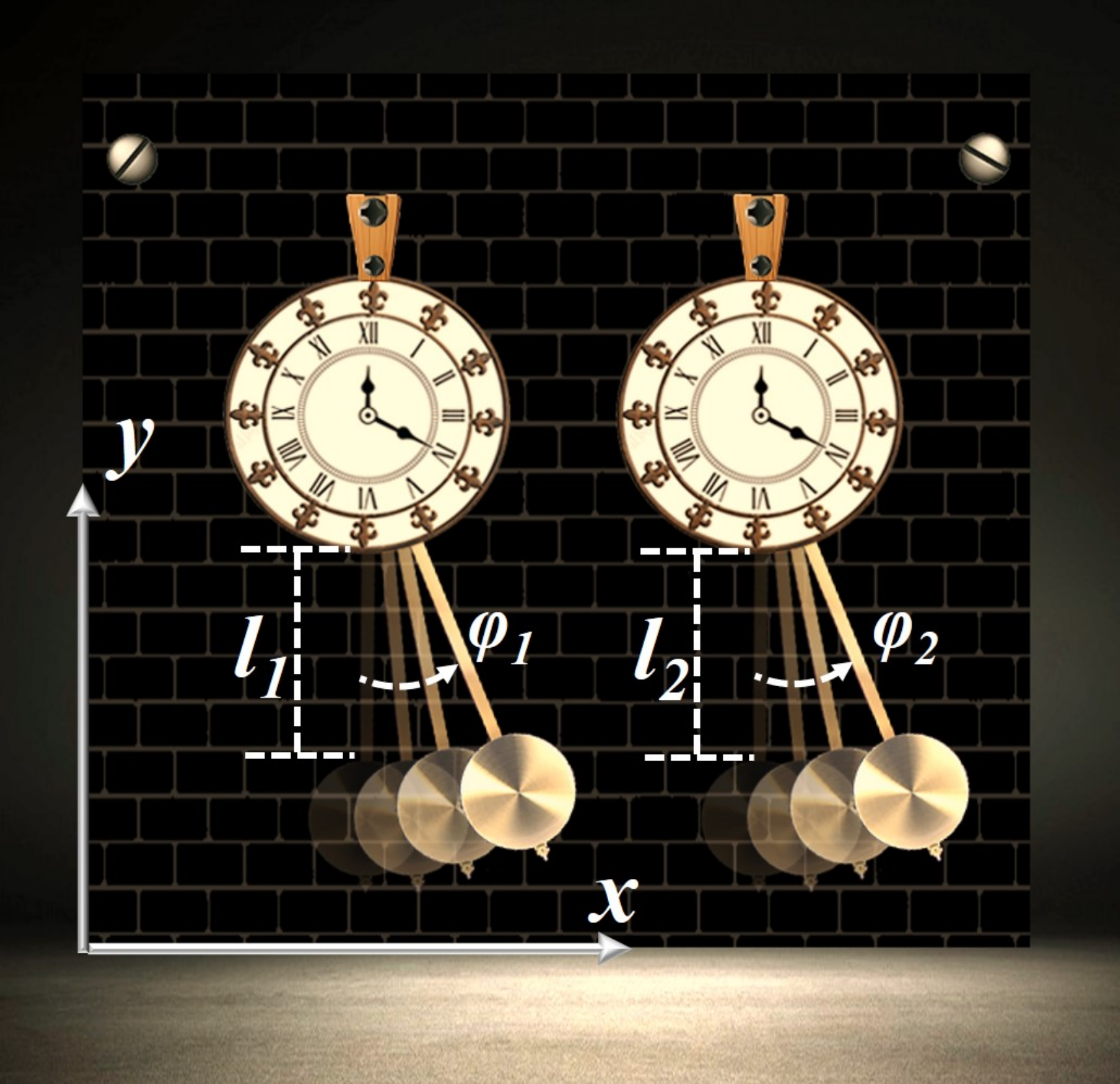}
\caption{Two pendulum clocks may be indirectly coupled via wall vibrations.  This system, known as Huygen's pendulums, is an example of base-mediated coupling, which highlights the role of the reservoir in dissipative interactions.}
\label{fig:huygensPendulums}
\end{center}
\end{figure}
\subsection{Base-Mediated Coupling: The Role of the Reservoir in Dissipative Interactions} \label{sec:basePendulums}
Dissipative interactions must be mediated by a reservoir, for example, the viscous fluid of the dashpot-coupled pendulums or an open photon bath in cavity magnonics.  Ultimately this means that dissipative interactions are indirect, and while the general features of dissipative coupling -- level attraction, linewidth repulsion and synchronization -- are already revealed in the dashpot-coupled system, the role of the reservoir is obscured by the phenomenological nature of the viscosity.  However, when pendulums interact via the vibrations of a common base, the reservoir plays a starring role.  Aptly known as base-mediated coupling, this phenomena is responsible for the widespread observation of synchronization in physical systems.

Base-mediated coupling was first described by Christiaan Huygens when he noted that two pendulum clocks, mounted on a common base, will swing at the same frequency with a 180$^\circ$ phase shift. \cite{HuygensBook, PikovskyBook}  Huygens' pendulum clocks, shown in \autoref{fig:huygensPendulums}, form a closed system, which includes two pendulums mounted to a wall.  The pendulums have no direct interaction.  However the wall acts as a mutual reservoir, allowing energy leaked by one pendulum to indirectly drive the second, resulting in an indirect, dissipative coupling.  Another example of this phenomena is the synchronization of metronomes placed on a freely moving base.  In this case, the centre-of-mass velocity of the system must be zero, leading to synchronization via an indirect interaction between the metronomes. \cite{Pantaleone2002}

\begin{figure} [t]
\begin{center}
\includegraphics[width=8.5 cm]{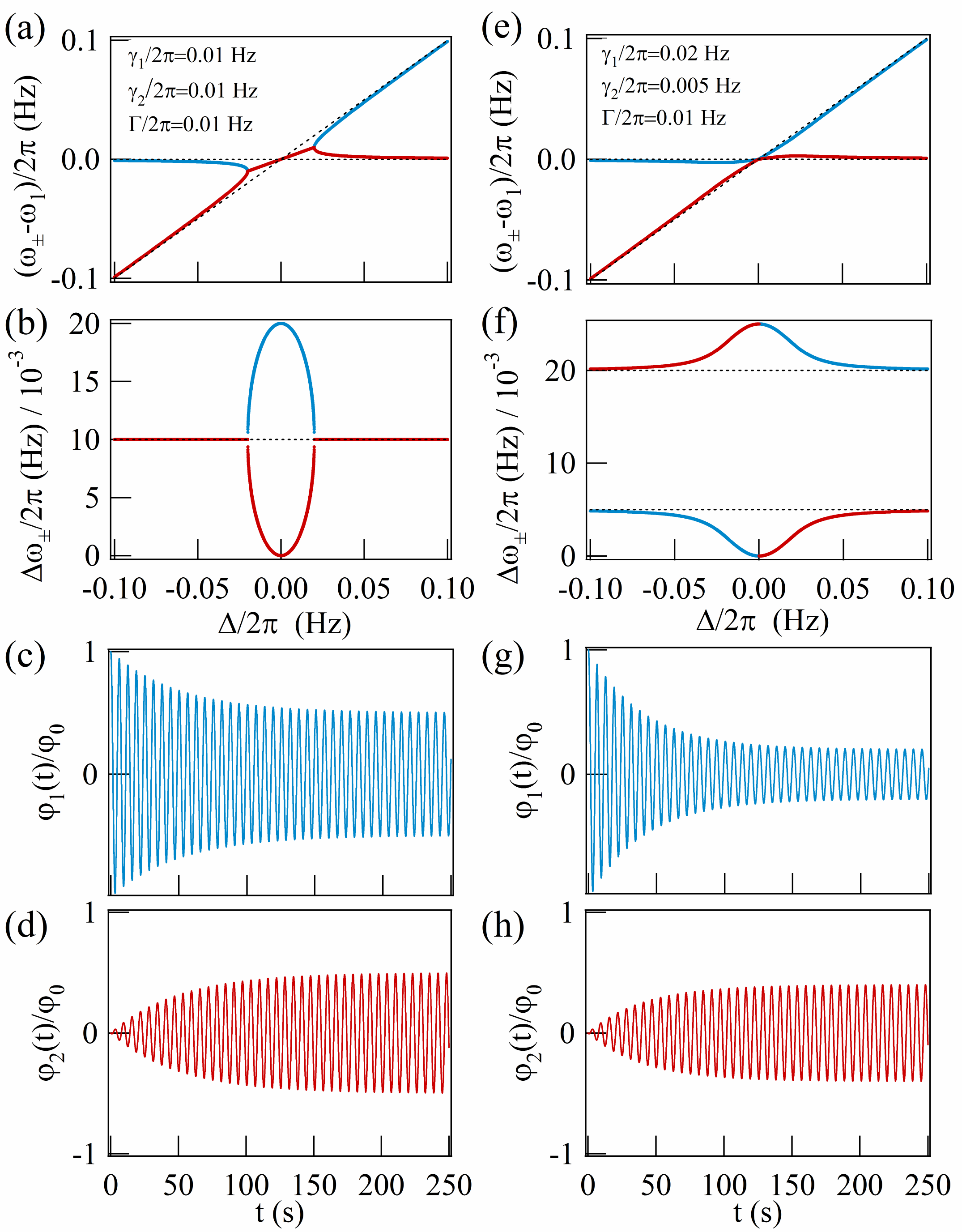}
\caption{Frequency and time behaviour of the base-mediated coupled pendulums.  Panels (a)$-$(d) were calculated using a dissipative coupling constant of $\Gamma/\omega_1 = 0.01$, with $\gamma_2/\gamma_1 = 1$, while panels (e)$-$(g) maintain $\Gamma/\omega_1= 0.01$ but with $\gamma_1/\gamma_2 = 4$. (a, e) Oscillation frequency $\omega_\pm$ and (b,f) linewidth $\Delta\omega_\pm$ calculated according to Eq. \eqref{Eq:eigenvalue_base}. The horizontal and diagonal dashed lines in (a, e), respectively, show the frequency of the uncoupled pendulums, while the horizontal dashed lines in (b, e) are the extrinsic damping rates of the pendulums caused by coupling to the base.  The oscillation and linewidth behaviour both indicate dissipative coupling.  (d, h) Time evolution of $\varphi_1(t)$ (c, g) and $\varphi_2(t)$ calculated according to Eqs. \eqref{Eq: Syn base}.  At large $t$, the two pendulums settle into an out-of-phase synchronized steady-state, and since this model does not include intrinsic damping, the pendulums oscillate, but their amplitudes do not decay after synchronization.}
\label{fig:baseResult}
\end{center}
\end{figure}

Huygen's pendulums can be analyzed by modelling the wall as a giant two-dimensional crystal with a wavevector dependent oscillation frequency.    The resulting eigenfrequencies are analogous to the dashpot coupled pendulums: 
\begin{equation}\label{Eq:eigenvalue_base}
\begin{split}
\widetilde{\omega}_{\pm}&= \dfrac{1}{2}[(\omega_1 + \omega_2) - i(\gamma_1+\gamma_2)] \\
&\pm\frac{1}{2}\sqrt{[(\omega_1-\omega_2)-i(\gamma_1-\gamma_2)]^2-4\gamma_1\gamma_2},
\end{split}
\end{equation}
where $\omega_{1,2}$ are again the uncoupled pendulum oscillation frequencies and $\gamma_{1,2}$ are the extrinsic damping parameters, which represent energy leakage from the pendulums to the wall.  Since the extrinsic damping is typically much greater than the intrinsic damping of each pendulum, $\gamma_{1,2} \gg \lambda_{1,2}$, the intrinsic damping has been neglected to simplify the results.  However, it should be noted that this is not a requirement of dissipative coupling in the base-mediated system. Comparing to Eq. \eqref{Eq:eigenvalue_dashpot} of the dashpot coupled pendulums, it is clear that the wall mediates an indirect dissipative coupling strength of $\Gamma = \sqrt{\gamma_1\gamma_2}$, and that the extrinsic damping provides the dominant decay channel for each mode.  The technical details of this model and a derivation of Eq. \eqref{Eq:eigenvalue_base} can be found in \hyperref[sec:appendixD]{Appendix D}.

The dispersion and linewidth are plotted in \autoref{fig:baseResult} for two sets of extrinsic damping parameters.  In the left panel, where $\gamma_1 = \gamma_2$, the behaviour is similar to the dashpot-coupled system, namely, level attraction in the oscillation frequency and level repulsion in the linewidth. In this case, there are two true exceptional points, at $\Delta = \pm 2 \Gamma$, corresponding to the bifurcation points of the dispersion in \hyperref[fig:baseResult]{Figs. 7 (a)} and \hyperref[fig:baseResult]{7 (b)}.  At these points, both the real and imaginary part of the complex eigenfrequencies coalesce.  Therefore, exceptional points may be engineered via base-mediated coupling, e.g., for highly sensitive detection techniques. \cite{Harder2017, Zhang2017a, Zhang2019, Tserkovnyak2020}

When the extrinsic damping rates are highly mismatched, the signatures of level attraction are less clear.  This effect is illustrated in the right panel of \autoref{fig:baseResult}, where $\gamma_1 = 4\gamma_2$, and occurs even though $\Gamma$ is unchanged.  In this case, the frequency dispersion shows a cross-like behaviour, and the bifurcation points are almost invisible.  This cross-like behaviour can also be observed in the spring-coupled system when $J^2<\lambda_1\lambda_2$, which means that it is difficult to distinguish the two effects based solely on the frequency dispersion.  However, the linewidth evolution of the base-mediated system still clearly shows repulsive behaviour, and therefore, the nature of the coupling mechanism can be confirmed regardless of damping mismatch.

 \begin{table*}[!t]
\def\arraystretch{1.5}
\begin{threeparttable}
\caption{Key features of the oscillator models considered in \autoref{sec:harmonic}.}
\begin{tabular}{>{\centering\arraybackslash}m{3cm}>{\centering\arraybackslash}m{4cm}>{\centering\arraybackslash}m{4cm}>{\centering\arraybackslash}m{4cm}}
\cline{1-4} \cline{1-4} \cline{1-4} \cline{1-4} \cline{1-4} \cline{1-4}
&\multicolumn{3}{c}{{\bf Pendulum System}}
\\ \cline{1-4} \cline{1-4} \cline{1-4}
&Spring-Coupled & Dashpot-Coupled & Base-Mediated \\
\cline{1-4} \cline{1-4} \cline{1-4}
\bf Coupling Type & Coherent &Dissipative & Dissipative \\
\bf Steady State Phase Shift& $\dfrac{\pi}{2}$ & $\pi$ & $\pi$ \\
\bf Dispersion Characteristic & Level repulsion with Rabi-like gap of $2J$ & Level attraction with EP-like separation of $4\Gamma$ & Level attraction with EP-like separation of $4\Gamma$\\
\bf Linewidth Characteristic & Crossing; Bounded by $\lambda_{1,2}$ & Gap of size $|\lambda_1 - \lambda_2|$ & Gap of size  $|\gamma_1 - \gamma_2|$ \\
\bf Time Domain Characteristic Near $\Delta = 0$ & Two-tone oscillations producing beat-like pattern & Two-tone decay producing synchronization & Two-tone decay producing synchronization \\
\cline{1-4} \cline{1-4} \cline{1-4}
\end{tabular}
\label{HarmonicComparison}
\end{threeparttable}
\end{table*}

Neglecting the driving term and for the initial conditions $\varphi_1(t=0)=\varphi_0$ and $\varphi_2(t=0)=0$, the time evolution of the base-mediated pendulums at $\omega_1=\omega_2$ is
\begin{subequations}
\label{Eq: Syn base}
\begin{align}
\varphi_1&=\varphi_0\left(\dfrac{\gamma_2}{\gamma_1+\gamma_2}\right)\left(1+\dfrac{\gamma_1}{\gamma_2}e^{-(\gamma_1+\gamma_2)t}\right)\cos(\omega_1t),\\
\varphi_2&=-\varphi_0\left(\dfrac{\sqrt{\gamma_1\gamma_2}}{\gamma_1+\gamma_2}\right)\left(1-e^{-(\gamma_1+\gamma_2)t}\right)\cos(\omega_1t).
\end{align}
\end{subequations}
The exponential decay term rapidly disappears resulting in the 180$^\circ$ out-of-phase synchronization originally observed by Huygens.  This synchronization is observed regardless of dissipation matching, as shown in \hyperref[fig:baseResult]{Figs. 7 (c)}, \hyperref[fig:baseResult]{7 (d)}, \hyperref[fig:baseResult]{7 (g)} and \hyperref[fig:baseResult]{7 (h)}.  However, when the dissipation rates are highly mismatched, the decay into the steady state is much faster and both amplitudes are reduced compared to the case of equal dissipation.  In the synchronized state, the pendulums oscillate, but their amplitudes no longer decay, indicating that the total losses due to direct coupling between the pendulums and the base is zero; any energy lost from the first pendulum can be completely and coherently transferred to the second pendulum.  Furthermore, the steady-state amplitudes of $\varphi_1$ and $\varphi_2$ are now unequal, with an amplitude ratio $\varphi_1/\varphi_2 \propto \sqrt{\gamma_2/\gamma_1}$, determined solely by the extrinsic damping to the common base.
\subsection{Summary of Coherent and Dissipative Harmonic Oscillators}
\hyperref[HarmonicComparison]{Table I} contains a summary of the key characteristics for the three case studies considered in this section: (1) Coherent, spring-coupled pendulums; (2) Dissipative, dashpot-coupled pendulums; and (3) Dissipative, base-mediated coupled pendulums.
\section{Coherent and Dissipative Cavity Magnonics} \label{sec:physics}
In this section, the theoretical and experimental details of coherent and dissipative cavity magnonics are outlined.  In the case of dissipative coupling, we first discuss an effective direct interaction, analogous to the dashpot-coupled pendulums.  We then highlight the role of the reservoir by examining a traveling-wave-mediated interaction, analogous to the base-mediated pendulums.  In each case key signatures and questions of device integration are addressed.
\subsection{Coherent Cavity Magnonics} \label{sec:coherent}
The canonical cavity magnonic device is a ferrimagnetic spin ensemble coupled to a single cavity resonance, for example, an yttrium-iron-garnet (YIG) sphere coupled to a microwave cavity mode, as shown in \figref[a]{fig:coherentFrequency}. The strongest spin-photon interaction will take place with the ferromagnetic resonance (FMR) mode, which includes the most spins, and therefore, at the lowest order, the effective Hamiltonian must contain a kinetic term for the photons and the FMR mode, plus all possible quadratic interactions,
\begin{equation}
H_{CC}=\hbar\omega_ca^\dag a+\hbar\omega_mb^\dag b+\hbar J(a+a^\dag) (b+b^\dag).
\end{equation}
Here $a^\dag$ ($a$) and $b^\dag$ ($b$) are the respective creation (annihilation) operators of the cavity and FMR modes, which have resonance frequencies $\omega_c$ and $\omega_m$, respectively.  $J$ is the coupling strength between the FMR and cavity resonance.  The interaction term will have time dependencies of the form, $e^{-i(\omega_c\pm\omega_m)t}$, but since typical experiments are performed near $\omega_m \approx \omega_c$, the rapidly oscillating $e^{-i(\omega_c+\omega_m)t}$ terms will have a small time averaged effect and can be neglected in the rotating wave approximation.\footnote{The rotating wave approximation amounts to keeping the component of the interaction whose time evolution (closely) co-rotates with the interaction picture eigenstates, dropping the counterrotating term. To physically understand this approximation, we can imagine a classical precessing dipole which represents transitions in a two-level system. A linearly oscillating driving field can be decomposed into two counterrotating fields. One of these fields will co-rotate with the dipole moment, applying a constant torque over many periods. However, the torque due to the counterrotating field component will reverse (with respect to the dipole precession) every period, and therefore, will have little average effect.  This description has been taken from Ref. \citenum{Harder2018}, which is adapted from the physical discussion in Ref. \citenum{SilvermanBook}.  A good discussion of the rotating wave approximation can also be found in Ref. \citenum{WallsBook}} Therefore,
\begin{equation}
H_{CC}=\hbar\omega_ca^\dag a+\hbar\omega_mb^\dag b+\hbar J(ab^\dag+a^\dag{}b).
\label{Eq:coherent_Hamiltonian}
\end{equation}
Equation \eqref{Eq:coherent_Hamiltonian} is essentially the Tavis-Cummings model, \cite{Tavis1968, Garraway2011} (the $N_s$ spin generalization of the single spin Jaynes-Cummings model \cite{Jaynes1963}) and can be extended to systems with multiple cavity resonances \cite{Hyde2016} and higher order magnon modes. \cite{Cao2014, Zhang2014, Lambert2015, Zhang2016, MaierFlaig2016}  While we have provided an effective field theory approach to the Hamiltonian of coherent cavity magnonics, microscopically, Eq. \eqref{Eq:coherent_Hamiltonian} is derived by quantizing the $\mathbf{M}\cdot\mathbf{B}$ Zeeman interaction between the magnetization, $\mathbf{M}$, of the magnetic sample and the cavity magnetic field, $\mathbf{B}$.  This approach is clearly described elsewhere in the literature, e.g., Refs. \citenum{Soykal2010, Huebl2013a, Goryachev2014, Harder2016b} and \citenum{Tabuchi2015}, and can be extended to exchange coupled systems. \cite{Cao2014} 

As the origin of $J$ is the Zeeman coupling between the cavity magnetic field and the magnetic sample, the coupling strength depends on the magnon and cavity mode volumes, $V_s$ and $V_c$, respectively, as $J \propto \sqrt{V_s/V_c}$.\cite{Huebl2013a, Zhang2014, Harder2016b} $V_s$ is proportional to the number of spins in the sample, $N_s$, and therefore $J\propto \sqrt{N_s}$, consistent with expectations from the Tavis-Cummings model. \cite{Tavis1968, Garraway2011}  Therefore, the coherent coupling strength is increased in large magnetic samples, \cite{Zhang2014, Bourhill2015} providing a simple approach to achieve strong coupling -- increasing sample size.  Moreover, the $\sqrt{N_s}$ enhancement of the coupling strength is a key motivation behind the exploration of cavity magnonics for quantum information applications. \cite{Soykal2010, Kurizki2015, LachanceQuirion2019} However, while useful for cavity magnonics exploration, large sample sizes are inappropriate for device integration.  Fortunately, the coupling also depends on the nature of the cavity mode, for example, regardless of sample size the coupling will be zero if the sample magnetization and cavity magnetic field polarization are perpendicular. \cite{Bai2016}  Therefore, it is useful to write $J \propto \sqrt{V_s/V_c} = \sqrt{N_s} K$, where $K$ is the filling factor, which describes how effectively the cavity magnetic field couples to the magnon mode. \cite{Bloembergen1954, Zhang2014, Bourhill2015, Floch2016}  Large coupling strengths can, therefore, also be achieved by increasing $K$, for example, by carefully selecting the cavity mode geometry, \cite{Bai2016} or by localizing the magnetic field within the sample volume via special 3D cavity designs \cite{Goryachev2014} or on-chip, lithographically defined cavities. \cite{Hou2019, Li2019a}  The latter method is particularly promising for on-chip integration of cavity magnonics.

\begin{figure} [t!]
\begin{center}
\includegraphics[width=8.9cm]{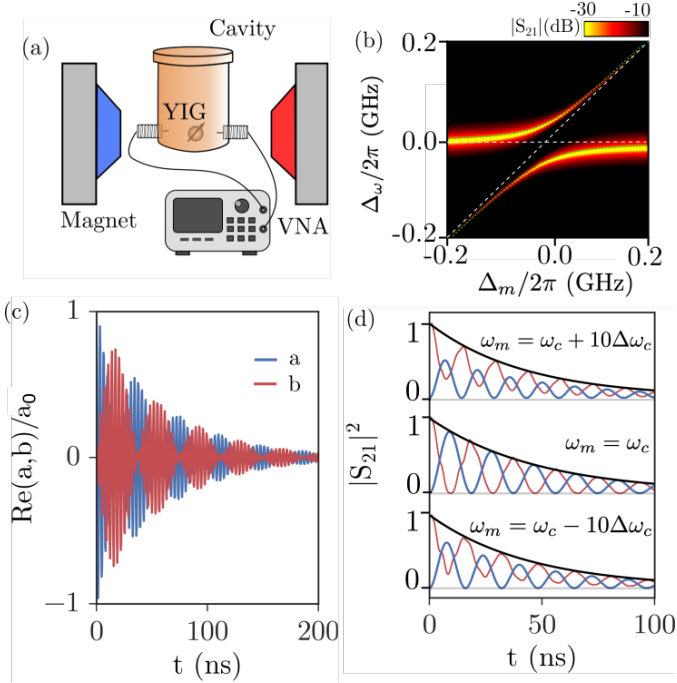}
\caption{(a) A coherent cavity magnonic experiment using an yttrium-iron-garnet sphere inside a 3D microwave cavity.  The magnetic resonance is controlled by an external bias field, and the microwave spectra is measured using a vector-network-analyzer. (b) The microwave transmission spectra of coherent cavity magnonics, plotted as a function of the cavity and magnon frequency detuning, $\Delta_\omega = \omega - \omega_c$ and $\Delta_m = \omega- \omega_m$.  Diagonal and horizontal dashed lines indicate the uncoupled magnon and cavity dispersion, respectively.  Key coherent features include an anticrossing, and linewidth evolution.  (c) Re$(a)/a_0$ and Re$(b)/a_0$ as a function of time, calculated according to Eq. \eqref{eq:basisChange}.  The rapid, GHz frequency oscillations, and MHz frequency beating pattern, determined by the scale of the generalized Rabi frequency,\cite{Zhang2014} mirror the behaviour of the spring-coupled pendulums.  The black decay curve is the sum, $|a/a_0|^2+|b/a_0|^2$, illustrating the total energy decay.  (d) The time domain transmission spectra, $|\text{S}_{21}|^2= |a/a_0|^2$, plotted as the solid red curve, time averages over the rapid oscillations.  Here, the blue curve indicates the 180$^\circ$ phase-shifted magnetization, $|b/a_0|^2$, calculated according to Eq. \eqref{eq:basisChange}.}
\label{fig:coherentFrequency}
\end{center}
\end{figure}

From Eq. \eqref{Eq:coherent_Hamiltonian} the equations of motion are
\begin{subequations}
\label{Eq:coherent_ME1}
\begin{align}
\dot{a} &= -\frac{i}{\hbar}\left[a,H\right] = -i\omega_c a + -i J b, \\
\dot{b} &= -\frac{i}{\hbar}\left[b,H\right]= -i \omega_m b - i Ja,
\end{align}
\end{subequations}
which describe the intrinsic behaviour of a closed system, i.e., without external coupling or intrinsic damping.  While these constraints are not realistic experimentally, the eigenvalues of Eq. \eqref{Eq:coherent_ME1} do accurately describe the coherent hybridized dispersion at strong coupling, which is insensitive to the damping.

Intrinsic damping can be included by replacing $\omega_c\rightarrow\widetilde{\omega}_c=\omega_c-i\beta$ and $\omega_m\rightarrow\widetilde{\omega}_m=\omega_m-i\alpha$ so that
\begin{subequations}
\label{eq:closed}
\begin{align}
\dot{a} &= -i\widetilde{\omega}_c a - i J b, \\
\dot{b} &= -i\widetilde{\omega}_m b - i J a.
\end{align}
\end{subequations}
Here $\beta$ and $\alpha$ are the cavity and magnon damping rates, respectively.\footnote{In this article, we have defined all damping parameters, e.g., $\alpha$ and $\beta$, to have units of frequency.  It is also common to normalize to the cavity frequency $\omega_c$ so that $\alpha_G = \alpha/\omega_c$ and $\beta = \omega_c/Q$, where $\alpha_G$ and $Q$ are the Gilbert damping parameter and cavity quality factor, respectively.}  Taking $a, b \propto e^{-i\omega t}$ and solving for $\omega$, the complex eignenfrequencies, $\widetilde{\omega}_{\pm}$, are determined to be
\begin{equation}\label{Eq:eigenvalue_CC}
\widetilde{\omega}_{\pm}= \dfrac{1}{2}\left[\widetilde{\omega}_c + \widetilde{\omega}_m
\pm\sqrt{(\widetilde{\omega}_c-\widetilde{\omega}_m)^2+4J^2}\right].
\end{equation}
Coherent cavity magnonics, therefore, mirrors the spring-coupled pendulums discussed in \autoref{sec:spring}, with level repulsion in the dispersion and a linewidth crossing in the damping, as shown in the microwave transmission spectra of \figref[b]{fig:coherentFrequency}.

In a fixed cavity experiment, illustrated in \figref[a]{fig:coherentFrequency}, $\omega_c$ is set by the cavity design, while $\omega_m$ is controlled by an external bias field.  When tuned to $\omega_c=\omega_m$, the dispersion reduces to $\omega_\pm=\text{Re}(\widetilde{\omega}_\pm) = \omega_c\pm|J|$, where the small effect of damping has been neglected.  \footnote{This is reasonable in the strong coupling regime, $J^2/\alpha/\beta \gg 1$, but is not appropriate at weak coupling \cite{Harder2017}}  The spin-photon interaction, therefore, produces a low energy state with increased stability, which is markedly different than the spring-coupled pendulums but is common in degenerate quantum systems, e.g., the formation of bonding orbitals in atomic systems.  While $\omega_+$ and $\omega_-$ have been defined as the high and low energy modes, respectively, the nature of the eigenvectors depends on the sign of the interaction.  Typically, $\alpha, \beta \ll \omega_{c,m}$, in which case the eigenvectors of Eq. \eqref{Eq:coherent_ME1}, $|+\rangle \propto \left(1/\sqrt{2}, 1/\sqrt{2}\right)$ and $|-\rangle \propto \left(1/\sqrt{2}, -1/\sqrt{2}\right)$, correspond to in-phase and $180^\circ$ out-of-phase motion, respectively. For $J>0$, as is the case for dipole coupling, the ground state corresponds to out-of-phase motion, $|-\rangle$.  However, if $J<0$ the in-phase $|+\rangle$ is the ground state.

To calculate the microwave transmission spectra one would measure experimentally, for example, using a vector-network-analyzer (VNA), we must allow for external feed line coupling into and out of the cavity.  As described in \hyperref[Sec:appendixE]{Appendix E} this can be accomplished through the input-output formalism, commonly associated with quantum optics, see, e.g., Refs. \citenum{Harder2018, WallsBook, Kusminskiy2019} and \citenum{Clerk2010}.  The resulting S-parameters for a two-port cavity are

\begin{subequations}
\label{Eq: coherent S-parameter}
\begin{align}
\text{S}_{21}&=\frac{d_\text{out}}{c_\text{in}} = \frac{i\sqrt{\kappa_c\kappa_d}(\omega-\widetilde{\omega}_m)}{(\omega-\widetilde{\omega}_c)(\omega-\widetilde{\omega}_m)-J^2}, \\
\text{S}_{11}&=\frac{c_\text{out}}{c_\text{in}} = -1+\frac{i\kappa_c(\omega-\widetilde{\omega}_m)}{(\omega-\widetilde{\omega}_c)(\omega-\widetilde{\omega}_m)-J^2},
\\
\text{S}_{22}&=\frac{d_\text{out}}{d_\text{in}} = -1+\frac{i\kappa_d(\omega-\widetilde{\omega}_m)}{(\omega-\widetilde{\omega}_c)(\omega-\widetilde{\omega}_m)-J^2},
\end{align}
\end{subequations}
and based on the symmetry of the system, $\text{S}_{21} = \text{S}_{12}$.  In Eqs. \eqref{Eq: coherent S-parameter}, $\kappa_{c,d}$ are, respectively, the extrinsic damping rates at ports 1 and 2 due to the feed line coupling, and $\widetilde{\omega}_c = \omega_c - i \beta_L = \omega_c - i \left[\beta + \left(\kappa_c + \kappa_d\right)/2\right]$ is the complex cavity resonance frequency with the loaded cavity damping $\beta_L$.  Therefore, experimentally the linewidth of the cavity resonance depends on both the intrinsic damping, due to losses in the cavity walls, and the extrinsic damping, due to the feed line coupling.  Thus, there is a tradeoff between amplitude and signal broadening, i.e., the S parameter amplitudes are proportional to $\kappa_{c,d}$, but $\kappa_{c,d}$ also contributes to the loaded damping which broadens the resonance.  For this reason, extrinsic dissipation is often seen as an inescapable nuisance in coherent cavity magnonics, although it plays a much more important role in dissipatively coupled systems as discussed in \autoref{sec:dissipative}.

A canonical cavity magnonic experiment and microwave transmission spectra are shown in \hyperref[fig:coherentFrequency]{Figs. 8 (a)} and \hyperref[fig:coherentFrequency]{8 (b)}, respectively.  A magnetic sample is placed inside a microwave cavity, and an external magnetic bias field is used to control $\omega_m$.  The transmission spectra is then detected using a VNA.  The data shown in panel (b) was collected using a 1-mm diameter YIG sphere and a 1.25 mm diameter, copper, cylindrical microwave cavity with a height of 29 mm.  Key features of coherent coupling, including level repulsion and linewidth crossing, can be clearly identified, in analogy with the spring-coupled pendulums.

The experimental apparatus shown in \figref[a]{fig:coherentFrequency} can also be used to probe the time domain behaviour via Fourier transform.  Since the external photon bath of the VNA is used to excite the cavity mode, $a|_{t=0} = a_0$ and $b|_{t=0} = 0$, where $a_0^2$ is proportional to the input microwave power.  With these initial conditions and the equations of motion, Eq. \eqref{eq:closed}, the photon and magnon states can be expressed as a linear combination of the normal modes (which oscillate as $e^{-i\widetilde{\omega}_\pm t}$),
\begin{align}
\begin{bmatrix}a\\b\end{bmatrix}& = a_0\begin{bmatrix} C_{11}&C_{12} \\ C_{21}& C_{22} \end{bmatrix} \begin{bmatrix}e^{-i\widetilde{\omega}_+t}\\e^{-i\widetilde{\omega}_-t}\end{bmatrix} \nonumber \\
&=a_0\begin{bmatrix} 1-C & C \\ \sqrt{C-C^2} &-\sqrt{C-C^2} \end{bmatrix} \begin{bmatrix}e^{-i\widetilde{\omega}_+t}\\e^{-i\widetilde{\omega}_-t}\end{bmatrix}, \label{eq:basisChange}
\end{align}
\noindent where
\begin{equation}
C = \frac{1}{2} \left[1-\frac{\widetilde{\omega}_c-\widetilde{\omega}_m}{\sqrt{(\widetilde{\omega}_c-\widetilde{\omega}_m)^2+4J^2}}\right].
\end{equation}
$C$ is approximately real since $\alpha, \beta \ll J \ll \omega_{c,m}$, and therefore,

\begin{equation}
\begin{split}
|\text{S}_{21}|^2 = \bigg|\frac{a}{a_0}\bigg|^2 = &~ C_{11}^2e^{-2t/\tau_+} +C_{12}^2e^{-2t/\tau_-}\\
&+2C_{11}C_{12} e^{-t/\tau_+}e^{-t/\tau_1}\cos(Rt).
\end{split}
\label{Eq: coherent time}
\end{equation}
Here, $R=\sqrt{(\omega_m-\omega_c)^2+4J^2}$ is the generalized Rabi frequency, and $\tau_\pm = 2\pi/\Delta\omega_\pm$ are the decay rates of the two normal modes.  Physically, the Rabi-like interference results from the fact that the VNA measurement perturbatively probes the cavity field, which is no longer an eigenstate of the hybridized system.

Re$(a/a_0)$ and Re$(b/a_0)$ are plotted as a function of time in \figref[c]{fig:coherentFrequency}, using typical experimental parameters. \footnote{The data used in \autoref{fig:coherentFrequency} are from Ref. \citenum{Match2019}.  The system parameters are $\omega_c/2\pi = 14.39$ GHz, $\Delta \omega_c = 2.14$ MHz, the Gilbert damping is $\approx 1 \times 10^{-4}$, the FMR resonance is determined by the Kittel dispersion $\omega_m/2\pi = \gamma \left(H+H_A\right)$, with a gyromagnetic ratio of $\gamma = 28 \times 2\pi \mu_0$ GHz/T and a shape anisotropy field of $\mu_0H_A = -44$ mT, and the coupling strength is $J = 26.5$ MHz.}  The decay timescale of ns corresponds to a GHz frequency cavity mode.  The behaviour here again mirrors the spring-coupled pendulums: $a$ and $b$ oscillate rapidly with a 180$^\circ$ phase shift, and a MHz frequency beating is observed at the generalized Rabi frequency.  These Rabi-like oscillations are the key features seen in the microwave transmission spectra, plotted as the solid red curve in \figref[d]{fig:coherentFrequency}.  Here, the rapid oscillations are time averaged away since the VNA probes $|\text{S}_{21}|^2$.   In panel (d), the dashed red curve indicates the 180$^\circ$ phase-shifted magnetization, $|b/a_0|^2$, calculated according to Eq. \eqref{eq:basisChange} and the black decay curve is the sum, $|a/a_0|^2+|b/a_0|^2$.  At $\omega_c = \omega_m$, $C_{11} = C_{12} = 1/2$ and $\tau_{+} = \tau_{-}$.  Therefore a complete extinction in the $|\text{S}_{21}|^2$-t spectrum is observed, resulting in an equal energy distribution between the spin and photon subsystems. \cite{Match2019}
\subsection{Dissipative Cavity Magnonics} \label{sec:dissipative}
In general, both coherent and dissipative interactions coexist in cavity magnonics.  However, it is possible to design systems such that one type of coupling, either coherent or dissipative, dominates.   The motivation to enhance or suppress certain interactions depends on the objective of a given device or experiment. For example, early cavity magnonics experiments emphasized quantum information and transduction applications, in which case a strong coherent coupling is typically advantageous.  Therefore, these experiments were implicitly designed to enhance coherent coupling, for example, by placing the magnetic sample at an RF magnetic field antinode.  Such placement enhances the Zeeman interaction, and, although unknown at the time, may actively suppress dissipative coupling.  

Since direct magnon-photon coupling, i.e., coupling via the Zeeman interaction, is coherent, dissipative coupling must be realized by introducing an indirect interaction.  This can be achieved in a variety of ways, generally classified as one-tone or two-tone experiments.\cite{Wang2020}  In one-tone experiments, only the photon mode is driven and the indirect interaction is realized by coupling both photon and magnon to a common auxiliary mode.  The nature of the auxiliary mode depends on the experimental system, for example, it could be a traveling-wave reservoir\cite{Rao2020} or an auxiliary cavity mode.\cite{Yu2019}  Regardless, indirect interactions in a one-tone platform can be understood through cooperative radiative damping, as described by the theory of reservoir engineering, \cite{Metelmann2015} which leads to a coupling strength that depends on the extrinsic dissipation of the photon and magnon.  Classically, this aligns with the picture of base-mediated coupling.  \footnote{It is worth noting that the origin of dissipative coupling has been discussed microscopically on the basis of $\mathcal{PT}$ symmetry in exchange coupled, \cite{Cao2019, Tserkovnyak2020} and spin-orbit torque systems \cite{Proskurin2020a}}
In two-tone experiments, both the photon and magnon are driven independently,\cite{Grigoryan2017, Boventer2019, Boventer2019b, Grigoryan2019} and the indirect interaction physically reflects the additional drive of the cavity mode via the independently driven magnons.  Both one and two-tone systems have an effective complex coupling strength, resulting in a mix of coherent and dissipative behaviour.  Importantly, the amplitude and phase of this complex coupling strength can be tuned, enabling systematic control between coherent and dissipative coupling.  The exact mechanism of tuning the complex coupling strength depends on the experimental platform used to realize the indirect interaction.  For example, coherent coupling may be explicitly suppressed at an RF magnetic field node,\cite{Harder2018} the traveling-wave phase delay may be changed by repositioning the magnetic sample,\cite{Rao2020} or the relative amplitude and phase of the two driving signals may be controlled.\cite{Boventer2019b}     

Beyond new applications, the discovery of dissipative coupling expanded our understanding of dissipation in the cavity magnonic system.  Although in the coherent system dissipation does play a role in the amplitude decay of Rabi oscillations, and the linewidth evolution can be used as a signature of coupling,  dissipation does not significantly impact the strongly coupled coherent dispersion, which is the key point of interest for applications of a cavity magnonic transducer.  Moreover, large dissipation reduces the cooperativity, which is an important figure of merit for coherent systems.  For these reasons, dissipation is often considered a nuisance in coherent cavity magnonics.  This explains why experiments focusing on coherent coupling typically employ quasi-closed cavities where both intrinsic dissipation, due to cavity losses and magnon decay, and extrinsic dissipation, due to environmental coupling, are small.  On the other hand dissipative coupling in cavity magnonics was first discovered by adopting open cavities, \cite{Harder2018a} where the extrinsic damping of the cavity mode, $\kappa_c \sim \left(0.01 - 1\right) \omega_c$, is much greater than the intrinsic damping.  In this section and in \autoref{sec:physicsTraveling}, we will focus on this ``dissipation induced" coupling, as opposed to the indirect two-tone induced interaction, but we must point out that both systems display level attraction, and importantly, the general features discussed here can be found in any system that displays level attraction.

The dissipative magnon-photon interaction is mediated by a mutual coupling of the magnon and cavity photon to a traveling-wave reservoir.  This results in an effective dissipative coupling strength $\Gamma = \sqrt{\kappa_c \kappa_m}$, where $\kappa_m$ is the extrinsic dissipation of the magnon.  Although the magnetic dipole interaction between the spin system and the environment is weak, and therefore $\kappa_m$ is small, a large $\Gamma$ can be achieved by increasing the extrinsic damping of the cavity mode.  This is why dissipative coupling, and level attraction, are commonly achieved using open cavities with large extrinsic dissipation and traveling-waves.  However, we again stress that both coherent and dissipative coupling exist simultaneously, and level attraction is not unique to open cavities.  For example, dissipative coupling may be realized via the broad antiresonance of a quasi-closed cavity\cite{Rao2019} and level attraction has been demonstrated in two-tone experiments, \cite{Grigoryan2017, Boventer2019, Boventer2019b, Grigoryan2019} which are not dissipatively coupled in the sense discussed here.

Phenomenologically, to account for both coherent and dissipative interactions the real spin-photon coupling $J$ in Eq. \eqref{Eq:coherent_Hamiltonian} must be replaced by a complex parameter, leading to the following cavity-magnonic Hamiltonian:
\begin{equation}
H=\hbar\widetilde{\omega}_ca^\dag a+\hbar\widetilde{\omega}_mb^\dag b+\hbar (J-i\Gamma)(a^\dag b+b^\dag a).
\label{Eq:H_dis_Wang}
\end{equation}
Here, $J$ and $\Gamma$ are real parameters that characterize the strength of coherent and dissipative interactions, respectively, and the complex frequencies $\widetilde{\omega}_c = \omega_c - i \beta_L = \omega_c - i \left(\beta + \kappa_c\right)$ and $\widetilde{\omega}_m = \omega_m - i \alpha_L = \omega_m - i \left(\alpha + \kappa_m\right)$ include both the intrinsic and extrinsic dissipation of the cavity and magnon, respectively.  Regardless of the detailed implementation, systems that display level attraction will have a Hamiltonian analogous to Eq. \eqref{Eq:H_dis_Wang}, which, therefore, serves as an important general case study.  The interaction in Eq. \eqref{Eq:H_dis_Wang} can be understood in several ways.  On one hand, it is natural to include a complex coupling constant in the effective Hamiltonian approach that originally led to Eq. \eqref{Eq:coherent_Hamiltonian}.  Alternatively, the complex interaction may be motivated phenomenologically through an electrodynamic approach, with Faraday's law and Amp\`ere's law accounting for the coherent interaction and Lenz's law generating a back-action that results in a dissipative coupling. \cite{Harder2018a}   These phenomenological motivations are most closely related to the dasphpot-coupled pendulums.  A more detailed approach, which explicitly accounts for the dissipation and aligns with the base-mediated pendulums, will be discussed in \autoref{sec:physicsTraveling}.

From the Hamiltonian of Eq. \eqref{Eq:H_dis_Wang}, the cavity mode and magnon equations of motion are determined to be
\begin{subequations}
\label{Eq:disspative_ME}
\begin{align}
\dot{a} &= -i\widetilde{\omega}_c a - i \left(J - i \Gamma\right) b, \\
\dot{b} &= -i\widetilde{\omega}_m b - i \left(J - i \Gamma\right) a.
\end{align}
\end{subequations}
Therefore, the complex eigenfrequencies are
\begin{equation}
\label{eq:dissipativeEigen}
\widetilde{\omega}_{\pm}= \dfrac{1}{2}\left[\widetilde{\omega}_c + \widetilde{\omega}_m
\pm\sqrt{(\widetilde{\omega}_c-\widetilde{\omega}_m)^2+4(J-i\Gamma)^2}\right],
\end{equation}
and using input-output theory,\cite{Wang2019b}
\begin{equation}
\text{S}_{21} = 1-\frac{i\kappa_c(\omega-\widetilde{\omega}_m)}{(\omega-\widetilde{\omega}_c)(\omega-\widetilde{\omega}_m)-(J-i\Gamma)^2}.
\label{eq:dissipativeS}
\end{equation}
Here, in analogy to previous definitions, $\widetilde{\omega}_c = \omega_c - i \beta_L = \omega_c - i \left(\beta + \kappa_c\right)$ and $\widetilde{\omega}_m = \omega_m - i \alpha_L = \omega_m - i \left(\alpha + \kappa_m\right)$.  These are similar to the results of \autoref{sec:coherent}, with the difference in $\text{S}_{21}$ due to the measurement configuration in typical open cavities which consist of waveguides galvanically coupled to a resonant structure, and therefore, a transmission resonance appears as a dip in $\text{S}_{21}$.\cite{Wang2019b}
\begin{figure} [t!]
\begin{center}
\includegraphics[width=8.5cm]{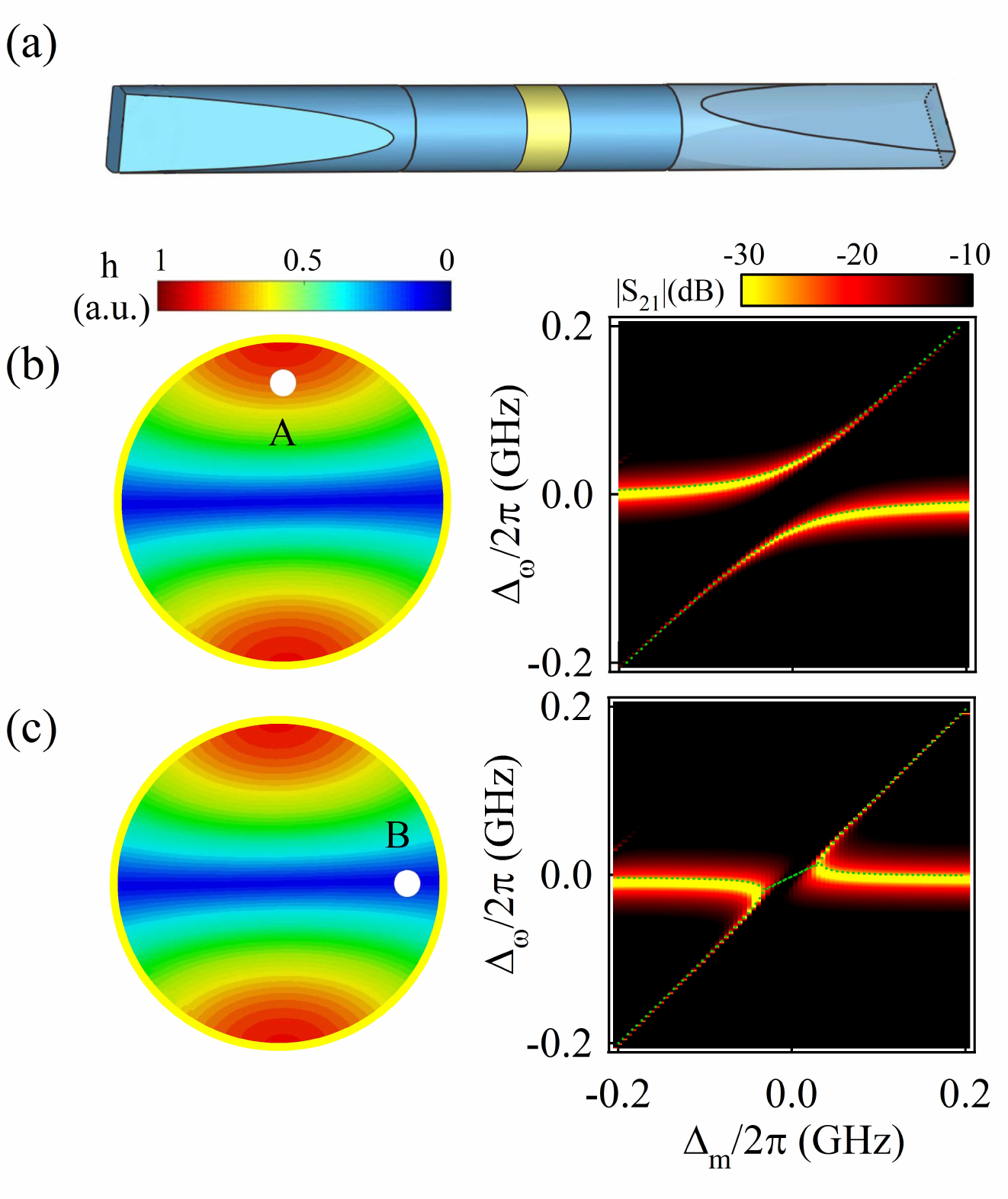}
\caption{(a) An open 1D Fabry-Perot-like cavity enables both coherent and dissipative cavity magnonics.  (b) When a YIG sample is placed at a magnetic field antinode, the coherent interaction dominates and level repulsion is observed in the transmission spectrum, plotted here as a function of $\Delta_\omega=\omega-\omega_c$ and $\Delta_m=\omega_m-\omega_c$.  (c) When the YIG is placed at a magnetic field node, the dissipative interaction dominates and level attraction is seen in the transmission spectra.  Reproduced with permission from Yao et al., Phys. Rev. B. \textbf{92}, 184407 (2015)\cite{Yao2015} and Harder et al., Phys. Rev. Lett. \textbf{121}, 137203 (2018).  Copyright 2015 and 2018 American Physical Society, respectively.}
\label{fig:dissipativeFrequency}
\end{center}
\end{figure}

An example of a dissipative cavity magnonic system is shown in \figref[a]{fig:dissipativeFrequency}.  This 1D Fabry-Perot-like cavity has two circular-rectangular transitions with a relative angle of $45^\circ$, supporting both a resonant and traveling-wave mode, and therefore, both coherent and dissipative coupling will be present.  However, by controlling the mode profile, it is possible to suppress either form of coupling.  A mapping of the cavity magnetic field strength is shown on the left side of panel (b) and (c).  When a YIG sphere is placed at the magnetic field antinode of position A, a direct coherent coupling dominates and level repulsion is observed in the VNA transmission spectrum as shown in \figref[b]{fig:dissipativeFrequency}.  However, at the magnetic field node, position B in \figref[c]{fig:dissipativeFrequency}, the coherent interaction is suppressed, indirect coupling via the traveling-wave dominates, and level attraction is observed.  

For purely dissipative coupling, the hybridized dispersion can be written as
\begin{equation}
\widetilde{\omega}_\pm = \frac{1}{2} \left[\widetilde{\omega}_c + \widetilde{\omega}_m \pm \widetilde{R}\right],
\end{equation}
where $\widetilde{R} = \sqrt{\widetilde{\Delta}^2 - 4 \Gamma^2}$ and $\widetilde{\Delta} = \widetilde{\omega}_c - \widetilde{\omega}_m$.  Compared to coherent coupling, the major change is that $J^2 \to -\Gamma^2$, and therefore, as the detuning becomes small, the effective Rabi frequency $\widetilde{R}$ becomes complex, even for small intrinsic damping.  This has a drastic effect on the hybridization behaviour, resulting in the transition to level attraction.  The global structure of the complex eigenfrequencies is best characterized by the branch points of $\widetilde{R}$, known as the exceptional points (EPs).  An EP represents a total degeneracy in $\widetilde{\omega}_\pm$, i.e., both the dispersion and linewidth are degenerate, which also leads to a coalescence of eigenmodes.  True EP's require $\alpha = \beta$ so that $\widetilde{\omega}_m = \widetilde{\omega}_c$ at zero detuning, and while this strict condition may not always be met in cavity magnonics, EP-like behaviour will be encountered as long as $\Gamma \ne 0$, though the magnitude of the effect will depend on $J/\Gamma$.  Furthermore, although EPs are more common in systems dominated by dissipative coupling, a branch point can also be reached in coherent cavity magnonic systems at weak coupling.\cite{Harder2017}

\begin{figure} [t]
\begin{center}
\includegraphics[width=8.5cm]{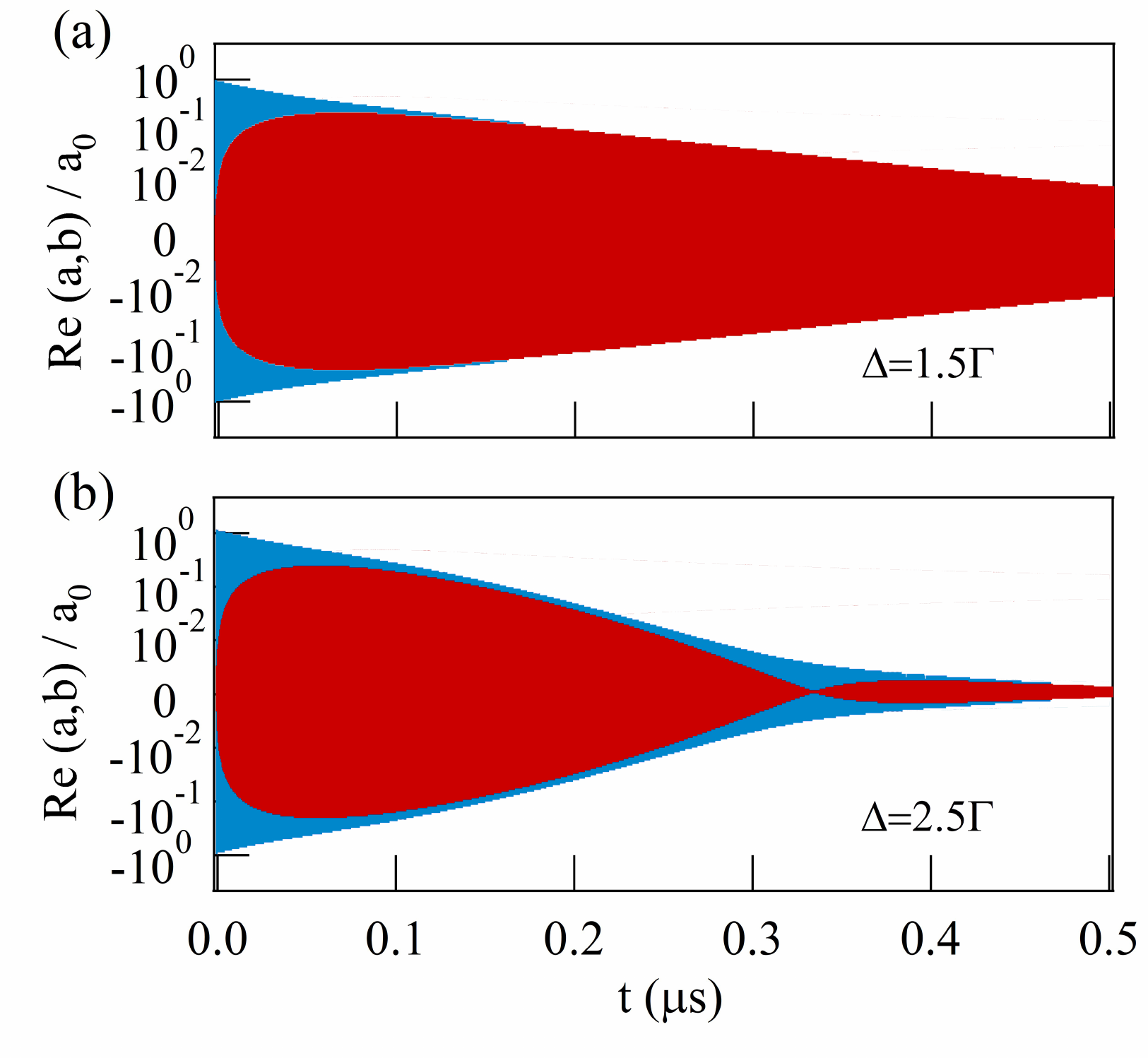}
\caption{Time evolution of a dissipative cavity magnonic system.  Red and blue curves are the envelopes of Re$(a)$ and Re$(b)$ calculated according to Eq. \eqref{eq:dissipativeTime}, respectively. The high frequency $(\omega_c + \omega_m)/2$ oscillations are not plotted.  (a) Between two EPs ($\Delta=1.5\Gamma$) $\widetilde{R}$ is imaginary and the system decays without beating. (b)  However, outside the EPs ($\Delta=2.5\Gamma$), $\widetilde{R}$ is real and Rabi oscillations are observed.  In this calculation, $\alpha = \beta$, $\kappa_m = 4\alpha$, $\kappa_c = \kappa_m = \Gamma$, and $\omega_c = 2 \Gamma$.}
\label{fig:dissipativeTime}
\end{center}
\end{figure}

The basis transformation of Eq. \eqref{eq:basisChange} can be used to examine the time evolution of purely dissipative coupling.   With the initial conditions $a(t)=a_0$ and $b(t)=0$,
\begin{subequations}
\label{eq:dissipativeTime}
\begin{align}
\frac{a}{a_0} &= \frac{1}{2} \left(1+\frac{\widetilde{\Delta}}{\widetilde{R}}\right)e^{-i\widetilde{\omega}_+ t} + \frac{1}{2} \left(1-\frac{\widetilde{\Delta}}{\widetilde{R}}\right)e^{-i\widetilde{\omega}_- t}, \\
\frac{b}{a_0} &= \frac{1}{2} \sqrt{1-\left(\frac{\widetilde{\Delta}}{\widetilde{R}}\right)^2}\left(e^{-i\widetilde{\omega}_+ t} -e^{-i\widetilde{\omega}_- t}\right).
\end{align}
\end{subequations}
When $\omega_m \approx \omega_c$, the real part of $\widetilde{R}$ is small, so that, $\omega_\pm \approx \left(\omega_c + \omega_m\right)/2$.  Therefore, both the cavity and magnon mode oscillate with a frequency of $(\omega_c+\omega_m)/2$ and the phase difference between $a(t)$ and $b(t)$ at late times is $\pi+\arcsin(\frac{\omega_m-\omega_c}{\sqrt{(\alpha_L - \beta_L)^2+4\Gamma^2}})$.  This means that at zero detuning, the magnon and photon oscillate 180$^\circ$ out-of-phase, similar to the base-mediated pendulums.

The time-evolution of the dissipative system is plotted in \autoref{fig:dissipativeTime} using Eqs. \eqref{eq:dissipativeTime} for $\alpha=\beta$ and $\kappa_m=\kappa_c$.  In this case, two EPs appear at $\Delta=\omega_m-\omega_c=\pm2\Gamma$ meaning $\widetilde{R}$ is imaginary for $|\Delta| < 2\Gamma$ and real for $|\Delta| > 2\Gamma$.  The time evolution decays without oscillation when $|\Delta| < 2\Gamma$, as shown in \figref[a]{fig:dissipativeTime} for $|\Delta| = 1.5\Gamma$. In this region, the two eigenmodes are synchronized.  However, when $|\Delta| > 2 \Gamma$, the real $\widetilde{R}$ produces an additional $\widetilde{R}/2$ oscillation.  The resulting beating pattern can be seen in \figref[b]{fig:dissipativeTime} when $|\Delta| = 2.5\Gamma$.
\subsection{Dissipative Cavity Magnonics with Traveling Photons} \label{sec:physicsTraveling}
Dissipative interactions in cavity magnonics are mediated by the mutual coupling of cavity and magnon modes to a common reservoir.  Therefore, going beyond the phenomenological picture of \autoref{sec:dissipative} requires the explicit introduction of the reservoir modes. \cite{Metelmann2015, Rao2020}  This approach uses the input-output formalism, analogous to the analysis of coherent coupling.  However, now, both the magnon and cavity mode must be coupled to a single external reservoir, whereas in the coherent system, the cavity mode alone was coupled to two independent external photon baths.  To clarify the physical picture, consider the experimental setup shown in \figref[a]{fig:travelingFrequency}.  Here, a transmission line connected between port 1 and 2 forms a traveling-wave reservoir, which couples to the cavity mode of a cross cavity and to the magnon modes in a YIG sphere.  To achieve purely reservoir-mediated coupling, the direct coherent interaction between the cavity and magnon is suppressed by shielding the cross cavity with a metal box.

\begin{figure} [t]
\begin{center}\
\includegraphics[width=8.5cm]{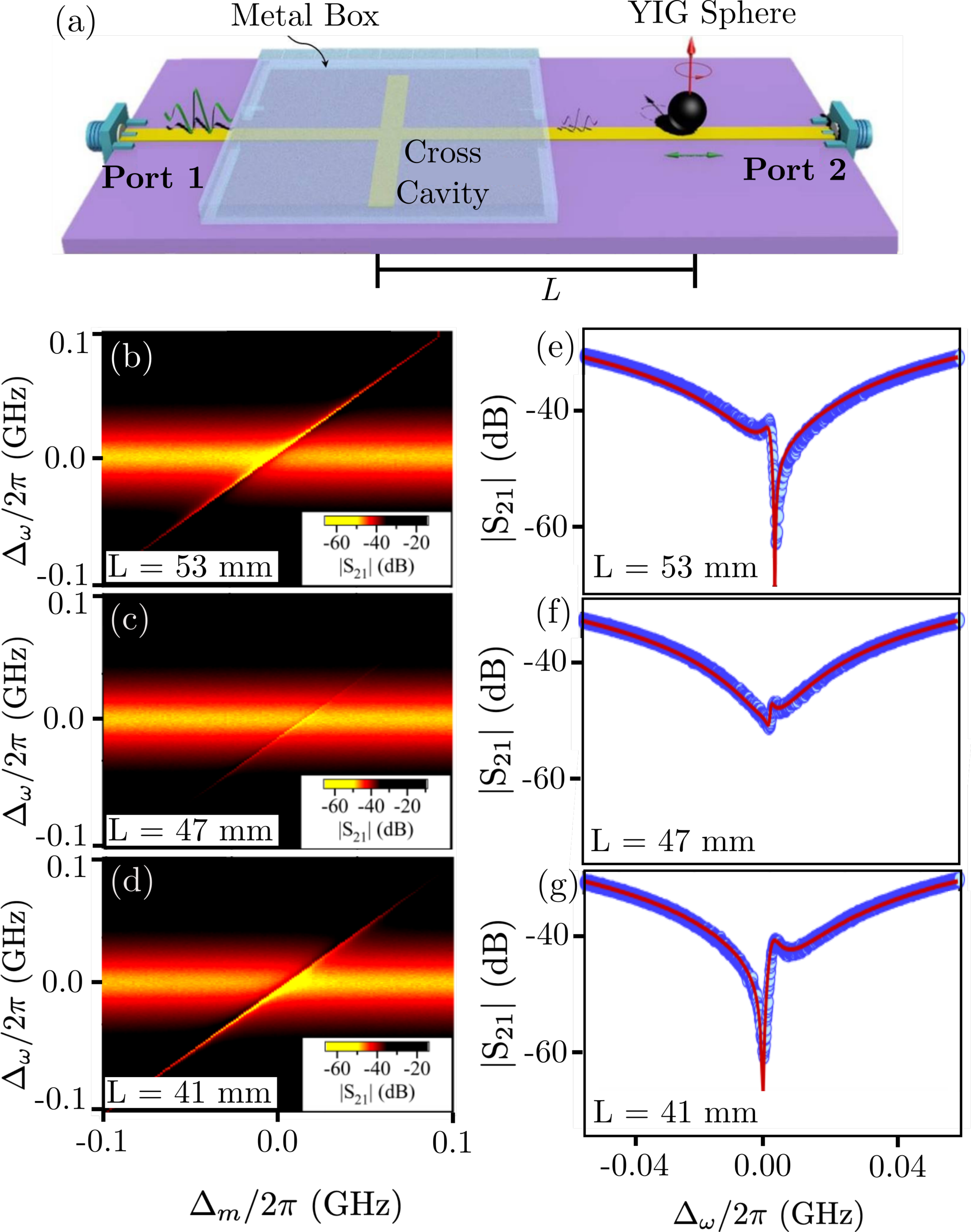}
\caption{(a) Schematic diagram of the experimental set-up. The cross cavity is shielded by a metal box, so that it can only interact with the magnon mode via the transmission line.  (b) Microwave transmission $|\text{S}_{21}|$ at a YIG-cross-cavity separation of $L = 53$ mm (c) $L = 47$ mm and (d) $L = 41$ mm, demonstrating cavity-magnon coupling due to the traveling-wave mediation.  The spectra shows a mix of dissapative and coherent characteristics with essentially no coupling at $L = 47$ mm where $\kappa_m \approx 0$.  These characteristics are confirmed by the two resonant dips in the line cuts at $\omega_m = \omega_c$ shown in (e) and (g), and the single broad resonance in (f).  The sharp resonance in (e) and (f) is due to a competition between the coherent and dissipative interactions which results in a zero damping condition where the linewidth of the hybridized mode approaches 0.  Reproduced with permission from Rao et al., Phys. Rev. B \textbf{101}, 064404 (2020).  Copyright 2020 American Physical Society.}
\label{fig:travelingFrequency}
\end{center}
\end{figure}

As derived in \hyperref[Sec:appendixF]{Appendix F}, the transmission spectrum for traveling-wave-mediated cavity magnonics is 
\begin{align}
&\text{S}_{21} = 1\nonumber \\
&-i\left[\frac{\kappa_m\left(\omega - \widetilde{\omega}_c\right)e^{2i\left(\phi + \theta\right)} + \kappa_c \left(\omega - \widetilde{\omega}_m\right) - 2 i \kappa_c \kappa_m e^{i\left(2\phi + \theta\right)}}{\left(\omega - \widetilde{\omega}_c\right)\left(\omega - \widetilde{\omega}_m\right) + \kappa_c \kappa_m e^{2i\phi}}\right].\label{eq:travelingSpectrum}
\end{align}
In this expression, $\phi$ is the traveling-wave phase delay between the cavity and magnon, which are spatially separated by a distance $L$.  In general, $\phi = k L$, where $k$ is the traveling-wave vector.  However, when focusing on the behaviour near $\omega_c \approx \omega_m$, $\phi$ is approximately $k$ independent.  The traveling-wave will also experience a phase shift when passing through a resonance.  This is characterized by the resonance phase $\theta$.  

The experiment shown in \autoref{fig:travelingFrequency} is, in effect, a two-tone experiment, \cite{Grigoryan2017, Boventer2019, Rao2020} since both the cavity and magnon modes are independently driven and the direct interaction is suppressed.  However, there is also a controllable, indirect, dissipative interaction due to the cooperative radiation damping of the cavity and magnon to the traveling-waves.  These effects all play a role in the observed transmission spectra.  In Eq. \eqref{eq:travelingSpectrum}, the $\kappa_c \left(\omega - \widetilde{\omega}_m\right)$ contribution is due to the direct driving of the cavity mode, which will be present in all cavity magnonic systems due to the driving of the cavity photons.  The additional $\kappa_m \left(\omega - \widetilde{\omega}_c\right)$ contribution results from the direct driving of the magnon mode.  Together, these two terms result in two-tone behaviour, with a phase delay controlled by the sample placement via $\phi$.\cite{Grigoryan2017, Boventer2019, Rao2020}  However, in addition to this two-tone behaviour, there is an indirect coupling which produces the $\kappa_c\kappa_m e^{i\left(2\phi+\theta\right)}$ term, since both the cavity and magnon modes experience radiative damping to the traveling-wave reservoir.  By controlling both the amplitude and phase of this indirect coupling, the nature of the hybridization can be tuned between level repulsion and level attraction, as we can see by examining the transmission resonance from Eq. \eqref{eq:travelingSpectrum}, which is characterized by a dip at 
\begin{equation}
\widetilde{\omega}_{\pm}= \dfrac{1}{2}\left[\widetilde{\omega}_c + \widetilde{\omega}_m
\pm\sqrt{(\widetilde{\omega}_c-\widetilde{\omega}_m)^2-4\kappa_c\kappa_m e^{2i\phi}}\right].
\end{equation}

In the experimental setup shown in \figref[a]{fig:travelingFrequency}, the YIG sphere position can be precisely controlled by a 3D-stage.  Moving the YIG along the axis of the transmission line tunes the traveling phase $\phi$ and, therefore, controls the nature and strength of the reservoir-mediated coupling.  To ensure there is no direct cavity-magnon interaction at all YIG positions, the cross-cavity is shielded by a metal box.  The $\Delta_\omega-\Delta_m$ spectra measured by a VNA at three different YIG positions is shown in \hyperref[fig:travelingFrequency]{Figs. 11 (b) - (d)}.  The left-right asymmetry (across the diagonal FMR dispersion) is the result of two-tone interference between the directly driven cavity and magnon.  However, such effects do not produce the transition from level repulsion to level attraction, which is due to the indirect traveling-wave-mediated coupling.  In general, the transmission shows a mixture of coherent and dissipative characteristics, with a broad cavity resonance due to the extrinsic damping.  When $\kappa_m \approx 0$, as shown in panel (c), both real and imaginary components of the coupling vanish and only the cavity mode is observed (the antiresonance at $\omega = \omega_m$ is clearly visible, but this is not indicative of cavity-magnon coupling. \cite{Harder2016, Rao2020})  \hyperref[fig:travelingFrequency]{Figures 11 (e) - (g)} show the microwave transmission as a function of $\Delta_\omega$ at $\omega_m=\omega_c$ for the same YIG-cavity separations as panels (b) - (d).  In general, two resonant dips indicate the indirect cavity-magnon coupling, with the single broad peak of panel (f) a result of $\kappa_m \approx 0$. The extremely sharp resonance observed in \hyperref[fig:travelingFrequency]{Figs. 11 (e)} and \hyperref[fig:travelingFrequency]{11 (g)}, which has a contrast ratio of up to 30 dB, is a result of competition between the real and imaginary components of the coupling, which leads to a zero damping condition where the hybridized linewidth approaches 0.\cite{Wang2019b}

\begin{figure} [t]
\begin{center}
\includegraphics[width=8.5cm]{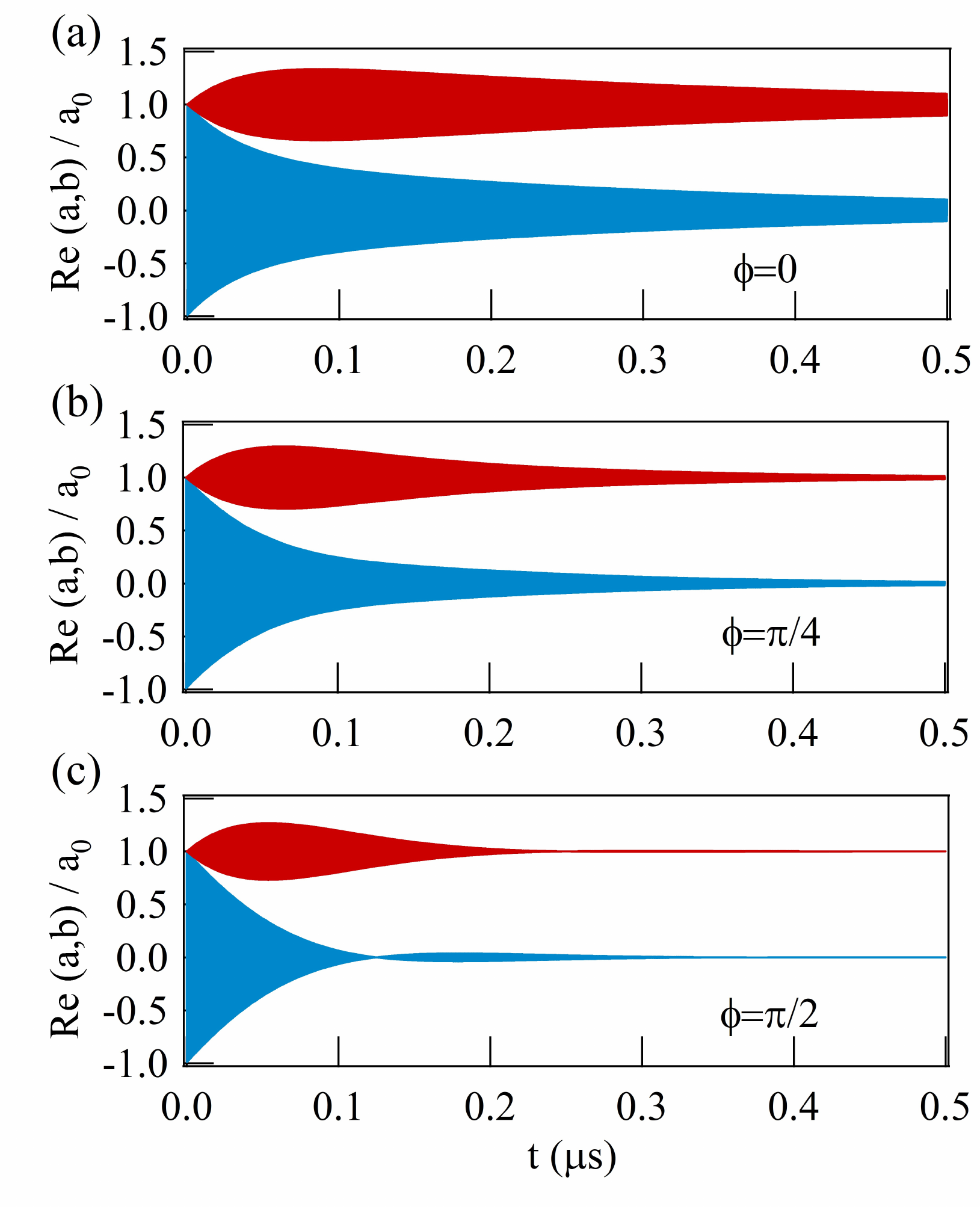}
\caption{Time evolution of a traveling-wave-mediated system for (a) $\phi=0$, (b) $\phi=\pi/4$, and (c) $\phi=\pi/2$. Blue and red curves are the envelopes of Re$(a)/a_0$ and Re$(b)/a_0$, respectively, calculated using Eq. \eqref{eq:travelingTE} with the rapid oscillations at $\omega_c$ excluded and the time evolution of the magnon offset by $1$ for clarity.  In this calculation, $\alpha = \beta$, $\kappa_m = 4\alpha$, $\kappa_c = \kappa_m = \Gamma$, and $\omega_c = 2 \Gamma$.}
\label{fig:travelingTime}
\end{center}
\end{figure}

The time evolution of $a(t)$ and $b(t)$ will be identical to the dissipative system of \autoref{sec:dissipative}, specifically Eq. \eqref{eq:dissipativeTime} for initial conditions $a(t)=a_0$ and $b(t)=0$ with $\Gamma \to \sqrt{\kappa_c \kappa_m}e^{i\phi}$.  Since $\phi$ controls the coherent/dissipative nature of the interaction, it is instructive to examine the $\phi$ dependence by taking $\Delta = 0$, $\alpha = \beta$ and $\kappa_m = \kappa_c$, in which case
\begin{subequations}
\label{eq:travelingTE}
\begin{align}
a(t)&= \frac{a_0}{2}e^{-(\beta+\kappa_c)t}e^{-i\omega_c t}(e^{G t}+e^{-G t}),\\
b(t)&= \frac{a_0}{2}e^{-(\beta+\kappa_c)t}e^{-i\omega_c t}(e^{G t}-e^{-G t}),
\end{align}
\end{subequations}
where $G=\kappa_c(\cos\phi+i\sin\phi)$ is the complex coupling constant.  The traveling phase has a two-fold effect.  First, the late time exponential decay rate is $\beta+\kappa_c(1-|\cos\phi|)$, and second, the time evolution has a slowly oscillating envelope at frequency $\kappa_c\sin\phi$.  Therefore, $\phi=(n+1/2)\pi$ corresponds to a purely coherent interaction and the decay rate reaches its maximum of $\beta+\kappa_c$, while the envelope oscillation frequency is maximized at $\kappa_c$.  On the other hand, $\phi = 2n\pi$ corresponds to a purely dissipative interaction with a minimum decay rate of $\beta$ and a minimum envelope oscillation frequency of 0 Hz.  These features are shown in the time evolution plots of \autoref{fig:travelingTime}.  In this figure, the blue and red curves are the envelopes of Re$(a)/a_0$ and Re$(b)/a_0$, respectively, calculated using Eq. \eqref{eq:travelingTE} with the rapid oscillations at $\omega_c$ excluded and the time evolution of the magnon offset by $1$ for clarity.  The minimum decay, slow oscillation is shown in \figref[a]{fig:travelingTime} for $\phi = 0$, while \figref[c]{fig:travelingTime} shows the maximum decay, fast oscillation at $\phi = \pi/2$.  In \figref[b]{fig:travelingTime}, $\phi = \pi/4$, and therefore, the coupling has equal coherent and dissipative contributions.  This $\phi$ controllability is advantageous for reservoir-based engineering of cavity magnonics.
\subsection{Summary of Coherent and Dissipative Cavity Magnonics}
Both coherent and dissipative interactions play an important role in cavity magnonics, resulting in distinct hybridization behaviour.  Coherent interactions are characterized by level repulsion in the dispersion and linewidth attraction, while dissipative coupling results in level attraction in the dispersion and linewidth repulsion.  Coherent coupling is the result of the magnetic dipole interaction between the cavity magnetic field and the spin ensemble, and therefore, the coherent coupling strength may be controlled by increasing the spin density in the cavity or enhancing the filling factor by careful cavity design.  Furthermore, strong coherent coupling can be achieved by minimizing intrinsic losses, for example, by using high quality cavities.  Therefore, coherent interactions are typically associated with closed-cavities, where external coupling is solely a means to perturbatively probe the linear response of the cavity-magnon system.  On the other hand, dissipative coupling is an indirect interaction mediated by the mutual coupling of the cavity and magnon mode to a common reservoir.  This requires both large extrinsic coupling rates and a traveling-wave reservoir.  Since the magnetic dipole interaction to the reservoir is typically weak, large dissipative coupling is normally achieved in open-cavities.  However, open-cavities support both coherent and dissipative coupling, and the cavity-magnon interaction will generally be a mix of the two unless measures are taken to suppress one interaction, for example, by placing the sample at a magnetic field antinode (node) in order to suppress the dissipative (coherent) interaction.  The ability to selectively engineer and control the nature and strength of cavity magnonic interactions is key to future device development.
\section{Applications of Coherent and Dissipative Cavity Magnonics} \label{sec:applications}
In this section, we summarize select applications driving the development of cavity magnonic technologies. 
\subsection{Transducers}  

Magnon-photon coupling may be used as a bridge between otherwise weakly interacting systems, enabling, for example, qubit-magnon coupling, \cite{Tabuchi2015, Tabuchi2014, LachanceQuirion2019} optical-to-microwave frequency conversion,\cite{Hisatomi2016, Lambert2019}  and magnon-phonon interactions.\cite{Zhang2015f,Li2018a}  As a transducer, the cavity-magnonic platform is highly versatile and controllable.  Both coherent and dissipative coupling may be realized via integration-friendly architectures, such as planar cavities and lithographically defined resonators.   Furthermore, strong coupling may be achieved even with nano-scale magnetic elements, \cite{Li2019a, Hou2019} active elements are not required, and the coupling strength (hence the efficacy of transduction) may be controlled, \cite{Zhang2014, Bai2016, Boventer2019, Yang2019} with functionality even at low cooperativities. \cite{Zhang2014, Liu2019}

\begin{figure} [t]
\begin{center}
\includegraphics[width=9cm]{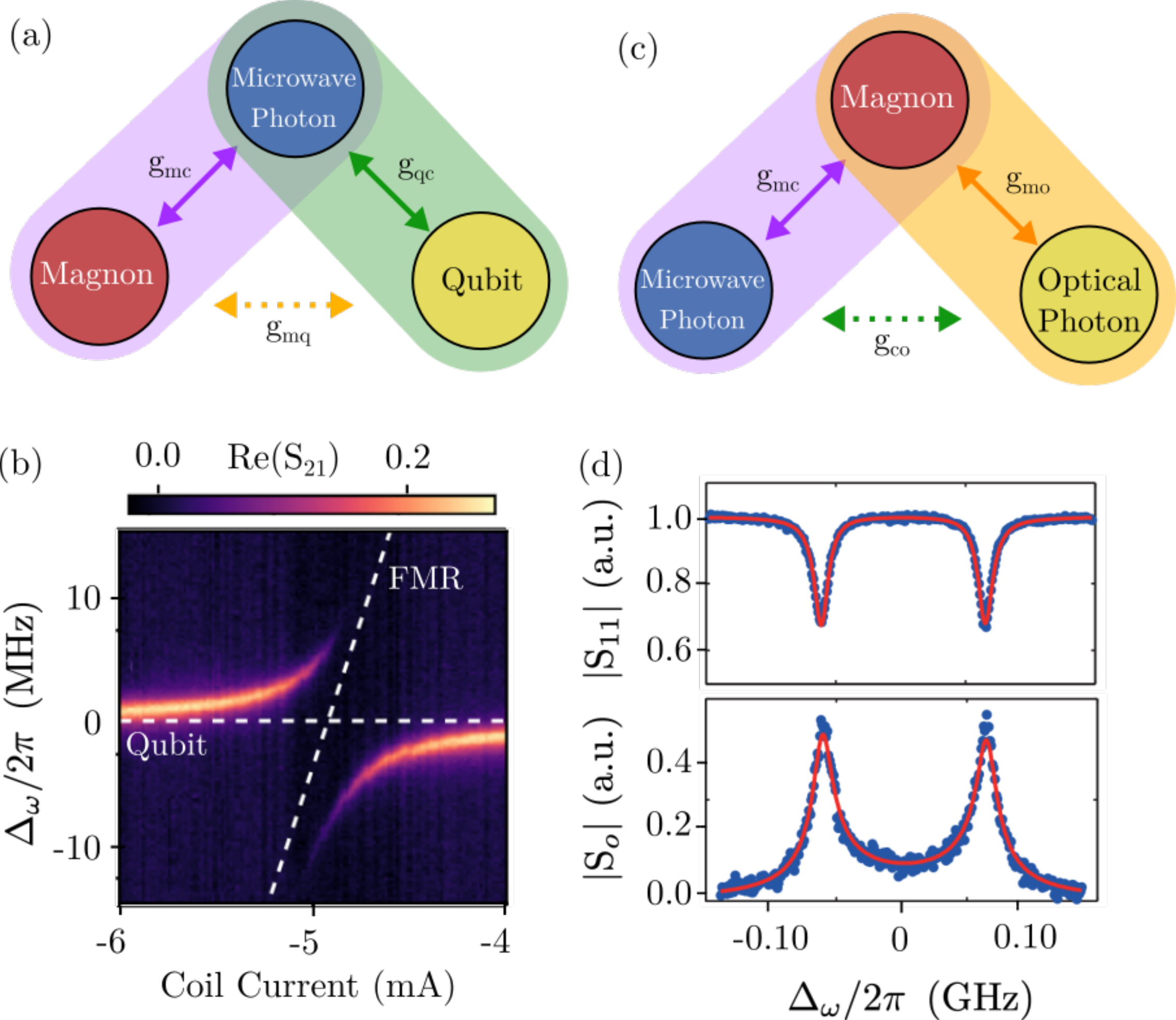}
\caption{(a) Cavity magnonic-transduction enables an indirect qubit-magnon interaction, opening the door to quantum magnonics. (b) Despite the indirect nature of the interaction, cavity transmission measurements have confirmed strong qubit-magnon coupling, which can be exploited for hybrid quantum information devices and quantum sensing of magnons. (c) Optical-to-microwave conversion can also be realized via cavity magnonic-transduction.  (d) Here, a reduction in the microwave reflectivity, $|\text{S}_{11}|$, is accompanied by the generation of optical photons, as detected by a heterodyne measurement using a high-speed photodiode, indicating efficient optical-microwave conversion.  Panels (a) - (c) are reproduced with permission from Lachance-Quirion et al., Appl. Phys. Express \textbf{12}, 070101 (2019).  Copyright 2019 Institute of Physics.  Panel (d) is reproduced with permission from Hisatomi et al., Phys. Rev. B \textbf{93}, 174427 (2016).  Copyright 2016 American Physical Society.}
\label{fig:transducer}
\end{center}
\end{figure}

As the basis of quantum magnonics, \cite{Kurizki2015, LachanceQuirion2019} cavity-magnonic transduction enables an indirect interaction between magnons and superconducting qubits, schematically illustrated in \figref[a]{fig:transducer}.  With their large electric dipole moments, superconducting qubits strongly couple to the electric field of a microwave cavity mode.  At the same time, the magnetic dipole moment of a magnetically ordered material will couple to the microwave magnetic field.  Therefore, when a superconducting qubit and a magnetically ordered material are strategically placed in the same microwave cavity, an indirect qubit-magnon coupling will form. \cite{Tabuchi2014, LachanceQuirion2017}  The characteristic anticrossing of \figref[b]{fig:transducer} experimentally illustrates the strong, coherent qubit-magnon interaction.  In this experiment, a transmon-type superconducting qubit was coupled to the FMR mode of a YIG sphere via a 3D microwave cavity.  The qubit was placed at a magnetic field node and shielded from stray magnetic fields, while the FMR resonance was controlled by a coil current.  A qubit-magnon coupling of $g_{mq} \approx 8$ MHz was achieved, corresponding to a large cooperativity of $C \approx 30$, aided by the narrow qubit/magnon linewidths of $\approx 1$ MHz.\cite{LachanceQuirion2017}  

Qubit-magnon transduction allows macroscopic, magnetically-ordered systems to be utilized within the framework of cavity quantum electrodynamics, enabling hybrid quantum circuits with complimentary qubit/magnon properties.  For example, while the harmonic nature of magnetostatic modes impedes the creation of non-classical states, the anharmonicity of the superconducting qubit enables encoding of quantum information. \cite{Tabuchi2014, LachanceQuirion2017}  Qubit-magnon coupling can also be exploited as a magnetostatic probe, with sensitivity below the single magnon level. \cite{Tabuchi2014, Morris2016}

As illustrated in \figref[c]{fig:transducer}, the magnon in a cavity-magnonic system may also be used as an intermediary between microwave and optical photons.\cite{Hisatomi2016, Lambert2019}  In certain ferrimagnetic materials, such as YIG, large spin-orbit coupling leads to a magneto-optical interaction, which indirectly couples magnons to optical electric fields.  On the other hand, the magnetic dipole interaction, which is negligible at optical frequencies, is large in the microwave regime, leading to an interaction between magnons and microwave frequency fields.  Therefore, the strong interactions of cavity magnonics can mediate an indirect interaction between microwave and optical frequency photons.

The optical-microwave transducer was demonstrated in a pioneering experiment by Hisatomi et al. \cite{Hisatomi2016}  The strong optical field of a 1550 nm continuous-wave laser was used to illuminate a YIG sphere inside a 3D microwave cavity.  YIG magnons, coherently driven via their hybridization with itinerate microwave photons, modulated the polarization of the drive field via the Faraday effect.  This generated two optical sidebands, centred around the laser frequency, which were detected by a heterodyne measurement using a high-speed photodiode.  As shown in \figref[d]{fig:transducer}, a dip in the microwave reflectivity (upper panel) corresponded exactly to a dip in the optical signal (lower panel), indicating the conversion from microwave to optical frequencies.  More generally, microwave-optical transduction may be realized via whispering-gallery-modes \cite{Zhang2015b, Haigh2015b, Osada2015, Lambert2019} and used to realize ultrafast magnetization control. \cite{Braggio2016}
\subsection{Memories}
Memory development has been a driving force in magnetism and spintronic research for decades.  From giant-magnetoresistance based HDDs, \cite{Fert2012} to spin-transfer-torque based RAM, \cite{Bhatti2017a} magnetic-based memory generally relies on magnetization manipulation and detection to implement read and write functionality.  Therefore, the controllability afforded by spin-photon hybridization is a natural fit for memory applications.  

\begin{figure}[t]
\begin{center}
\includegraphics[width=9cm]{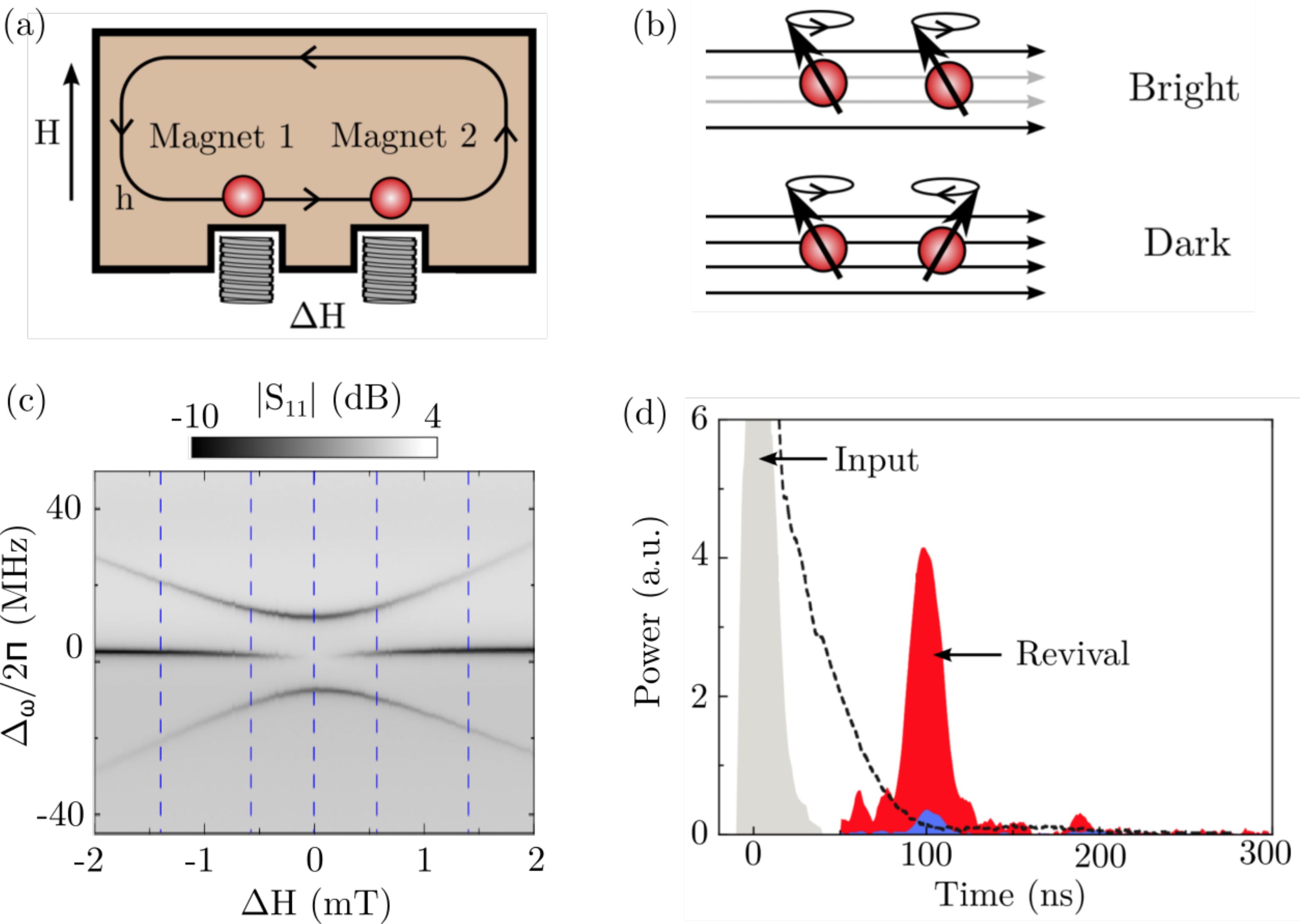}
\caption{(a) A simple device used to create magnon dark modes. Here, two identical YIG spheres are coupled to the magnetic field of a 3D microwave cavity.  The magnon resonance frequencies can be detuned by applied a field gradient, $\Delta H$, via small coils below each sphere.  (b) In a bright mode, the magnons hybridize in-phase and couple to the cavity mode, while in a dark mode, the out-of-phase magnon hybridization decouples from the cavity. (c) When the magnon resonances are detuned, three hybidized modes are observed in the reflectivity spectra, indicating coherent coupling between each magnon and the cavity mode.  However, when $\Delta H = 0$, the magnon dark mode decouples and only two hybridized states can be observed.  (d) Demonstration of a magnon gradient memory.  Without magnon-photon hybridization, the 15 ns microwave input pulse will quickly decay (dashed curve).  However, when hybridization is allowed, the long lived dark mode results in a strong revival peak at 100 ns.  Reproduced with permission from Zhang et al., Nat. Commun. \textbf{6}, 8914 (2015).  Copyright 2015 Nature Research.}
\label{fig:memory}
\end{center}
\end{figure}

The first cavity magnonic memory architecture exploited strong coherent coupling between multiple YIG spheres to realize long-lived magnon dark modes. \cite{Zhang2015g}  The basic device idea is sketched in \figref[a]{fig:memory}.  Here, two identical YIG spheres are coupled to a microwave magnetic field, $h$, in a 3D microwave cavity, and the magnon resonance frequencies are biased by an external, static magnetic field $H$.  A field gradient $\Delta H$, applied via small coils below each YIG sphere, allows individual control of the magnon resonance frequencies.  As illustrated in panel (b), the coherent coupling of this cavity magnonic device produces both bright and dark magnon states.  In the bright state, the hybridized magnon modes precess in-phase, and couple to the cavity field, while in the dark state, the magnon modes precess out-of-phase and decouple from the cavity field.  This phase behaviour can be directly probed by analyzing the antiresonance structure of the microwave spectra. \cite{Harder2016}

The reflection spectra as a function of the field gradient $\Delta H$ is shown in \figref[c]{fig:memory}.  When the magnon resonance frequencies are detuned, i.e., $\Delta H \ne 0$, there are three hybridized modes, indicating that each YIG sphere is coherently coupled to the cavity.  However, when the magnon modes are brought on resonance by removing the field gradient, i.e., setting $\Delta H = 0$, the absorption of the central mode disappears.  The remaining two modes are the result of hybridization between the bright state and the microwave cavity, while the dark state decouples from the cavity mode.  This means that the bright state can be used to transfer information to/from photons, while the dark state is ideal for storing information, due to its long lifetime.

In a practical device, temporal dark modes, not eigenmodes, are used.  These temporal modes periodically convert into bright modes, thereby removing the need for fast magnetic field manipulation while still enhancing the lifetime of the hybridized state.  Data illustrating information storage in such a device is shown in \figref[d]{fig:memory}.  First, a 15 ns microwave pulse was injected into the cavity.  With the external bias field turned off, the magnon and cavity mode are strongly detuned and hybridization does not occur, resulting in an exponential decay of microwave power (dashed curve).  However, when hybridization is enabled, the microwave pulse couples to the magnon bright mode before quickly converting into a magnon dark mode.  This greatly extends the lifetime of the microwave pulse, as indicated by the strong revival peak at 100 ns.  This work highlights the potential of cavity magnonics and spin-photon hybridization for memory devices.
\subsection{Cavity-Mediated Spin-Spin Interactions}
Spin-spin interactions play an important role in magnetic devices.  For example, spin currents may be manipulated via the exchange interaction in magnetic bilayers, \cite{Silsbee1979, Slonczewski1996, Berger1996} and hybridized magnetostatic modes have been exploited for magnonics. \cite{Chumak2015}  With this in mind, a key application of cavity magnonic transduction is to generate cavity-mediated spin-spin interactions with enhanced functionality, for example, long distance coherent control. \cite{Lambert2015b, Bai2017, ZareRameshti2018, Janssonn2020}  

\begin{figure}[t]
\begin{center}
\includegraphics[width=9cm]{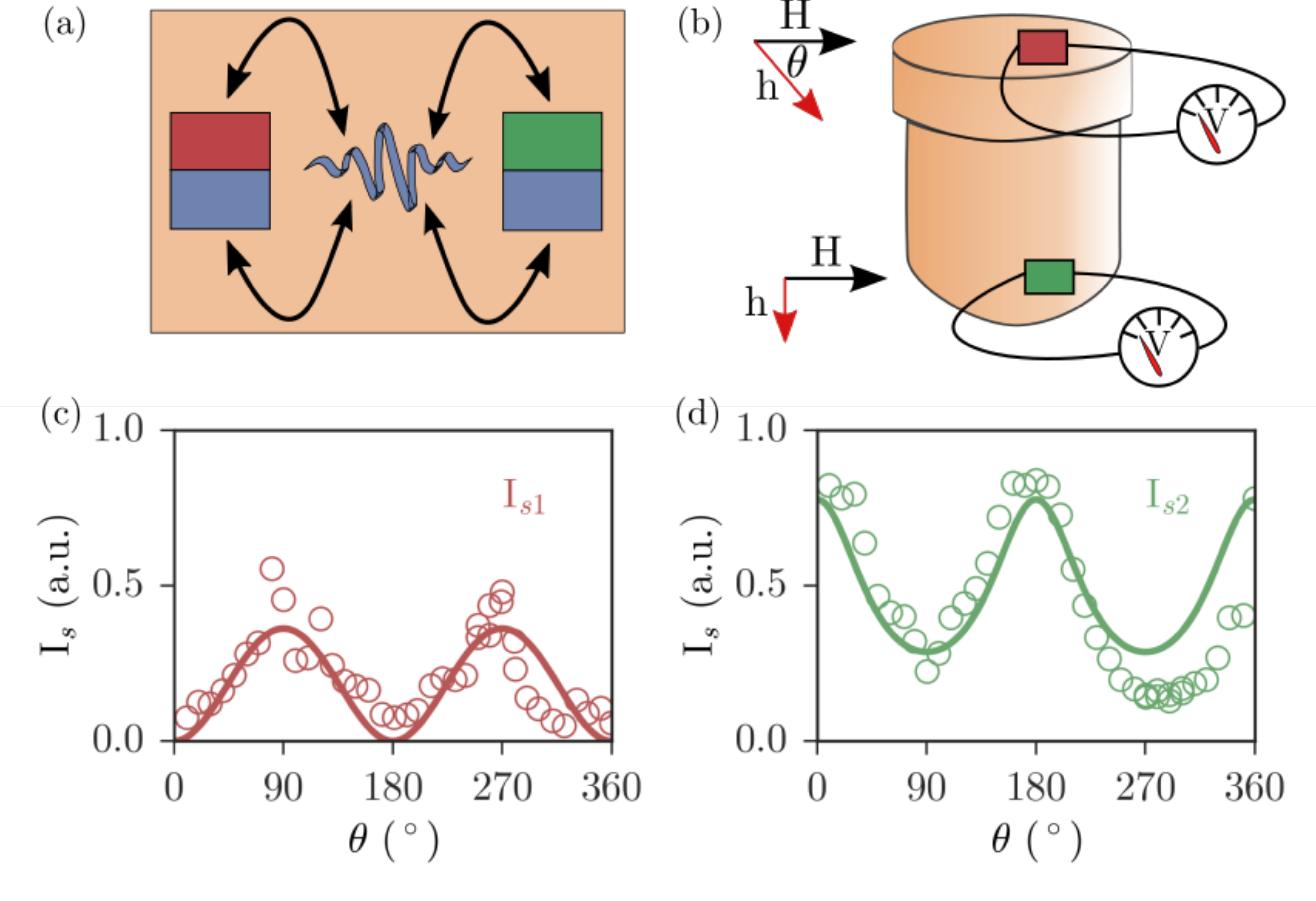}
\caption{(a) Cavity-mediated spin-spin interactions form when multiple magnetic devices hybridize with a single cavity.  The versatility of such interactions -- they can be realized by both coherent and dissipative coupling and are agnostic to device details -- combined with their controllability, leads to myriad applications. (b) Setup demonstrating non-local spin current control.  Two spatially separated magnetic bilayers are coupled to a single cavity field.  The local field orientation is controlled, and the spin current is measured in each device, demonstrating both (c) local and (d) non-local control.  Panels (c) and (d) are reproduced with permission from Bai et al., Phys. Rev. Lett. \textbf{118}, 217201 (2017).  Copyright 2017 American Physical Society.}
\label{fig:spinspin}
\end{center}
\end{figure}

\hyperref[fig:spinspin]{Figure 15 (a)} illustrates the idea of cavity-mediated spin-spin interactions.  Here, a common cavity field acts as an intermediary between spins in two separate magnetic devices.  Due to its non-local nature, this coupling can overcome the inherently short range of exchange, dipole or spin-orbit interactions.  For example, in the experiment by Lambert et al. \cite{Lambert2015b} two YIG spheres were separated by 1.4 cm and coupled to a coaxial transmission line, with an avoided crossing indicated that the separated magnetostatic modes were strongly coupled, even at large cavity detuning.  

By combining the non-local nature of cavity-mediated interactions with the controllability of spin-photon coupling, it is possible to remotely control spin currents using the experimental setup shown in \figref[b]{fig:spinspin}. \cite{Bai2017} In this device, two YIG/Pt bilayers are placed at opposite ends of a 3D microwave cavity, with an arbitrarily large separation (in the data of panel (c) and (d), the separation was approximately 3 cm).  The spin-photon coupling of one bilayer is controlled by rotating the sample, which changes the orientation between the local microwave magnetic field (at that sample) and the static magnetic field. \cite{Bai2016}  The spin current in each device was locally monitored via electrical detection, \cite{Bai2015}  and as shown in \figref[c]{fig:spinspin}, will oscillate sinusoidally with the microwave magnetic field orientation in the rotated bilayer.  While this local control is useful in its own right, the spin current in the distant bilayer, which is not directly manipulated, is also found to oscillate sinusoidally, as shown in panel (d).  Such a non-local interaction can be exploited in spintronic applications for the long range control of spin currents, which are typically limited by the micrometer spin diffusion length.\cite{Bai2017} 

Though the experimental demonstrations mentioned above exploited coherently coupled ferrimagnets, cavity-mediated spin-spin interactions are a general phenomena across cavity magnonic platforms. A number of recent theoretical proposals highlight the potential of such cavity-mediated interactions.  For example, it has been suggested that multiple drive fields could be used to selectively control between coherent and dissipative spin-spin coupling, \cite{Grigoryan2019} and a large coupling is expected even between antiferromagnets and ferromagnets. \cite{Johansen2018}  Hybridization may also enhance magnon-magnon entanglement, providing a mechanism to manipulate quantum steering of magnons and to probe the magnetic damping of individual sublattices, \cite{Zheng2020} and could be exploited to realize macroscopic superconducting spintronics. \cite{Janssonn2020}  Moreover, combined with the potential to generate non-classical photon states and non-linear magnetic interactions, one may envisage using non-local spin-spin interactions for spintronic applications analogous to work that has been done to non-locally couple qubits \cite{Sillanpaa2007} or quantum dots. \cite{Nicoli2018} 
\subsection{Nonreciprocal Transport and Isolators}
Nonreciprocal electromagnetic propagation plays an important role in information processing, enabling sensitive signal detection and processing by reducing reflection induced noise.\cite{Caloz2018}  To realize nonreciprocal behaviour, a system must break time-reversal symmetry, which is traditionally achieved at microwave frequencies using ferrites. \cite{LaxBook} However, broad application of nonreciprocity requires control of the isolation bandwidth combined with large isolation ratios, which is technically challenging.  In this regard, the flexibilty and tunablity of the cavity magnonic platform has proven beneficial, leading to large bandwidth isolators, $\approx 0.5$ GHz, \cite{Zhang2020a} exceeding 60 dB isolation. \cite{Zhu2020}  Moreover, nonreciprocity has been achieved in both coherent \cite{Yu2020, Zhu2020, Zhang2020a} and dissipative \cite{Wang2019b, Qian2020} cavity magnonics.  

\begin{figure}[b]
\begin{center}
\includegraphics[width=8.5cm]{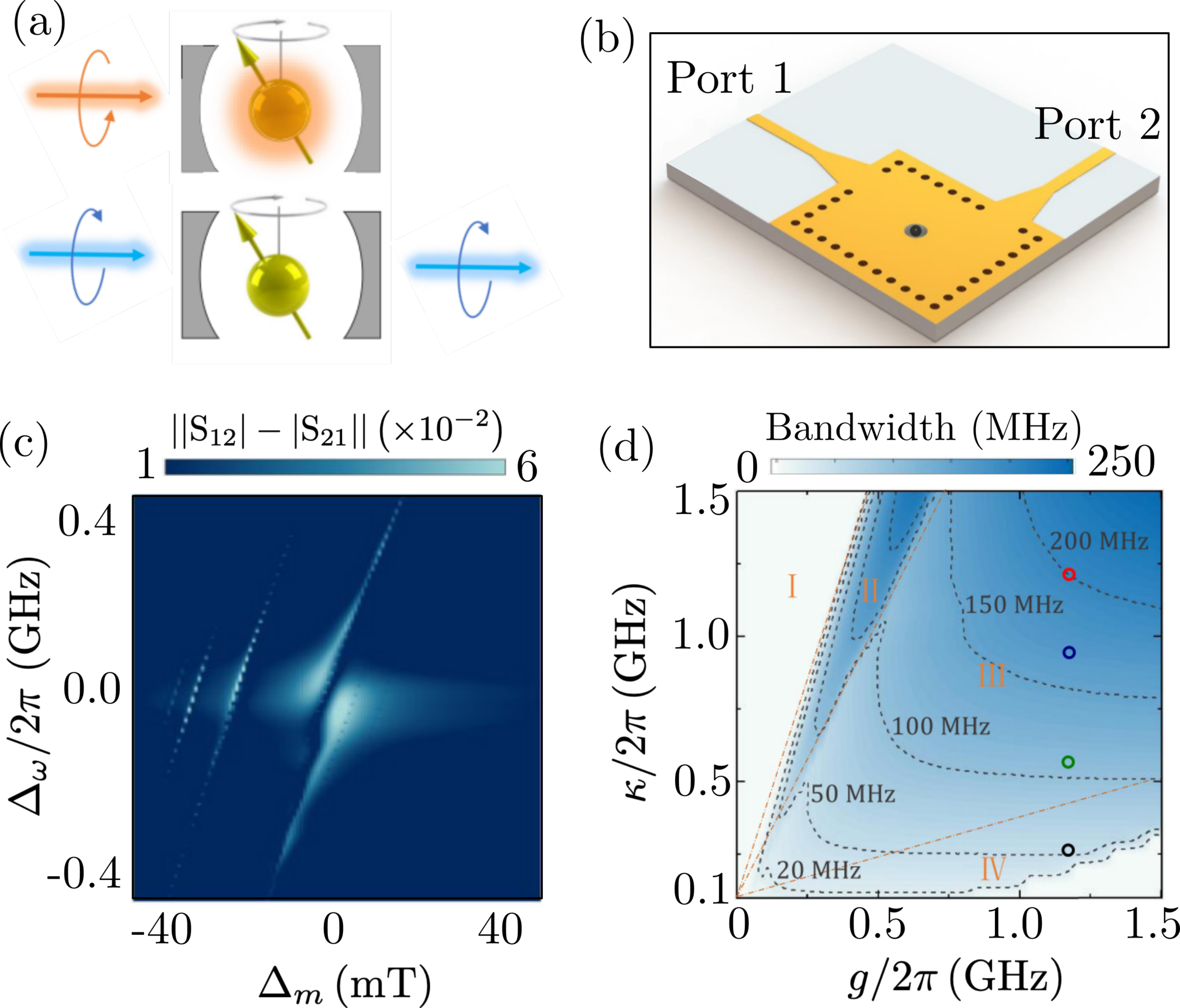}
\caption{Chiral coupling leads to nonreciprocal behaviour in coherent systems.  (a) Hybridization only occurs between magnons and photons of the same chirality (top).  When the photon chirality is reversed, at fixed magnon chirality, the coupling is effectively turned off (bottom).  (b) Chiral cavity modes can be realized in a substrate integrated waveguide.  A high-dielectric-constant substrate is sandwiched between two copper layers, and a YIG sphere is placed on top.  Metalized vias (black dots) connect the upper and lower copper layers.  The vias and ports are designed to create and control the chiral cavity modes.  (c) Nonreciprocal behaviour in the hybridized microwave spectra.  (d) Theoretically achievable bandwidth at 20 dB isolation ratio in a three-port superconducting resonator.  Panels (a) - (c) are reproduced with permission from Zhang et al., Phys. Rev. Appl. 13, \textbf{1} (2020).  Copyright 2020 American Physical Society.  Panel (d) is reproduced with permission from Zhu et al., Phys. Rev. A \textbf{101}, 43842 (2020).  Copyright 2020 American Physical Society.}
\label{fig:nrCoherent}
\end{center}
\end{figure}

In coherent systems, time-reversal symmetry may be broken via chiral magnon-photon coupling.    As illustrated in \figref[a]{fig:nrCoherent}, hybridization will only occur between magnons and photons of the same chirality.  Therefore, if the photon chirality is reversed, and the magnon chirality fixed, the magnon-photon interaction will disappear.  As a result, by designing a cavity which supports orthogonal microwave chiralities, i.e., clockwise and counterclockwise circularly polarized photons, nonreciprocal behaviour can be realized and controlled.  

\begin{figure}[t]
\begin{center}
\includegraphics[width=8.5cm]{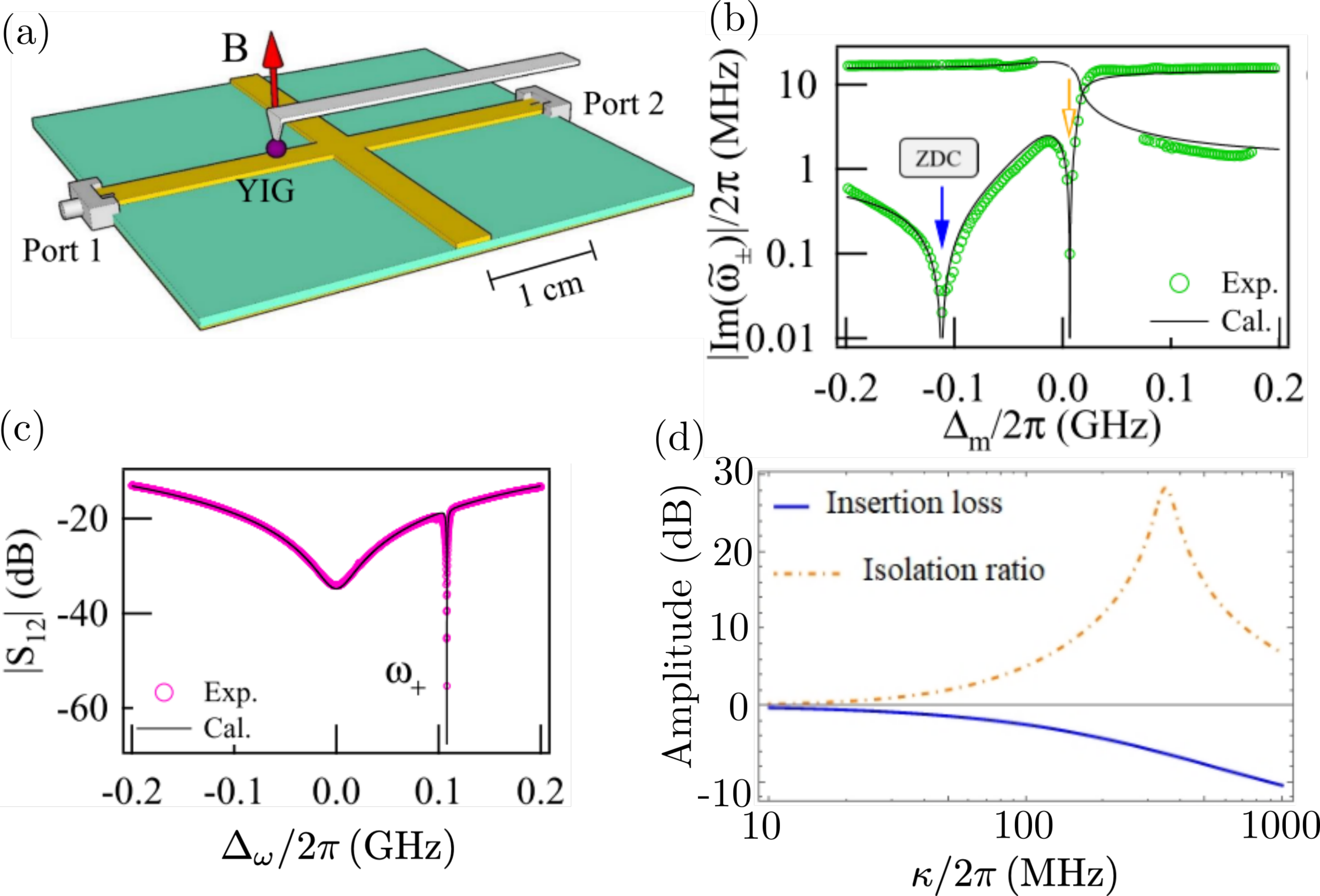}
\caption{Interference between dissipative and coherent coupling leads to nonreciprocal behaviour in open systems.  (a) In an open system, such as this cross-cavity, hybridization is due to both coherent and dissipative coupling.  By controlling the YIG position, for example, with a displacement cantilever, the ratio of coherent to dissipative coupling strength can be controlled.  (b) At a certain magnon-cavity detuning, known as the zero-damping-condition, the hybridized linewidth approaches zero.  (c) At the ZDC, the transmission spectra are highly non-reciprocal with $|\text{S}_{12}| \left(\omega_+\right) \to 0$ and $|\text{S}_{21}|>0$.  (d) This nonreciprocal behaviour leads to large and controllable isolation ratios with low insertion loss.  Reproduced with permission from Wang et al., Phys. Rev. Lett. \textbf{123}, 127202 (2019).  Copyright 2019 American Physical Society.}
\label{fig:nrDissipative}
\end{center}
\end{figure}

A coherent cavity magnonic system that displays nonreciprocal behaviour is shown in \figref[b]{fig:nrCoherent}.  In this device, a high dielectric constant substrate is sandwiched between two copper layers.  The top and bottom copper layer are connected by metalized vias, and a YIG sphere is placed in the centre.  By controlling both the via and port design, chiral cavity modes can be realized. \cite{Zhang2020a}  The nonreciprocal behaviour of this device is illustrated in \figref[c]{fig:nrCoherent}, with large differences between $\text{S}_{12}$ and $\text{S}_{21}$.  The nonreciprocity is largest near zero detuning due to the strong hybridization, and by controlling the coupling strength, a nonreciprocity bandwidth of nearly 0.5 GHz can be achieved. \cite{Zhang2020a}

Coherent nonreciprocity has been proposed in other cavity magnonic devices as well.  For example, it has been predicted that an array of YIG spheres in a toroidal cavity would generate a unidirectional, highly coherent photon beam,\cite{Yu2020} and a three-port superconducting resonator has been proposed to achieve isolation ratios in excess of 60 dB with insertion losses below 0.05 dB. \cite{Zhu2020}  In all cases, the key advantage compared to other nonreciprocal platforms is the controllability available in cavity magnonics.  For example, Zhu et al. \cite{Zhu2020} predict that the isolation ratio and bandwidth can be systematically tuned by controlling the coupling strength and external dissipation rate, as illustrated in \figref[d]{fig:nrCoherent}.  

Interference between coherent and dissipative coupling may also be exploited to break time-reversal symmetry. \cite{Wang2019b, Qian2020}  Generally, open cavities, such as the cross cavity illustrated in \autoref{fig:nrDissipative}, support both standing and traveling-waves and, therefore, exhibit both coherent and dissipative interactions.  By controlling the ratio of coherent to dissipative coupling, for example, by controlling the YIG position, a ``zero-damping condition (ZDC)" can be achieved, where the intrinsic damping of the hybridized mode goes to 0, as shown in \figref[b]{fig:nrDissipative}.  At the ZDC, the on-resonance transmission spectra is highly nonreciprocal, with $|\text{S}_{21}\left(\omega_-\right)| = |\text{S}_{12}\left(\omega_+\right)| = 0$ and $|\text{S}_{12}\left(\omega_-\right)| = |\text{S}_{21}\left(\omega_+\right)| > 0$.  This is shown for the upper branch (i.e., $\omega_+$) in \autoref{fig:nrDissipative}.  This behaviour has been exploited to demonstrate highly flexible and effective isolation, which may also be controlled by tuning the external damping rates as shown in \figref[d]{fig:nrDissipative}. 
\subsection{Enhanced Sensing}
Sensing techniques are generally based on the response of a system to external perturbations.  Therefore, a non-linear response, for example, near a singularity, can be exploited to enhance sensitivity.  In cavity magnonics, two types of singularities have been observed: a bound state in the continuum (BIC) \cite{Yang2020a} and an exceptional point (EP). \cite{Harder2017, Zhang2017a}   

Since cavity magnonic systems are inherently dissipative, they are non-Hermitian and, as we can see from Eq. \eqref{eq:dissipativeEigen}, the eigenspectrum contains a branch-point.  In fact, this is the mathematical definition of an EP: the branch-point in the eigenspectrum of a non-Hermitian system.  Although non-Hermiticity generally leads to complex eigenvalues, a real eigenspectrum still exists when the system is $\mathcal{PT}$-symmetric. \cite{Bender1998}  This realization provides an important physical interpretation of the EP: it is a signature of the phase transition between a $\mathcal{PT}$-symmetry preserved and $\mathcal{PT}$-symmetry broken phase.  In coherent cavity magnonics, this phase transition can be reached by balancing gains and losses, e.g., for coherent coupling ($\Gamma = 0$), the eigenspectrum is real provided $\alpha_L = \beta_L$. \cite{Cao2019}  On the other hand, dissipatively coupled systems are actually anti-$\mathcal{PT}$ symmetric, however, EPs still exist. \cite{Yang2020a, Tserkovnyak2020}

\begin{figure}[t]
\begin{center}
\includegraphics[width=8.5cm]{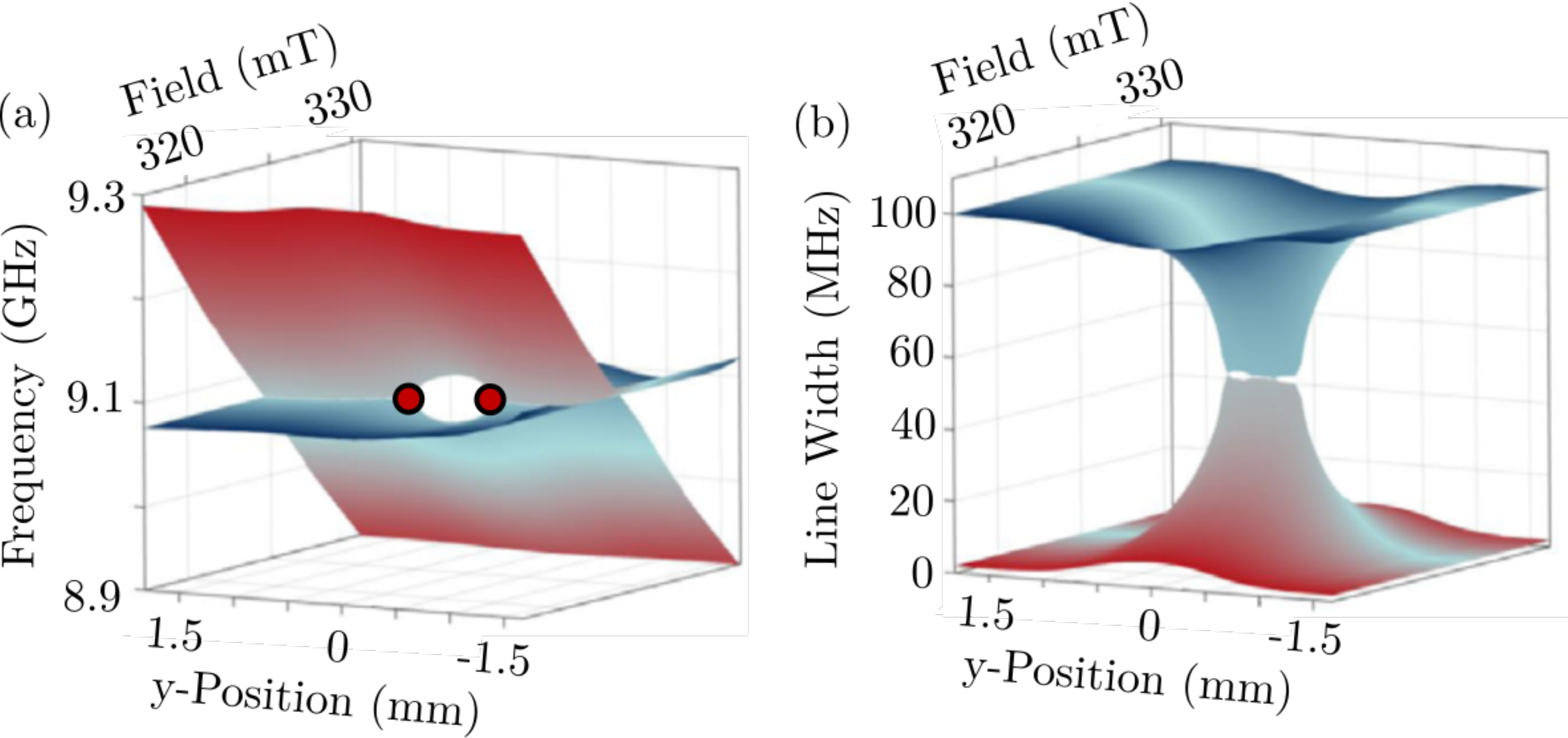}
\caption{(a) The real and (b) imaginary components of the hybridized eigenspectrum as a function of frequency, field and YIG position.  In this experiment, a YIG sphere was coupled to a high-dielectric constant printed circuit board microwave cavity.  In these panels, the YIG position was moved along a single axis, which controlled the coupling strength, and hence the dispersion.  In panel (a), two EPs are marked by red circles.  At these points the system is very sensitive to magnetic field perturbations.  Reproduced with permission from Zhang et al., Phys. Rev. Lett. \textbf{123}, 237202 (2019).  Copyright 2019 American Physical Society.}
\label{fig:EP}
\end{center}
\end{figure} 

The application of EPs to enhance sensing can be best understood by examining the complex eigenspectrum, as shown in \autoref{fig:EP}.  These data were collected using a YIG sphere coupled to a high-dielectric constant printed circuit board microwave cavity, and the magnon-photon coupling strength was controlled by systematically moving the YIG sample. \cite{Zhang2019}  Panels (a) and (b) show the real and imaginary parts of the dispersion, respectively.  Near the two EPs, indicated by red circles in panel (a), the eigenspectrum is extremely sensitive to changes in the external magnetic field.  It has been suggested that this behaviour could be exploited for room temperature magnetometry as sensitive as superconducting quantum interference devices. \cite{Yu2020b, Ebrahimi2020} Other control parameters have also been used to manipulate cavity-magnonic EPs, \cite{Harder2017, Zhang2017a, Yang2020a} and higher order exceptional points, occurring when many eigenstates coalesce, further enhance the sensitivity. \cite{Zhang2019, Cerjan2019}  Furthermore, the creation of exceptional surfaces in cavity magnonics \cite{Zhang2019} could be further exploited for enhanced sensitivity with increased flexibility, as has been done in photonic systems. \cite{Qin2020, Zhong2019}  

In addition to EPs, cavity-magnonic systems may also contain BICs, and in fact, both may exist in the same device.  A BIC can be created when a confined mode is embedded inside the radiation continuum and cannot radiate away.  In cavity magnonics, this has been realized by dissipatively coupling two anti-resonances.\cite{Yang2020a}  Near the BIC, the hybridized group velocity becomes zero, which could be used for slow light applications. 

The quantum magnonic platform, i.e., magnon-qubit coupling mediated by cavity photons, \cite{Tabuchi2015, Tabuchi2014, LachanceQuirion2019} is also promising for sensing applications.  Allowing a qubit to couple with the same cavity field as the magnon creates a dispersive qubit-magnon interaction, introducing the necessary nonlinearity to study quantum effects in magnonics.  This has been used to perform single magnon detection by entangling a qubit and magnetostatic mode \cite{LachanceQuirion2020} and by probing the qubit coherence, a technique in which the sensitivity is actually inversely proportional to the magnon-linewidth. \cite{Wolski2020}  These examples underscore not only the rich application potential of quantum magnonics to quantum information processing, but also the potential to explore a unique set of physical questions at the single magnon level, including magneto-optical effects \cite{Hisatomi2016} and axion-like dark matter searches. \cite{Barbieri1989, Barbieri2017, Flower2019, Crescini2020a}

\section{Outlook: Open Cavity Magnonics} \label{sec:outlook}

This tutorial article outlines the basics of spin-photon hybridization -- where the light-matter interaction between magnetic materials and electrodynamic cavities leads to unique, emergent behaviour.  While the history of coherent spin-photon coupling can be traced back to the 1953 work of Artman and Tannenwald, \cite{Artman1953} in the 2010s, new themes, perspectives, implementations and applications have grown this seed into the diverse, modern field of cavity magnonics.  

Motivated by quantum information and transduction applications, early cavity magnonic research focused on closed microwave cavities, and coherent coupling.  To achieve strong coupling, spatial mode overlap between the cavity microwave field and the magnon mode was essential, hence, magnetic materials were deliberately placed inside cavities and at the anti-node of the microwave magnetic field.  The development of the coherent cavity-magnonic platform opened many paths to discovery and development, such as microwave-to-optical frequency conversion, \cite{Hisatomi2016, Lambert2019} and novel memory architectures. \cite{Zhang2015g}  In these early works, dissipation was typically considered a drawback which led to amplitude decay of Rabi oscillations and linewidth evolution, but did not influence the coherent dispersion.  However, the introduction of open cavities, where the extrinsic dissipation is much larger than the intrinsic damping, has highlighted the fact that dissipation does significantly more, and such open cavity systems are expected to play a major role in the future development of cavity magnonics.

One reason for the interest in open cavity magnonic systems is that they display dissipative coupling, which, due to traveling-wave-mediated interactions, \cite{Yao2019b} leads to level attraction. \cite{Harder2018}  This hybridization is distinctly different from the level repulsion of coherent interactions.  Study of dissipative cavity magnonics has also revealed the important role of singularities, such as exceptional points and an unconventional bound state in the continuum. \cite{Yang2020a}  Near such singularities, small perturbations dramatically alter the system response, providing a route to sensitive detection techniques.  Furthermore, given the similarities between open cavity magnonics and other open cavity based hybrid quantum systems, one could devise ultra-sensitive, broadband measurement techniques by exploiting inevitable environmental dissipation.  In addition, by utilizing the interference between coherent and dissipative interactions, \cite{Wang2019b} cavity magnonics exhibits a nonreciprocal response, which can be used to design novel microwave isolators and circulators, and recent work\cite{Rao2020} has shown that well-separated cavity and magnon modes can be coupled via cooperative damping even without any spatial mode overlap.  Taken together, it seems that the roadmap for open cavity magnonics is currently being drawn, and there is much new and exciting territory to be explored.   For additional perspectives on the future of cavity magnonics, see, for example, Refs. \citenum{LachanceQuirion2019, Hu2020} and \citenum{Barman2021}.

\section*{Acknowlegements}
\noindent
This work was funded by NSERC Discovery Grants and NSERC Discovery Accelerator Supplements (C.-M.H.). We thank all members and alumni of the Dynamic Spintronics Group at the University of Manitoba, for their contributions.  B.M.Y wa supported by National Natural Science Foundation of China under Grant No. 11974369.  M.H. was supported by Institute Research Funds from the British Columbia Institute of Technology.

\begin{appendix}
\section{Method of Averaging} \label{Sec:appendixA}

In this appendix, we outline how the method of averaging can be used to analyze the time-evolution of coupled oscillations.  

In terms of the generalized coordinates $\varphi_{1,2}$, the equations of motion for two coupled oscillators have the form

\begin{subequations}
\label{Eq:EOM}
\begin{align}
\ddot{\varphi}_{1} &+ 2\lambda_{1} \dot{\varphi}_{1} + \omega_{1}^2\varphi_{1}  - f_{1}(\varphi_{1},\dot{\varphi}_{1}) \nonumber \\
&=  2J_1\omega_1(\varphi_2-\varphi_1)+2\Gamma_1(\dot{\varphi}_{2}-\dot{\varphi}_{1}), \\
\ddot{\varphi}_{2} &+ 2\lambda_{2} \dot{\varphi}_{2}+\omega_{2}^2\varphi_{2} -f_{2}(\varphi_{2},\dot{\varphi}_{2}) \nonumber   \\
&= 2J_2\omega_2(\varphi_1-\varphi_2)+2\Gamma_2(\dot{x}_{1}-\dot{\varphi}_{2}).
\end{align}
\end{subequations}

The left hand side of Eqs. \eqref{Eq:EOM} include a linear restoring force, with frequency $\omega_{1,2}$, and a linear frictional force, with damping coefficient $\lambda_{1,2}$. The $f_{1,2}$ terms account for any nonlinear restoring or frictional forces due to non-isochronous and self-oscillation effects. Terms on the right hand side of Eqs. \eqref{Eq:EOM} describe coupling between the two oscillators.  The first terms, proportional to the difference between the coordinates, describe a coherent interaction with coupling constant $J_{1,2}$, while the second terms, proportional to the difference in the velocities, describe a dissipative interaction with coupling constant $\Gamma_{1,2}$.  Both coupling terms will vanish if the states of the two oscillators coincide, i.e., if $\varphi_1=\varphi_2$ and $\dot{\varphi}_1=\dot{\varphi}_2$.  The coupled system described by Eqs. \eqref{Eq:EOM} does not have a general analytical solution for all parameters. However, in many physically meaningful situations, $\lambda_{1,2}\ll{}J_{1,2}$ and $\Gamma_{1,2}\ll\omega_{1,2}$ in which case approximate analytical expressions may be found.  \cite{LandauBookMechanics}

Equation \eqref{Eq:EOM} can be simplified by moving to a rotating reference frame with $\omega_{r\!e\!f}=(\omega_1+\omega_2)/2$ and a slowly-varying envelope function $a_{1,2}(t)$ such that $\varphi_{1,2}=[a_{1,2}(t)e^{-i\omega_{r\!e\!f}t}+a_{1,2}^*(t)e^{i\omega_{r\!e\!f}t}]/2$.  For a linear system, where $f_1=f_2=0$, we obtain
\begin{equation}\label{Eq:2x2matrix}
\begin{pmatrix}\dot{a}_1\\\dot{a}_2\end{pmatrix}\simeq-i\begin{pmatrix}-\Delta/2+G_1-i\lambda_1&-G_1\\
  -G_2&\Delta/2+G_2-i\lambda_2\end{pmatrix}\begin{pmatrix}a_1\\a_2\end{pmatrix}.
\end{equation}
Here, $\Delta=\omega_2-\omega_1$ and $G_{1,2}=J_{1,2}-i\Gamma_{1,2}$, respectively, denote the frequency detuning and the total complex coupling strength.

In the rotating reference frame, the complex eigenfrequencies of the hybridized modes, $\widetilde{\omega}_\pm$,  deduced from Eq.
\eqref{Eq:2x2matrix} are
\begin{align}\label{Eq:eigenfrequency}
\widetilde{\omega}_{\pm} - \omega_{r{\!}e{\!}f}&=\dfrac{1}{2}[(G_1-i\lambda_1)+(G_2-i\lambda_2)]\nonumber\\
&\pm\frac{1}{2}\sqrt{[\Delta-(G_1-i\lambda_1)+(G_2-i\lambda_2)]^2+4G_1G_2}.
\end{align}
There are two time-scales associated with this system: a fast oscillation at frequency $\omega_{r{\!}e{\!}f}$, and a much slower drifting of the amplitude and phase.  This allows the time evolution of the dynamical system to be written as $\varphi_{1,2} = A_{1,2}(t)\cos[\omega_{r{\!}e{\!}f}t+\theta_{1,2}(t)]$, where $A_{1,2}$ and $\theta_{1,2}$ are slowly varying functions of time.  Therefore, in the rotating reference frame, Eqs. \eqref{Eq:EOM} can be written as
\begin{equation}
\label{Eq:EOM_p}
\ddot{\varphi}_{1,2} + \omega_{r{\!}e{\!}f}^2\varphi_{1,2}+h_{1,2}=0,
\end{equation}
where $h_{1,2}=2\lambda_{1,2}\dot{\varphi}_{1,2}+(\omega_{1,2}^2-\omega_{r{\!}e{\!}f}^2)\varphi_{1,2}-2J_{1,2}\omega_{1,2}(\varphi_{2,1}-\varphi_{1,2})-2\Gamma_{1,2}(\dot{\varphi}_{2,1}-\dot{\varphi}_{1,2})$ are small perturbations.  For such systems, the method of averaging \cite{BogliubovBook, MinorskyBook} can be applied, in which we average the variables $\varphi_{1,2}$ over the oscillation period to eliminate fast oscillations and observe the qualitative behaviour of  $A_{1,2}$ and $\theta_{1,2}$,
\begin{subequations}
\label{Eq: TA_method}
\begin{eqnarray}
\frac{dA_{1,2}}{dt}&=&\langle h_{1,2}\sin(\tau)\rangle, \\
\frac{d\theta_{1,2}}{dt}&=&\langle h_{1,2}\cos(\tau)\rangle, \label{Eq: TA_method_phase}
\end{eqnarray}
\end{subequations}
where $\langle*\rangle$ denotes a time average over the period $T = 2\pi/\omega_{r{\!}e{\!}f}$ and $\tau=\omega_{r{\!}e{\!}f}t+\theta_{1,2}$ for $h_{1,2}$, respectively.  Since $A_{1,2}$ and $\theta_{1,2}$ are approximately constant over $T$, time averaging $h_{1,2}$ leads to:

\begin{widetext}
\begin{subequations}
\label{eq:pendEnv}
\begin{align}
\frac{dA_1}{dt}&=-(\lambda_1+\Gamma_1) A_1-\frac{A_2J_1\omega_1\sin(\theta_1-\theta_2)}{\omega_{r\!e\!f}}
                           +A_2\Gamma_1\cos(\theta_1-\theta_2), \\
\frac{dA_2}{dt}&=-(\lambda_2+\Gamma_2) A_2+\frac{A_1J_2\omega_2\sin(\theta_1-\theta_2)}{\omega_{r\!e\!f}}
                            +A_1\Gamma_2\cos(\theta_1-\theta_2), \\
\frac{d(\theta_1-\theta_2)}{dt}&=-\Delta+\frac{J_1\omega_1-J_2\omega_2}{\omega_{r\!e\!f}}
                          +\frac{(A_1^2\omega_2J_2-A_2^2\omega_1J_1)\cos(\theta_1-\theta_2)}{A_1A_2\omega_{r\!e\!f}}
                           -\left(\frac{A_1\Gamma_2}{A_2}+\frac{A_2\Gamma_1}{A_1}\right)\sin(\theta_1-\theta_2). \label{eq:pendEnvc}
\end{align}
\end{subequations}
\end{widetext}

\section{Equations of Motion for Spring Coupled Pendulums} \label{Sec:appendixB}
Referring to \autoref{fig:coherentPendulum}, in the linear regime, the oscillations are small, and therefore, the kinetic energy $T$ can be written using a small angle approximation as
\begin{equation}
\label{eq:pendT}
T = \frac{1}{2}{ml_1^2}\dot{\varphi}_1^2 + \frac{1}{2}{ml_2^2} \dot{\varphi}_2^2.
\end{equation}
The total potential energy of the coupled pendulums has both gravitational and spring contributions, taking the form
\begin{equation}
\label{eq:pendU}
\begin{split}
U &= mgl_1(1-\cos\varphi_1) + mgl_2(1-\cos\varphi_2) \\
 &+ \frac{k}{2}(l\sin\varphi_1-l\sin\varphi_2)^2,
\end{split}
\end{equation}
where the potential energy is referenced to zero at $\varphi_1 = \varphi_2 = 0$.  In the linear regime, $\sin \varphi_{1,2}\approx\varphi_{1,2}$ while $\cos\varphi_{1,2}\approx1-\varphi^2_{1,2}/2$, and therefore, the Lagrangian is
\begin{equation}
\label{eq:pendL}
\begin{split}
\mathcal{L} &=T-U \\
&=\frac{1}{2}{ml_1^2} \dot{\varphi}_1^2 + \frac{1}{2}ml_2^2 \dot{\varphi}_2^2 - \frac{1}{2}mgl_1\varphi_1^2- \frac{1}{2}mgl_2\varphi_2^2\\
&- \frac{1}{2}kl^2(\varphi_1-\varphi_2)^2.
\end{split}
\end{equation}

Dissipation can be included as a velocity proportional non-conservative force via the Rayleigh dissipation function, \cite{LandauBookMechanics}
\begin{equation}
\label{eq:Rayleigh2}
\mathcal{F} = \lambda_1 m l_1^2 \dot{\varphi}^2_1 + \lambda_2 m l_2^2 \dot{\varphi}^2_2,
\end{equation}
where $\lambda_1$ and $\lambda_2$ characterize the intrinsic damping rate of the pendulums.  Therefore, the equations of motion for $\varphi_{1,2}$ are determined according to the generalized Euler-Lagrange equations,
\begin{equation}
\label{eq:E-L}
\dfrac{d}{dt}\dfrac{\partial \mathcal{L}}{\partial\dot{\varphi}_{1,2}} - \dfrac{\partial \mathcal{L}}{\partial {\varphi}_{1,2}} +\dfrac{\partial \mathcal{F}}{\partial \dot{\varphi}_{1,2}} = 0,
\end{equation}
to be
\begin{equation}
\label{eq:spring_motion}
\ddot{\varphi}_{1,2} + 2\lambda_{1,2}\dot{\varphi}_{1,2} + \omega_{1,2}^2 \varphi_1 - 2 J_{1,2}\omega_1(\varphi_{2,1}-\varphi_{1,2})=0,
\end{equation}
where $\omega_{1,2} = \sqrt{g/l_{1,2}}$, are the uncoupled oscillation frequencies and $J_{1,2}=kl^2/(2m\omega_{1,2}l_{1,2}^2)$ are the coupling strengths. The first three terms in Eqs. \eqref{eq:spring_motion} describe the independent oscillations of the two pendulums, while the fourth term directly couples the motion of the two pendulums.  Clearly, the spring coupled pendulums are an example of pure coherent coupling described by Eq. \eqref{Eq:EOM}.  
\section{Equations of Motion for Dashpot Coupled Pendulums} \label{Sec:appendixC}

Referring to \autoref{fig:dissipativePendulum}, in the linear regime, the Lagrangian for the dashpot coupled pendulums is
\begin{equation}\label{eqn-3b}
\begin{split}
\mathcal{L} &=T-U \\
&=\frac{1}{2}{ml_1^2} \dot{\varphi}_1^2 + \frac{1}{2}ml_2^2 \dot{\varphi}_2^2 - \frac{1}{2}mgl_1\varphi_1^2- \frac{1}{2}mgl_2\varphi_2^2.
\end{split}
\end{equation}
For this system, there is no potential energy associated with coupling.  Instead, the dashpot introduces a velocity proportional force in Rayleigh's dissipation function,
\begin{equation} \label{Eq:Rayleigh2b}
\mathcal{F} = \lambda_1ml_1^2\dot{\varphi}^2_1 +\lambda_2 ml_2^2\dot{\varphi}^2_2 + \nu m(\dot{\varphi}_1 - \dot{\varphi}_2)^2.
\end{equation}
Here, the kinematic viscosity $\nu$ acts as a proportionality constant, characterizing the coupling between the two pendulums via the dissipative force.  The generalized Euler-Lagrange equations can then be obtained analogously to \hyperref[Sec:appendixB]{Appendix B},
\begin{equation}
\label{eq:emDash}
\ddot{\varphi}_{1,2} + 2\lambda_{1,2}\dot{\varphi}_{1,2} + \omega_{1,2}^2 \varphi_{1,2} - 2 \Gamma_{1,2}(\dot{\varphi}_{2,1}-\dot{\varphi}_{1,2})=0,
\end{equation}
where $\Gamma_{1,2}=\nu/l_{1,2}^2\ll \omega_{1,2}$ describes the coupling strength.  Since the dashpot induced coupling appears in the angular velocity term, this system is said to be dissipatively coupled according to the discussion in \hyperref[Sec:appendixA]{Appendix A}.
\section{Equations of Motion for Base-Mediated Coupling} \label{sec:appendixD}
Huygens' pendulum clocks are shown in \autoref{fig:huygensPendulums}, forming a closed system, which includes two pendulums mounted to a wall.  The wall can be treated as a giant two-dimensional crystal with one brick in the primitive cell.  The displacement of one brick will cause the whole wall to vibrate, and therefore, the wall can be modelled as an oscillating system with a wavevector ($k$) dependent frequency dispersion, $\omega_k$.  Therefore the kinetic energy of the system is
\begin{equation}\label{Eqn: T1}
  T = \dfrac{1}{2}ml_1^2 \dot{\varphi_1}^2 + \dfrac{1}{2}ml_2^2 \dot{\varphi_2}^2 + \dfrac{1}{2}\sum_{k}m_k\dot{x}_k^2,
\end{equation}
and the potential energy is,
\begin{equation}\label{Eqn: U1}
\begin{split}
 U  & = \frac{1}{2}mgl_1\varphi_1^2 + \frac{1}{2}mgl_2\varphi_2^2 + \sum_{k}\dfrac{1}{2}c_kx_k^2 \\
 & + \sum_{k}\dfrac{1}{2}c_{1k}(l_1\varphi_1 - x_k)^2 + \sum_{k}\dfrac{1}{2}c_{2k}(l_2\varphi_2-x_k)^2.
\end{split}
\end{equation}
Here, $m_k$ is the effective mass of the wall, $x_k$ is the displacement, and $c_k$, $c_{1k}$  and $c_{2k}$ are the force constants which couple the displacement to the pendulum motion. Neglecting intrinsic damping for simplicity (which is justified since the extrinsic damping to the base is typically much larger than the intrinsic damping of each pendulum), the equations of motion read

\begin{subequations}
\label{eq:HuygenInitial}
\begin{align}
&ml_1^2\ddot{\varphi}_1 + mgl_1\varphi_1 + \sum_{k}c_{1k}l_1(l_1\varphi_1 - x_k) = 0,  \\
&ml_2^2\ddot{\varphi}_2 + mgl_2\varphi_2 + \sum_{k}c_{2k}l_2(l_2\varphi_2 - x_k) = 0,  \\
&m_k\ddot{x}_k + c_k x_k + c_{1k}(l_1\varphi_1-x_k) + c_{2k}(l_2\varphi_2-x_k) = 0.
\end{align}
\end{subequations}
Equations \eqref{eq:HuygenInitial} can be rewritten as
\begin{subequations}
\label{Eqn: Base ME}
\begin{eqnarray}
\ddot{\varphi}_1 + \omega_1^2\varphi_1 - \sum_{k}g^\prime_{1k}x_k = 0, \label{Eqn: Base MEa} \\
\ddot{\varphi}_2 + \omega_2^2\varphi_2 - \sum_{k}g^\prime_{2k}x_k = 0, \label{Eqn: Base MEb} \\
\ddot{x}_k + \omega_k^2x_k + g_{1k}\varphi_1 + g_{2k}\varphi_2 = 0,  \label{Eqn: Base MEc}
\end{eqnarray}
\end{subequations}
where $\omega_{1,2} = \sqrt{g/l_{1,2}}$ are the oscillation frequencies of the pendulums, $\omega_k  = \sqrt{c_k/m_k}$ are the vibration frequencies of the wall, and the small frequency shift induced by the pendulum-wall coupling, $\omega_{ik}^2 = c_{ik}/m$, has been neglected.  $g_{ik} = c_{ik}l_i/m_k$ and $g^\prime_{ik}=m_kg_{ik}/(ml_i^2)$ characterize the interaction between the individual pendulums and the wall.  Therefore, while there is no direct coupling between the two pendulums, they are indirectly coupled via their common base, i.e., the wall.

Solving Eq. \eqref{Eqn: Base MEc} the wall motion follows:
\begin{align}\label{Eq: base motion}
x_k&=B_k\cos(\omega_k{t}+\xi_k)  \nonumber \\
&-\frac{\cos(\omega_k{t})}{\omega_k}\int_{t_0}^{t} [g_{1k}\varphi_1(t^\prime)+g_{2k}\varphi_2(t^\prime)]\sin(\omega_k t^\prime)dt^\prime \nonumber \\
&+\frac{\sin(\omega_k{t})}{\omega_k}\int_{t_0}^{t}[g_{1k}\varphi_1(t^\prime)+g_{2k}\varphi_2(t^\prime)]\cos(\omega_k t^\prime)dt^\prime.
\end{align}
Here, the first term describes the intrinsic wall oscillations, while the second and third terms are perturbations due to the pendulum motion.  In the zero coupling limit, the pendulums oscillate sinusoidal with resonance frequency $\omega_{1,2}$, and therefore, at zeroth order in $g^\prime_{ik}$, $\varphi_{1,2} = A_{1,2}\cos(\omega_{1,2} t+\theta_{1,2})$.  This zeroth order solution can be substituted into Eq. \eqref{Eq: base motion} to evaluate the wavevector sums in Eqs. \eqref{Eqn: Base MEa} and \eqref{Eqn: Base MEb}.  To first order, $A_{1,2}$ and $\theta_{1,2}$ will be constants, and due to the rapid oscillations at frequency $\omega_{1,2}$, the integrals can be performed by time averaging over one period, $0 < t < 2\pi/\omega_{1,2}$.  Furthermore, the sum over wavevectors will be dominated by the resonance at $\omega_k = \omega_{1,2}$ and in the first Markov approximation, the coupling constants are independent of $k$.  Therefore,
\begin{subequations}
\begin{align}
\sum_kg^\prime_{1k}x_k&=g^\prime_{1}\sum_k B_k\cos(\omega_k{t}+\xi_k) \nonumber \\
&+\frac{g_1^\prime g_1}{2\omega_1^2} A_1\sin(\omega_1 t+\theta_1)+\frac{g_1^\prime g_2}{2\omega_2^2} A_2\sin(\omega_2 t+\theta_2), \\
\sum_kg^\prime_{2k}x_k&=g^\prime_{2}\sum_k B_k\cos(\omega_k{t}+\xi_k) \nonumber \\
&+\frac{g_2^\prime g_1}{2\omega_1^2} A_1\sin(\omega_1 t+\theta_1)+\frac{g_2^\prime g_2}{2\omega_2^2} A_2\sin(\omega_2 t+\theta_2).
\end{align}
\end{subequations}
Defining $\gamma_{1,2} = g_{1, 2}^\prime g_{1, 2}/4\omega_{1,2}^3$, these sums can be written as
\begin{subequations}
\begin{align}
\sum_k g_{1k}^\prime x_k &= g_1^\prime \sum_k B_k \cos\left(\omega_k t + \xi_k\right) - 2 \gamma_1 \dot{\varphi}_1 \nonumber \\
&- 2 \sqrt{\gamma_1 \gamma_2}\left(\frac{l_2}{l_1}\right) \dot{\varphi}_2, \\
\sum_k g_{2k}^\prime x_k &= g_2^\prime \sum_kB_k \cos\left(\omega_k t + \xi_k\right) - 2 \gamma_2 \dot{\varphi}_2 \nonumber \\
&- 2 \sqrt{\gamma_1 \gamma_2}\left(\frac{l_1}{l_2}\right) \dot{\varphi}_1.
\end{align}
\end{subequations}
Therefore, the equations of motion for the base-mediated pendulums are
\begin{align}
\label{Eq: base coupling}
\ddot{\varphi}_{1,2} &+ \omega_{1,2}^2 \varphi_{1,2} - g_{1,2}^\prime \sum_k B_k \cos\left(\omega_k t+\xi_k\right)\nonumber\\
&+ 2 \gamma_{1,2} \dot{\varphi}_{1,2} +2 \sqrt{\gamma_1\gamma_2} \left(\frac{l_{2,1}}{l_{1,2}}\right)\dot{\varphi}_{2,1}= 0.
\end{align}
The first line in Eq. \eqref{Eq: base coupling} describes an undamped pendulum driven by the intrinsic wall oscillations, while the effects of the reservoir-mediated interaction appear in the second line; the first term is an extrinsic damping via energy leakage to the wall and the second term is an indirect coupling between the two pendulums, which occurs when energy leaked by the first pendulum coherently drives the second pendulum via the wall.

To determine the complex eigenfrequencies, Eq. \eqref{Eq: base coupling} can be compared to the dashpot-coupled equations of motion in Eq. \eqref{eq:emDash}.  To first order in the coupling, $l_1 \sim l_2$, and therefore, $\Gamma \to -\sqrt{\gamma_1\gamma_2}$.  The intrinsic damping is zero since it has been neglected here, and $\Gamma_{1,2} = \gamma_{1,2}$ for the extrinsic damping.  Furthermore, since the complex eigenfrequencies are intrinsic to the system, they are independent of the driving term, leading to Eq. \eqref{Eq:eigenvalue_base}.
\section{Input-Output Theory for a Gap-Coupled Two-Port Cavity} \label{Sec:appendixE}
To probe the linear response of cavity magnonics, e.g., through microwave transmission measurements using a vector-network-analyzer (VNA), external coupling must be introduced.  This can be handled theoretically via the input-output formalism by including an external photon bath, which couples to the cavity mode as outlined below.  For a general introduction to the input-output formalism, see Refs. \citenum{WallsBook} and \citenum{Clerk2010}; for a detailed application to cavity magnonics, see Refs. \citenum{Harder2018} and \citenum{Kusminskiy2019}.

In this appendix, we focus on the discussion of a two-port cavity directly excited through the gap between the cavity and the feed lines. In this case, the cavity transmission measurement also includes an input and output port, and therefore, two external photon baths and is described by the Hamiltonian
\begin{align}
\label{eq:coherentExternalH}
H &= \hbar \widetilde{\omega}_c a^\dag a + \hbar \widetilde{\omega}_m b^\dag b + \hbar J \left(a b^\dag + a^\dag b\right) \nonumber \\
&+\hbar  \int \omega_k c_k^\dag c_k dk + \hbar \int \omega_k d_k^\dag d_k dk  \nonumber \\
&+ \hbar \int \lambda_c \left(a c_k^\dag + a^\dag c_k\right)dk+ \hbar \int \lambda_d \left(a d_k^\dag + a^\dag d_k\right)dk.
\end{align}
In this equation: (1) The first line is just the spin-photon Hamiltonian of Eq. \eqref{Eq:coherent_Hamiltonian}; (2) The second line contains the kinetic terms for the bath photons at ports 1 and 2, with creation operators $c_k^\dag$ and $d_k^\dag$, respectively; and (3)  The third line contains the interaction terms between the cavity photons and the external baths, which have coupling rates $\lambda_c$ and $\lambda_d$ at ports 1 and 2, respectively.  Here, the rotating wave approximation has been applied to all interactions, and the integrals are taken over all bath modes, $-\infty < k < \infty$.  Using the commutation relations $[c_k, c_{k^\prime}^\dag] = \delta \left(k-k^\prime\right)$ and $[c_k,c_{k^\prime}]$ = 0 (analogous for $d_k$), the equations of motion for the bath modes are
\begin{subequations}
\begin{align}
\dot{c}_k &= -\frac{i}{\hbar}\left[c_k, H\right] = -i\omega_k c_k - i \lambda_c a,\\
\dot{d}_k &= -\frac{i}{\hbar}\left[d_k, H\right] = -i\omega_k d_k - i \lambda_d a,\
\end{align}
\end{subequations}
which have the integral solutions
\begin{subequations}
\label{eq:2portIntegralSolutions}
\begin{align}
c_k\left(t\right) &= e^{-i \omega_k\left(t-t_0\right)} c_k\left(t_0\right) - i \int_{t_0}^t \lambda_c a e^{-i\omega_k\left(t-t^\prime\right)} dt^\prime, \\
d_k\left(t\right) &= e^{-i \omega_k\left(t-t_0\right)} d_k\left(t_0\right) - i \int_{t_0}^t \lambda_d a e^{-i\omega_k\left(t-t^\prime\right)}dt^\prime,
\end{align}
\end{subequations}
where $c_k\left(t_0\right)$ and $d_k\left(t_0\right)$ are the initial states of the bath modes at $t_0 < t$. From Eq. \eqref{eq:coherentExternalH}, the equations of motion for the cavity resonance and magnon are
\begin{subequations}
\begin{align}
\dot{a} &= -i\widetilde{\omega}_c a - i J b - i \int \lambda_k c_k dk - i  \int \lambda_d d_k dk, \\
\dot{b} &= - i \widetilde{\omega}_r b - i J a.
\end{align}
\end{subequations}
Since the external modes only couple to the cavity photon, its equation of motion is directly modified, while the magnon equation of motion is only affected indirectly through the changes to $a$.  With a mode independent external coupling (the first Markov approximation), $\lambda_{c,d}$ can be taken outside the $k$ integrals, and therefore, substituting the integral solutions of Eqs. \eqref{eq:2portIntegralSolutions}, the equation of motion for the cavity resonance is
\begin{align}
\label{eq:eomCoherent}
\dot{a}&=-i\widetilde{\omega}_ca -iJb-i\sqrt{2\pi} \lambda_c c_\text{in}-\pi \lambda_c^2 a - i\sqrt{2\pi} \lambda_d d_\text{in} - \pi \lambda_d^2 a.
\end{align}
Here, $c_\text{in}$ and $d_\text{in}$ are the input fields, defined as
\begin{subequations}
\begin{align}
c_\text{in} &= \frac{1}{\sqrt{2\pi}} \int e^{-i \omega_k\left(t-t_0\right)} c_k\left(t_0\right) dk, \\
d_\text{in} &= \frac{1}{\sqrt{2\pi}} \int e^{-i \omega_k\left(t-t_0\right)} d_k\left(t_0\right) dk,
\end{align}
\end{subequations}
which are just wavepackets formed by the time evolution of the $c_k\left(t_0\right)$ [$d_k\left(t_0\right)$] modes to time $t$. The external coupling introduces a new source of dissipation, the $\pi \lambda_{c,d}^2a$ terms in Eq. \eqref{eq:eomCoherent}, and therefore, it is convenient to define the extrinsic damping rates $\kappa_{c,d} = 2\pi \lambda_{c,d}^2$, and redefine the cavity resonance as $\widetilde{\omega}_c \to \omega_c - i \beta_L = \omega_c - i \left[\beta + \left(\kappa_c + \kappa_d\right)/2\right]$.  With these substitutions, the coupled equations of motion for the cavity photon and magnon, including the effects of the external baths, are
\begin{subequations}
\label{eq:coherentLagevin}
\begin{align}
\dot{a}&=-i\widetilde{\omega}_ca -iJb-i\sqrt{\kappa_c} c_\text{in} - i\sqrt{\kappa_d} d_\text{in}, \label{eq:coherentLagevinPhoton}\\
\dot{b} &= - i \widetilde{\omega}_r b - i J a.
\end{align}
\end{subequations}
These are the quantum Langevin equation for canonical coherent cavity magnonics.

To calculate the transmission spectra, note that the integral solution for $c_k$ can also be written in terms of a late time state $c_k(t_1)$ at $t_1 > t$,
\begin{equation}
c_k\left(t\right) = e^{-i \omega_k\left(t-t_1\right)} c_k\left(t_1\right) - i \int_{t}^{t_1} \lambda_c a e^{-i\omega_k\left(t-t^\prime\right)} dt^\prime,
\end{equation}
which defines the output field,
\begin{equation}
c_\text{out}(t) = \frac{1}{\sqrt{2\pi}} \int e^{-i \omega_k\left(t-t_1\right)} c_k\left(t_1\right).
\end{equation}
Carrying out the same procedure that led to Eq. \eqref{eq:coherentLagevinPhoton} yields the time reversed Langevin equation, relating $a$ and $b$ to $c_\text{out}$ and $d_\text{in}$, which can then be combined with Eq. \eqref{eq:coherentLagevinPhoton} to determine the input-output relation for port 1,
\begin{equation}
\label{eq:inputOutput1}
c_\text{in} = c_\text{out} + i \sqrt{\kappa_c} a.
\end{equation}
Physically, this just means that the input field is either reflected at the port or enters the cavity.  Taking the same approach for port 2 leads to
\begin{equation}
\label{eq:inputOutput2}
d_\text{in} = d_\text{out} + i \sqrt{\kappa_d} a.
\end{equation}
Equations \eqref{eq:coherentLagevin}, \eqref{eq:inputOutput1} and \eqref{eq:inputOutput2} can be used to determine the reflection and transmission parameters of Eq. \eqref{Eq: coherent S-parameter}.
\section{Microwave Transmission for Dissipative Cavity Magnonics Mediated by Traveling Photons } \label{Sec:appendixF}
In this appendix, we derive the microwave transmission presented in Eq. \eqref{eq:travelingSpectrum} of the main text.   The Hamiltonian for the traveling-wave-mediated system is similar to Eq. \eqref{eq:coherentExternalH},
\begin{align}
H&=\hbar\widetilde{\omega}_c a^\dag a+\hbar\widetilde{\omega}_m b^\dag b+\hbar\int\omega_k p_k^\dag p_k dk \nonumber \\
&+\hbar \int \lambda_c(a p_k^\dag+a^\dag p_k)dk + \hbar \int \lambda_m e^{i\phi}(b p_k^\dag+b^\dag p_k)dk,
\label{eq:Ht}
\end{align}
where $p_k^\dag$ is the creation operator for the traveling-wave mode, $\widetilde{\omega}_c=\omega_c-i\beta$ and $\widetilde{\omega}_m=\omega_m-i\alpha$.  The first line in Eq. \eqref{eq:Ht} contains the kinetic terms for the cavity mode, magnon mode and traveling-wave, while the second line describes the interaction between the traveling-wave and the cavity and magnon modes, with real valued coupling strength $\lambda_c$ and $\lambda_m$, respectively.  These coupling strengths can also be thought of as the extrinsic dissipation of the cavity and magnon modes to the traveling-wave reservoir, hence our choice of notation.  Here, only the lowest order interaction terms are kept and the rotating wave approximation is used.  The cavity and magnon are spatially separated by a distance $L$, and therefore, the traveling-wave has a phase delay between these two locations, characterized by $\phi$ in Eq. \eqref{eq:Ht}.  In general, $\phi = k L$, however, if focusing on the behaviour near $\omega_c \approx \omega_m$, then $\phi$ is approximately $k$ independent.

Using the commutation relations $[p_k, p_{k^\prime}]=\delta(k-k^\prime)$ and $[p_k,p_{k^\prime}]=0$, the equation of motion for the traveling-wave is
\begin{equation}
\dot{p}_k =-\frac{i}{\hbar}[p_k,H]=-i\omega_k p_k-i\lambda_m e^{i\phi} b-i\lambda_c a,
\label{eq:dpk}
\end{equation}
and therefore,
\begin{align}
p_k(t)&=e^{-i\omega_k(t-t_0)} p_k(t_0) \nonumber \\
&-i \int_{t_0}^t\left(\lambda_me^{i\phi}{b}+\lambda_c a\right)e^{-i\omega_k(t-t^\prime)}dt^\prime,
\label{eq:pk}
\end{align}
where $p_k(t_0)$ is the initial state of the traveling-wave at $t_0 < t$.  Combining Eqs. \eqref{eq:Ht} and \eqref{eq:pk}, and taking $\lambda_{c,m}$ independent of $k$ in the first Markov approximation, the quantum Langevin equations for the two modes are
\begin{subequations}
\begin{align}
\dot{a}&=-i\widetilde{\omega}_c a-2\pi(\lambda_c^2 a+\lambda_m \lambda_c e^{i\phi} b)-i\sqrt{2\pi}\lambda_c p_\text{in} \nonumber  \\
\dot{b}&=-i\widetilde{\omega}_m b-2\pi(\lambda_m^2 b+\lambda_m\lambda_c e^{i\phi} a)-i\sqrt{2\pi}\lambda_m e^{i(\phi+\theta)}p_\text{in},
\label{eq:travelingME1}
\end{align}
\end{subequations}
where
\begin{equation}
p_\text{in}(t)=\frac{1}{\sqrt{2\pi}}\int e^{-i\omega_k(t-t_0)} p_k(t_0)dk
\end{equation}
is the input field through the transmission line. In addition to the traveling phase $\phi$, the phase of the input field will shift after passing through a resonance.  This is characterized by the resonance phase $\theta$, i.e., $\theta=0$ for $\omega_k\ll\omega_c$, $\theta=90^\circ$ at $\omega_k=\omega_c$, and $\theta=180^\circ$ for $\omega_k\gg\omega_c$.  Redefining the extrinsic damping rates $\kappa_c = 2\pi\lambda_c^2$ and $\kappa_m= 2\pi\lambda_m^2$ and the complex frequencies to include the extrinsic damping, $\widetilde{\omega}_c = \omega_c - i \beta_L = \omega_c - i \left(\beta + \kappa_c\right)$ and $\widetilde{\omega}_m = \omega_m - i \alpha_L = \omega_m - i \left(\alpha + \kappa_m\right)$, the equations of motion for the cavity and magnon modes are
\begin{subequations}
\label{eq:travelingME2}
\begin{align}
\dot{a} &= - i \widetilde{\omega}_c a - \sqrt{\kappa_c \kappa_m} e^{i \phi} b - i \sqrt{\kappa_c} p_\text{in} \label{eq:travelingME2a}\\
\dot{b} &= - i \widetilde{\omega}_m b - \sqrt{\kappa_c \kappa_m} e^{i \phi} a - i \sqrt{\kappa_m} e^{i\left(\phi+\theta\right)}p_\text{in}.
\end{align}
\end{subequations}
Therefore, the cavity-magnon coupling occurs in two ways: (1) An indirect traveling-wave-mediated interaction with coupling strength $\Gamma = \sqrt{\kappa_c\kappa_m}e^{i\phi}$, which indicates that the traveling phase will influence the coherent/dissipative nature of the interaction; and (2) A direct independent driving of the cavity and magnon modes, by $\sqrt{\kappa_c}p_\text{in}$ and $\sqrt{\kappa_m}e^{i(\phi+\theta)}p_\text{in}$, respectively, which results in two-tone interference. \cite{Boventer2019, Rao2020}

Analogous to \hyperref[Sec:appendixE]{Appendix E}, the traveling-wave equation of motion can also be solved by defining the output state at $t_1 > t$,
\begin{equation}
p_\text{out}(t)=\frac{1}{\sqrt{2\pi}}\int e^{-i\omega_k(t-t_1)}{p}_k(t_1)dk.
\label{eq:pout}
\end{equation}
Using $p_\text{out}$ to determine the time reversed Langevin equations and combining with Eq. \eqref{eq:travelingME2a} yields the input-output relation,
\begin{equation}
\label{eq:travelingInputOutput}
p_\text{out} = p_\text{in} - i \sqrt{\kappa_c} a -i \sqrt{\kappa_m}e^{i\left(\phi+\theta\right)} b.
\end{equation}
Finally, combining Eqs. \eqref{eq:travelingME2} and \eqref{eq:travelingInputOutput} leads to the transmission spectrum of Eq. \eqref{eq:travelingSpectrum}.
\end{appendix}

\bibliographystyle{apsrev4-1}
\bibliography{mainText.bbl}

\end{document}